\documentclass{jfm}

\usepackage{graphicx}
\usepackage{epstopdf, epsfig}
\usepackage{amsmath}
\usepackage{amssymb}
\usepackage{color}
\usepackage{subfigure}
\usepackage{url}

\makeatletter
\let\saved@includegraphics\includegraphics
\AtBeginDocument{\let\includegraphics\saved@includegraphics}
\renewenvironment*{figure}{\@float{figure}}{\end@float}
\makeatother

\usepackage[makeroom]{cancel}
\usepackage{qtree}
\usepackage[normalem]{ulem}

\usepackage[resetlabels]{multibib}
\newcites{Supp}{References}

\renewcommand{\div}{\nabla \cdot}

{\vskip 2pt \begin{compactitem}[#1]\setlength{\itemsep}{2pt}}
{\end{compactitem}\vskip 2pt}

\newcommand{\be}{\begin{equation}}
\newcommand{\ee}{\end{equation}} 
\newcommand{\lb}{\label}
\newcommand{\OL}{\overline}
\newcommand{\wh}{\widehat}

\newcommand{\ba}{{\bf a}}

\newcommand{\bk}{{\bf k}}

\newcommand{\br}{{\bf r}}
\newcommand{\bu}{{\bf u}}

\newcommand{\bx}{{\bf x}}

\newcommand{\bS}{{\bf S}}

\newcommand{\wt}{\widetilde}

\newcommand{\grad}{{\mbox{\boldmath $\nabla$}}}

\shorttitle{Scales in RT turbulence}
\shortauthor{D. Zhao, R. Betti and H. Aluie}

\title{Scale interactions and anisotropy in Rayleigh-Taylor turbulence}

\author{Dongxiao Zhao\aff{1,3}
  ,
  Riccardo Betti\aff{1,2}
 \and Hussein Aluie\aff{1}
 \corresp{\email{hussein@rochester.edu}}}

\affiliation{\aff{1}Department of Mechanical Engineering and Laboratory for Laser Energetics,
University of Rochester, Rochester, New York 14627, USA
\aff{2}Department of Physics, University of Rochester, Rochester, NY 14627, USA
\aff{3} UM-SJTU Joint Institute, Shanghai Jiao Tong University, Shanghai 200240, China}

\begin{document}

\maketitle

\begin{abstract}
We study energy scale-transfer in Rayleigh-Taylor (RT) flows by coarse-graining in physical space without Fourier transforms, allowing scale analysis along  vertical direction. Two processes are responsible for kinetic energy flux across scales: baropycnal work $\Lambda$, due to large-scale pressure gradients acting on small-scales of density and velocity, and deformation work $\Pi$, due to multi-scale velocity. Our coarse-graining analysis shows how these fluxes exhibit self-similar evolution that is quadratic-in-time, similar to RT mixing layer. We find that $\Lambda$ is a conduit for potential energy, transferring energy non-locally from the largest scales to smaller scales in the inertial range where $\Pi$ takes over. In 3D, $\Pi$ continues a persistent cascade to smaller scales, whereas in 2D $\Pi$ re-channels the energy back to larger scales despite the lack of vorticity conservation in 2D variable density flows. This gives rise to a positive feedback loop in 2D-RT (absent in 3D) in which mixing layer growth and the associated potential energy release are enhanced relative to 3D, explaining the oft-observed larger $\alpha$ values in 2D simulations. Despite higher bulk kinetic energy levels in 2D, small inertial scales are weaker than in 3D. 
Moreover, the net upscale cascade in 2D tends to isotropize the large-scale flow, in stark contrast to 3D. Our findings indicate the absence of net upscale energy transfer in 3D-RT as is often claimed; growth of large-scale bubbles and spikes is not due to ``mergers'' but solely due to baropycnal work $\Lambda$.
\end{abstract}


\section{Introduction}
The Rayleigh-Taylor (RT) instability occurs when a heavy fluid is on top of a light fluid in the presence of gravity, or more generally when a light fluid is accelerated against a heavy fluid \citep{Rayleigh83, Taylor50}. It is of both fundamental interest and practical importance in many fields and applications. For example, RT instability plays a leading role in the propagation of the thermonuclear flame front in supernova explosions  \citep{Supernova00}, and  is a main obstacle towards achieving ignition in inertial confinement fusion (ICF)  \citep{Betti16}.

Any reasonable person would probably agree that modeling such RT unstable 3D systems using 2D simulations is inferior to using 3D simulations. Yet, what aspects of the dynamics may be misrepresented in a 2D simulation? How might such misrepresentation affect the overall modeled system and conclusions? Answers to these questions are not readily clear, which perhaps explains why 2D-RT modeling 
still being relied upon due to its lower computational costs. This holds to varying degrees in the fluids literature \citep[e.g.][]{wei2017novel,zhang2018weakly,bestehorn2020rayleigh}, the astrophysics literature \citep[e.g.][]{Kotakeetal2018,just2018core,chen2020gas}, among several others. For example, in plasma physics research, 2D simulations are still the main ``work horse'' for experimental design in inertial fusion to this day \citep[e.g.][]{hopkins2018toward,weber2020mixing,anderson2020effect,meaney2020carbon,carpenter2020temperature}.
ICF modelers aim to predict or postdict results from shots (\textit{i.e.} experiments) in the laboratory. Given finite computational resources, the choice is often 
to include more physical processes in simulations at the expense of conducting them in 2D, despite the recognition that 3D modeling is preferable. Such decisions are often made in large part because the trade-offs between 2D and 3D flow physics are not as clear as the trade-offs between whether or not to include laser deposition physics, hohlraum physics, or resolve effects such as from the ``tent'' or ``stalk-mount'' \citep{Clarketal16}. A main aim of this work is to highlight the stark differences between 2D- and 3D-RT flow physics.

Based on constant-density incompressible homogeneous 2D turbulence \citep{Kraichnan67}, it is sometimes argued that due to vorticity (or enstrophy) conservation, we should expect a net upscale cascade in 2D-RT \citep[e.g.][]{Cabot06POF}. However, RT flows do not conserve vorticity, even in 2D, due to baroclinicity.  Therefore, extending such an argument \citep{Kraichnan67} from constant-density homogeneous flows to inhomogeneous flows with variable density remains dubious. Why should we expect 2D-RT flows to yield an upscale cascade if the underlying cause for such a phenomenon, namely vorticity conservation, does not hold in variable density flows? One contribution of this work is to demonstrate, via a direct measurement of the cascade, that a net upscale cascade does indeed exist in 2D-RT flows with significant density variation, despite the lack of vorticity conservation. Moreover, we show that such an upscale cascade leads to a spurious positive feedback loop that is absent from 3D-RT.

Even in 3D-RT, there is considerable confusion on whether kinetic energy is transferred upscale or downscale. In many studies discussing RT flows in 3D, there is a pervasive notion that the generation of successively larger bubbles following the presence of smaller ones is due to an upscale cascade of energy which causes small bubbles to ``merge'' into larger ones. This seems to pervade several research areas, such as studies motivated by ICF (e.g. \cite{Shvartsetal1995,Alonetal1995,Oferetal1996,Oronetal2001,Zhou03}), or magnetic fusion \citep[e.g.][]{Douglasetal1998},
or astrophysics (e.g. \cite{Joggerstetal2010,Porthetal2014}), or fundamental fluid dynamics (e.g. \cite{Abarzhi1998,Chengetal2002,Cabot06POF}). According to such a narrative, the  small-scales are the source of energy feeding and sustaining the large-scales, even in 3D. A main result of our paper is measuring the net energy transfer in 3D-RT directly and showing that it is downscale rather than upscale. We show that the emergence of successively larger scales in 3D-RT is driven not by an upscale cascade but by baropycnal work.

\subsection{Brief Overview of RT Flows}
RT instability driven flows occur when a heavy fluid is accelerated against a light fluid and is manifested by the formation of upward rising bubbles of the light fluid and downward sinking spikes of the heavy fluid. 
There are several valuable reviews on the topic \citep{Sharp84, Kull91,Abarzhi10b, Zhou17-1, Zhou17-2,Boffetta17, Livescu20}.

RT flows are usually categorized either as \emph{single-mode},
in which the initial perturbation consists of a single sinusoidal wave spanning the domain, or as \emph{multi-mode}, in which the initial perturbation has a broad spectrum.
During the early linear stage of evolution, single-mode RT grows exponentially before nonlinear effects set in and its growth rate plateaus \citep{Layzer55,Goncharov02}, then undergoes further acceleration \citep{Tie12} due to vorticity production \citep{Ramaprabhu12,Bian20}.
In contrast, multi-mode RT is expected to evolve as a set of uncoupled modes during the linear stage before mode-coupling sets in due to 
nonlinear interactions between different scales \citep{Dimonte04}.
Multi-mode RT becomes turbulent with a wide range of scales, characterized by a quadratic-in-time self-similar growth of the mixing layer between the heavy and light fluids \citep[e.g.][]{Ristorcelli04}.


The self-similar stage is dominated by growth of the largest structures present at any given time. The self-similar dynamics has been divided into two  categories, depending on the initial perturbation spectrum \citep{Glimm88,Shvartsetal1995,Dimonte04,Dimonte05}: (i) when long-wavelength modes grow directly from perturbations present at the initial time. This category is sometimes called ``bubble competition,'' and is believed to be sensitive to initial conditions \citep{Alphagroup,Zhou17-1}. (ii) when long-wavelength modes are absent at the initial time and grow from the interaction of short-wavelength perturbations present initially. This self-similar category is often called ``bubble merger,'' and is believed to be insensitive to initial conditions \citep[e.g.][]{Ramaprabhu05}. In both categories, growth of the RT mixing layer between the heavy and light fluids is quadratic-in-time $\sim t^2$ of the mixing layer, but differing in the growth proportionality constant, $\alpha$. Whether describing single-mode or multi-mode RT, the primary focus of past studies has been the RT growth rates at different times \citep[e.g.][]{Alphagroup}, which is of foremost importance in applications such as fusion. Our focus here is rather different and aimed at understanding the energy pathways across scales.



\subsection{Energy Scale-Transfer in Boussinesq RT Turbulence}
RT turbulent flows are unsteady, inhomogeneous, and anisotropic, and it is unclear that the traditional cascade picture from Kolmogorov's theory holds \citep{Alexakis18,zhou2021turbulence}. A motivation of this paper is to measure the cascade in RT flows directly, without having to rely on traditional tools from turbulence theory, such as Fourier analysis or structure functions, which rest on assumptions that do not hold in RT turbulence.

Several previous works \citep{Zhou01, Chertkov03, Ristorcelli04, Boffetta09, Zhou16, Livescu10} 
have made valuable contributions to addressing this problem, the majority of which focused on low Atwood number incompressible RT flows described by the Boussinesq model. In this model,  density variation is negligible except in the buoyancy term of the momentum equation, and the associated velocity field is solenoidal.

An important phenomenological theory on Boussinesq RT turbulence was proposed by \citet{Chertkov03}, but had to assume a (quasistationary) constant kinetic energy flux to small scales and a balance between buoyancy and the time derivative of velocity.
Subsequently, \citet{Boffetta09} carried out 3D Boussinesq RT simulations and analyzed the energy budget equations in Fourier space, where a Fourier transform was applied in the homogeneous (horizontal) directions. Simulations of 2D Boussinesq RT flows were also analyzed using both Fourier analysis \citep{Boffetta12} and a filtering approach in physical space \citep{Zhou16} similar to what we do here for the more general (non-Boussinesq) problem. \citet{Boffetta12} and \citet{Zhou16} both concluded that there is an inverse energy cascade and a forward enstrophy cascade in 2D Boussinesq RT flows.

Moreover, \citet{Boffetta12} studied
the 3D to 2D transition in high aspect ratio RT simulations, with a domain size $L_x = 32 L_y=\frac{1}{2}L_z$, and inferred the presence of a transition scale. Using Fourier analysis in the homogeneous directions, \citet{Boffetta12} concluded that the flow has 2D-like evolution at (horizontal) scales larger than the transition scale and seems to follow a Bolgiano-Obukhov phenomenology, characterized by an inverse energy cascade, while smaller scales evolve as in 3D following the Kolmogorov-Obukhov phenomenology, characterized by a forward energy cascade.

These studies have established a self-consistent picture of phenomenological Boussinesq RT turbulence. Note that the Boussinesq description of RT fails in the presence of large density ratios, 
which restricts its application to a relatively limited class of flows. Generalizing  this framework to the full RT problem is non-trivial \citep{Livescu07} and our paper may be viewed as a contribution towards that goal.

\subsection{Variable Density Flows: Which ``Length-scale''?}
Many flows which can exhibit RT instability, ranging from astrophysics to ICF to multiphase flow applications, are characterized by large density ratios, $O(10)-O(10^{20})$ \citep{Zhao18} for which the Boussinesq description is invalid.
In these applications, RT flows have to be investigated in a variable density (VD) flow setting, where the kinetic energy is a non-quadratic quantity due to the explicit density dependence, unlike in Boussinesq flows where kinetic energy is quadratic. 

A ``length-scale'' is not an independent entity but is associated  with the specific flow variable being analyzed \citep{Zhao18}. This is often lost in our discussion of the `inertial range' or the 'viscous range' of length-scales in turbulence as if they exist independently of a flow variable, which in constant density (and Boussinesq) turbulence is the velocity field \citep{Kolmogorov41}. Therefore, in constant-density turbulence, when discussing inertial \emph{scales} or a $k^{-5/3}$ scaling of the energy spectrum, $k$ is associated with the velocity field. What is the corresponding variable whose scales are appropriate to analyze in VD turbulence?

In constant density (and Boussinesq) turbulence, the velocity spectrum and the energy spectrum are the same. This is no longer the case in variable density turbulence \citep{Sadek18}. Moreover, in variable density flows there are many possible ways to construct a ``spectrum'' in $k$-space having units of energy per wavenumber. These different ``spectra'' correspond to the Fourier transform (or scale decomposition) of different variables and are not equivalent. Which of these is meaningful, if any, remains unclear. 

\citet{Zhou01} proposed an RT turbulence (and also of the related Richtmyer–Meshkov instability) phenomenology and predicted a power-law scaling for the energy spectrum, although it was not specified which variable's length-scales are to be analyzed. \citet{CookZhou02} introduced a new variable $\boldsymbol{w} \equiv \rho^{1/2} \bu$ in RT flows, borrowed from \citet{kida1990energy},
where $\rho$ is density and $\bu$ is the fluid velocity, effectively treating kinetic energy as a quadratic term. \citet{CookZhou02} Fourier analyzed $\boldsymbol{w}$ in the homogeneous directions from 3D-RT simulations and found that the net energy transfer is from large to small (horizontal) scales, but with significant backscatter in $k$-space. A follow-up work by \citet{Cabot04} also analyzed energy transfer in Fourier space across scales associated with $\boldsymbol{w}=\rho^{1/2} \bu$ in the homogeneous (horizontal) directions.

Several previous works in variable density flow \citep{Alphagroup,Cabot06, LawrieDalziel2011JFM} studied the budgets of kinetic, potential, and internal energy, but these were only for the bulk flow and not as functions of scale as we do in this paper. Other studies of RT turbulence within the full variable density (non-Boussinesq) flow setting \citep{Livescu07, Schilling10} performed Reynolds averaging on DNS data, and analyzed the time evolution of the ensemble mean kinetic energy across the mixing width, but did not analyze energy transfer across \emph{length-scales}. 

The above works on energy transfer in RT were either restricted to analyzing scales in the the homogeneous directions via Fourier analysis, or relied on statistical (Reynolds or Favre) averaging procedures. The former lacks information on the energy transfer in the vertical (inhomogeneous) direction, which is essential in the RT dynamics. The latter lacks information about the dynamics (including the cascades) as a function of scale. 

A series of past studies \citep{Aluie11c,Aluie12,Aluie13,EyinkDrivas17a,Zhao18} developed the coarse-graining approach \citep{Eyink05,EyinkAluie09} for variable-density flows to probe the energy scale pathways in inhomogeneous and anisotropic systems. The coarse-graining formalism is the same as that used in large-eddy simulation (LES) modeling of turbulent flows. However, while the coarse-grained equations analyzed in these studies coincide (to a considerable extent) with those employed in LES, their use was for rather different purposes. In LES, plausible but uncontrolled closures are adopted for the subscale terms, whereas the primary aim of coarse-graining as used in the aforementioned studies is to develop several exact estimates and a general physical understanding 
of the subscale terms. Another difference is that LES generally takes the length-scale parameter
$\ell$ to be a fixed length of the order of the ``integral scale'' $L$. The primary interest in coarse-graining is rather to probe \emph{all scales in the flow}, including limits of small $\ell\ll L$.
While our paper was undergoing journal revisions, an important work toward this effort by \citet{Saenzetal2021filtering} was published, which aims to connect coarse-graining with Reynolds averaging in 3D homogeneous variable density flows, and showed that such flows undergo a downscale cascade.

Despite the important work done previously, we are still lacking a thorough understanding of the RT instability dynamics at different scales, including the cascades, anisotropy, and inhomogeneity. Addressing these gaps is a main thrust of our paper. In addition,  notions of the cascade and the inertial range that we will address below lay a theoretical foundation for the practical development of LES for RT flows.

\subsection{Outline of this Paper}
The paper is organized as follows. In section \ref{sec:stat_results}, we describe high resolution turbulent RT simulations in 2D and 3D,
and discuss basic flow properties. In section \ref{sec:detailed_balance}, we analyze the energy pathways across scales, including the kinetic energy cascades, and coupling to potential and internal energy. 
In section \ref{sec:split_direction}, we examine the anisotropy of RT cascades in physical space. We summarize our findings and conclude the paper in section \ref{sec:conclude}. 
We include two appendices which show that our results are insensitive to the compressibility levels, either due to the initial setup (Appendix \ref{sec:append_2DRT_Ma}) or due to the governing equations by repeating our analysis to data from the two-species incompressible RT model (Appendix \ref{sec:2-species}). A separate document found on the journal's website contains Supplementary Material on the i) efficiency of energy fluxes (Supplementary Material, section \ref{suppsec:flux_efficiency}), ii) anisotropy of energy fluxes obtained with anisotropic kernels (Supplementary Material, section \ref{suppsec:scaleanisotropy}), iii) filtering spectra for different scale-decompositions of kinetic energy (Supplementary Material, section \ref{suppsec:append_spectra}), iv) insensitivity of our results to initial conditions (Supplementary Material, section \ref{suppsec:append_2DRT}), and v) discussion of another formulation of the energy budget commonly used in the LES literature (Supplementary Material, section \ref{suppsec:pathway_comparison}).


\section{Preliminaries and Simulations} \label{sec:stat_results}
\subsection{Governing equations} \label{sec:governing_eq}
In this paper we primarily focus on what we call the single-species two-density RT model, which is governed by the fully compressible Navier-Stokes equations. In Appendix \ref{sec:2-species}, we repeat our main analysis on the more widely used two-species incompressible Navier-Stokes model \citep{Cook01,Livescu07} for comparison.

The dynamics is governed by the continuity equation (\ref{eq:continuity}), the momentum equation (\ref{eq:momentum}), and the total energy equation (\ref{eq:total}). The internal (\ref{eq:internal}) and kinetic energy (\ref{eq:kinetic}) equations
are also shown for convenience.  
\begin{eqnarray}  \label{eq:basic_equations}
    &\partial_t\rho+\partial_j(\rho u_j)=0 \label{eq:continuity}\\
    &\partial_t(\rho u_i)+\partial_j(\rho u_i u_j)=-\partial_i P+\partial_j\sigma_{ij}+\rho F_i \label{eq:momentum}\\
     &\partial_t (\rho E) + \partial_j (\rho E u_j)=-\partial_j (Pu_j)+\partial_j \left[2\mu u_i \left(S_{ij}-\frac{1}{d} S_{kk}\delta_{ij}\right)\right]\nonumber \\
    &-\partial_j q_j +\rho u_i F_i \label{eq:total}\\
    &\partial_t (\rho e)+\partial_j(\rho e u_j)=-P\partial_j u_j+2\mu\left(|S_{ij}|^2-\frac{1}{d}|S_{kk}|^2\right)-\partial_j q_j \label{eq:internal} \\
&\partial_t\left(\rho \frac{|\boldsymbol{u}|^2}{2}\right)+\partial_j \left\{\left(\rho \frac{|\boldsymbol{u}|^2}{2}+P\right)u_j-2\mu\left(u_i S_{ij}-\frac{1}{d}u_j S_{kk}\right)\right\}= \nonumber \\ 
    &P\partial_j u_j-2\mu\left(|S_{ij}|^2-\frac{1}{d}|S_{kk}|^2\right)+\rho u_i F_i \label{eq:kinetic} 
\end{eqnarray}
Here, $\rho$ is the density field, $\bu$ is the velocity field, $P$ is pressure, $e$ is the specific internal energy (or  internal energy per unit mass),
$E=e+|\bu|^2/2$ is total energy per unit mass. $S_{ij}=(\partial_i u_j+\partial_j u_i)/2$ is the strain rate tensor,
while $\sigma_{ij}=2\mu (S_{ij}-d^{-1}S_{kk}\delta_{ij})$ is the deviatoric viscous stress tensor in $d$ space dimensions, where $\mu$ is dynamic viscosity. 
$\bf{F}$ is the external acceleration, $\boldsymbol{q}$
is the heat flux satisfying Fourier's law $\boldsymbol{q}=-\kappa \nabla T$, where $\kappa$ is thermal conductivity and $T$ is temperature.
The ideal gas law $P=\rho R T$ is adopted to close the set of equations, with $R$ the specific gas constant.

The governing equations are solved in 2D and 3D rectangular domains in Cartesian coordinates. The boundary conditions are periodic in the horizontal directions and no-slip rigid walls in the vertical $z$-direction. 
Following \citet{Bian20}, the density is initialized such that $\rho_0(z)$ is uniform in the $z$-direction on either side of the interface. Hydrostatic equilibrium requires that away from the interface, the initial (background) pressure varies as, $P_0\sim \rho_0 g z$. For our single fluid case here, using the ideal gas equation of state yields that the background temperature gradient is constant and equal on both sides of the interface. Thus, the initial conditions represent a particular case of the analysis of \citet{gerashchenko2016}, with $d T_0/dz= -g/R$. Away from the interface, when thermal conductivity coefficient is constant as in our case here, a constant temperature gradient implies that the heat conduction term vanishes in the energy equation, so that the initial conditions are also in thermal equilibrium.

In this work, we choose viscosity $\mu$ to be spatially constant. In most practical applications, $\mu$ varies with temperature, sometimes strongly such as in ICF. Previous work by \citet{Aluie13} and \citet{Zhao18} has shown that the effects of a variable viscosity are still guaranteed to be localized to the smallest scales if the Reynolds number remains sufficiently high, thereby allowing for an inertial range of scales to exist, which is the main focus of this paper. We also choose the Prandtl number $Pr=c_p \mu/\kappa=1$, where again $\kappa$ is spatially constant, which is often not the case, especially in ICF \citep[e.g.][]{Huasen18}. Our choices are similar to those adopted in previous studies of the fundamental hydrodynamics of RT \citep[e.g.][]{Cabot06,reckinger2016comprehensive,Wielandetal2019}. They are meant to simplify the problem to focus on the fundamental study of the cascades in RT flows, although we acknowledge ongoing  research on the potentially important roles of temperature dependent viscosity and heat diffusion in some applications \citep[e.g.][]{Weberetal2014}.

\subsection{Numerical method} \label{sec:Numerical_method}
The pseudo-spectral
method \citep{Patterson71} is adopted in the periodic directions, in which spatial derivatives and time integration are performed in Fourier space, while nonlinear terms are calculated in physical space. The fast Fourier transform (FFT) provides an efficient way to convert the fields between physical and Fourier spaces.
Aliasing errors in nonlinear terms are removed by the 2/3 dealiasing rule \citep{Patterson71}.
For the vertical direction with wall boundaries, we use a compact finite difference scheme that is 6th order inside the domain and 5th order at the boundaries \citep{Lele92,Brady19}. In addition, the compact filtering scheme is performed in the inhomogeneous direction to filter out unphysical high wavenumber 
oscillations \citep{Lele92,Brady19}. Fourth order explicit Runge-Kutta method is used to integrate in time.

At the interface, as density changes between the heavy and light regions, the temperature gradient can no longer be constant. The energy equation then has non-zero time derivative at the initial time,
$\partial_t (\rho e) = \partial_j (\kappa \partial_j T_0)$,
which results in the generation of acoustic waves at initial time that can reflect off the top and bottom walls. We have carried controlled tests by including a sponge layer at the top and bottom boundaries to damp these transient waves \citep{Bian20}, and could not detect any effect on the RT evolution. Even without the sponge layer, these initial acoustic waves are quickly damped at very early times, long before the flow becomes turbulent.

\subsection{Analysis method} \label{sec:Coarse-graining}
In fluid dynamics, coarse-graining provides a natural and versatile framework to understand scale 
interactions \citep{Leonard75,MeneveauKatz00,Eyink05}.  For any field $\ba(\bx)$, a coarse-grained or (low-pass) filtered field,
which contains modes at scales $\gtrsim\ell$, is defined in $n$-dimensions as
\begin{align}
\OL \ba_\ell(\bx) = \int d^n\br~ G_\ell(\br)\, \ba(\bx+\br),
\label{eq:filtering}
\end{align}
where $G(\br)$ is a normalized convolution kernel and $G_\ell(\br)= \ell^{-n} G(\br/\ell)$ is a 
dilated kernel with width comparable to $\ell$.
The scale decomposition in (\ref{eq:filtering}) decomposes the field $\ba(\bx)$ into a large scale ($\gtrsim\ell$) component $\OL \ba_\ell$,
and a small scale ($\lesssim\ell$) component, captured by the residual $\ba'_\ell=\ba-\OL \ba_\ell$.
The coarse-graining framework is a general method to analyze unsteady, anisotropic, and inhomogeneous flows,
such as the Rayleigh-Taylor instabilities studied here, which conventional turbulence analysis techniques are unable to handle.
More extensive discussions of the framework and its utility can be found in many references 
\citep{Piomellietal91,Germano92,Meneveau94,Eyink95prl,AluieEyink09,Riveraetal14,FangOuellette16}.
In what follows, the subscript $\ell$ may be dropped from filtered variables when there is no risk of ambiguity.

Here, we use a Gaussian filter kernel at scale $\ell$ \citep{Piomellietal91}:
\begin{align} \label{eq:Gaussian_kernel_physical}
    G_\ell({|\bx|})=\left(\frac{6}{\pi \ell^2}\right)^{n/2}e^{-\frac{6}{\ell^2}|\bx|^2}.
\end{align}
in dimensions $n=1,2,3$. Its Fourier transform is:
\begin{align} \label{eq:Gaussian_kernel_Fourier}
    \widehat{G}_\ell({|\bk|})=e^{-\frac{\ell^2}{24}|\bk|^2}.
\end{align}
\begin{figure}
\centering 
\begin{minipage}[b]{1.0\textwidth}  
\centering
\subfigure
{\includegraphics[height=1.8in]{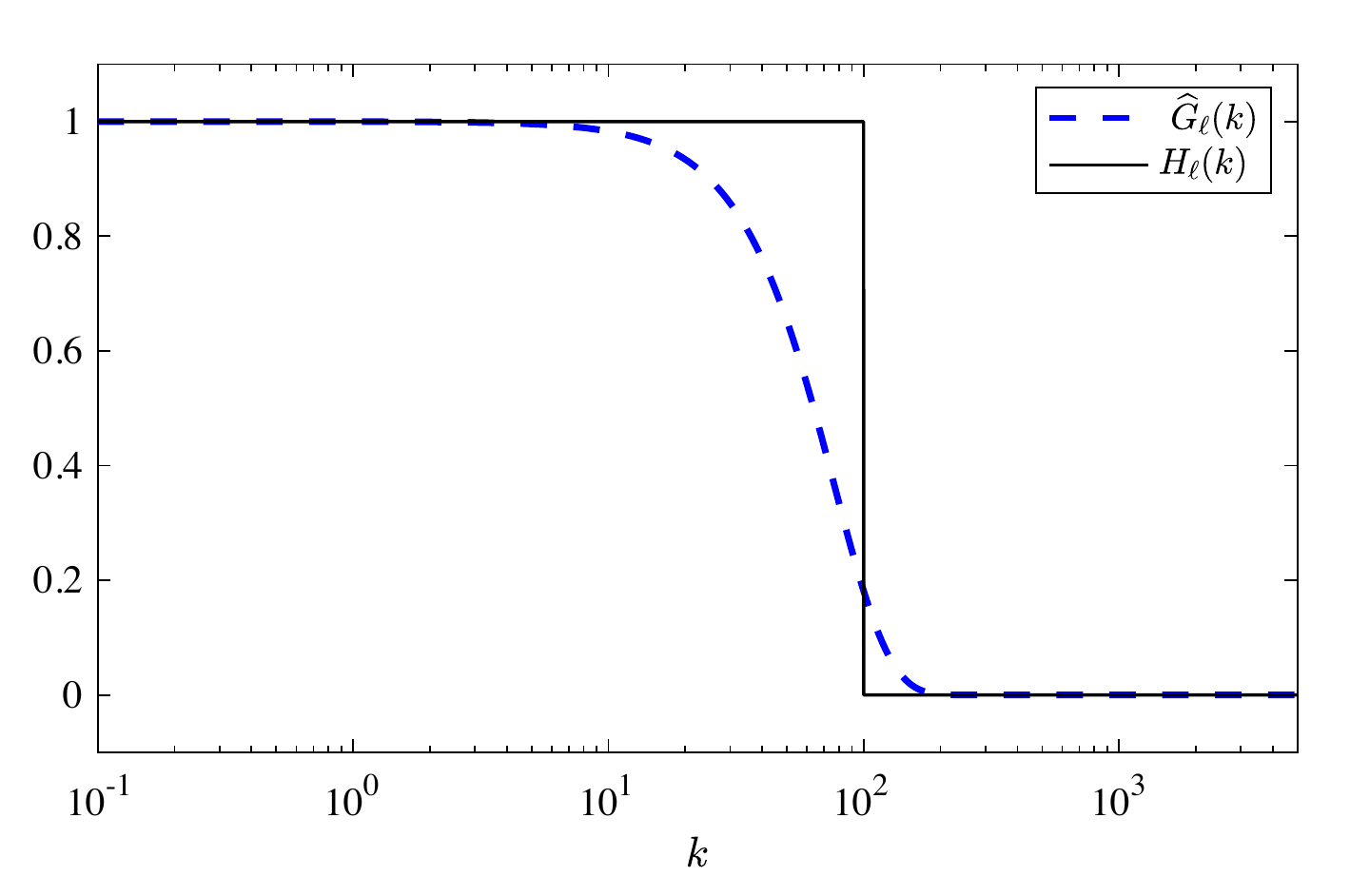}} 
\caption{\footnotesize{The Gaussian kernel $\widehat{G}_\ell({|\bk|})$ in equation \eqref{eq:Gaussian_kernel_Fourier} compared to a sharp-spectral cut-off, ${H}_\ell({|\bk|})$, in Fourier space, with cutoff wavenumber $k_\ell=100$ and $\ell = L_z/k_\ell=6.4/100$. The horizontal axis is in logarithmic scale.} \label{fig:GaussianKernel}}
\end{minipage}
\end{figure}

Figure \ref{fig:GaussianKernel} plots the Gaussian kernel in $k$-space and compares it to sharp-spectral truncation. Note that a Fourier analysis is not possible in the non-homogeneous (vertical) direction. A detailed discussion on the different kernels can be found in \cite{Riveraetal14} (see Figures 12, 13 in that paper and associated discussion). The Gaussian kernel form in equation \eqref{eq:Gaussian_kernel_physical} has been used in several prior studies \citep{Piomellietal91,Wangetal18} due to its numerical discretization advantages (see \citet{John12}, page 30).

In simulations with non-periodic boundary conditions, such as our RT flow with rigid walls at the top and bottom,
filtering near the boundary requires a choice for the fields beyond the boundary.
In this paper, the choice is to extend the domain beyond the physical boundaries with values compatible with the boundary 
conditions \citep{Aluieetal2018_jpo}. Filtering near the wall is then performed on this extended data. For our RT problem specifically, the velocity is kept zero beyond the walls, the density field is kept constant (zero normal gradient), and the extended pressure field
satisfies the hydrostatic condition $dP/dz=-\rho g$. The coarse-graining operation can then be performed at every point in the flow domain 
without requiring complicated inhomogeneous filters, which do not commute with spatial derivatives.

Analyzing the energy cascade and inertial range dynamics in constant-density turbulence centers on studying the large scale kinetic energy (KE), $\frac{1}{2}|\OL{\bu}_\ell|^2$. Since KE in variable density flows, $\frac{1}{2}\rho|\bu|^2$, is non-quadratic, its large scale definition is not as straightforward. Several different definitions have been proposed in the literature, including $\OL{\rho}_\ell|\OL{\bu}_\ell|^2/2$ \citep{chassaing1985alternative,BodonyLele05,Burton11,KarimiGirimaji17}, and $|\OL{(\sqrt{\rho}\bu)}_\ell|^2/2$ \citep{kida1990energy,CookZhou02,Wangetal18}. However, as was shown in \citet{Zhao18}, the above two definitions can fail to satisfy the `inviscid criterion' in flows with large density variations, leading to viscous contamination at all scales, and preventing the unravelling of inertial range dynamics. The `inviscid criterion' requires that viscous contributions be negligible at large scales. \citet{Aluie13} proved mathematically, and \citet{Zhao18} showed numerically, that the Favre decomposition, $\frac{1}{2}\OL{\rho}_\ell|\wt{\bu}_\ell|^2$, where 
\be\wt{\bu}=\OL{\rho\bu}_\ell/\OL{\rho}_\ell
\ee 
is the Favre filtered velocity \citep{favre1958further,Aluie13,EyinkDrivas17a}, satisfies the inviscid criterion even in the presence of arbitrarily large density variations. In the current paper, we shall adopt the Favre decomposition when analyzing scale dynamics.

\subsection{Simulation results}
\begin{table}
  \begin{center}
\def~{\hphantom{0}}
  \begin{tabular}{lcccccccccc}
    Label & Grid size & $L_x$ & $L_y$ & $L_z$ &$\mu$ & $\mathcal{A}$ &$Gr$ &$Re$ &$\eta/\Delta x$ & $Ma$ \\
3D1024 & $1024\times 1024 \times 2048$  & 3.2 & 3.2 & 6.4 & $3\times 10^{-5}$ & 0.5 & 15.07 & 13854 & 0.70 & 0.45 \\ 
3D512high & $512\times 512 \times 1024$  & 1.6 & 1.6 & 3.2 & $4.5\times 10^{-5}$ & 0.8 & 7.44 & 4582 & 0.80 & 0.28 \\ 
3D512low & $512\times 512 \times 1024$  & 3.2 & 3.2 & 6.4 & $4.5\times 10^{-5}$ & 0.15 & 27.38 & 8027 & 0.49 & 0.55 \\ 
2D4096 & $4096\times 8192$ & 3.2 & -- & 6.4 & $1.5\times 10^{-5}$ & 0.5 & 0.94 & 44562 & 2.25 & 0.45 \\
2D2048high & $2048 \times 4096$  & 3.2  & --  & 6.4 & $4.5\times 10^{-5}$ & 0.8 & 0.93 & 25766 & 2.36 & 0.39  \\ 
2D2048low & $2048 \times 4096$  & 3.2 & -- &  6.4 & $4.5\times 10^{-5}$ & 0.15 & 0.43 & 12397 & 2.99  & 0.55\\
2D2048 & $2048 \times 4096$ & 3.2 & -- &  6.4 & $1.5\times 10^{-5}$ & 0.5 & 7.53 & 44052 & 0.52 & 0.45 \\
2D1024 & $1024\times 2048$ & 3.2 & -- &  6.4 & $1.5\times 10^{-5}$ & 0.5  & 60.3 & 38002 & 0.28 & 0.45 \\
2D1024lowM & $1024\times 2048$ & 3.2 & -- &  6.4 & $1.5\times 10^{-5}$ & 0.5  & 59.38 & 41949 & 0.26 & 0.13 
  \end{tabular}
  \caption{Parameters of the 2D and 3D RT simulations conducted in this paper. All simulations have a spatially uniform dynamic viscosity $\mu$ 
    and Prandtl number $Pr=c_p \mu/\kappa=1$, with thermal conductivity $\kappa$ and specific heat capacity $c_p$ at constant pressure.
      $L_x,L_y,L_z$ are the domain lengths in three directions, and $\mathcal{A}$ is the Atwood number.
    Gravitational acceleration $g=1$ in the $-z$ direction. The mesh Grashof number is $Gr=2\mathcal{A}g\langle \rho\rangle^2 \Delta x^3/\mu^2$,
       the Kolmogorov scale is  $\eta=\mu^{3/4}/(\epsilon^{1/4}\langle\rho\rangle ^{3/4})$, and the Reynolds number is $Re=\langle |\bu|^2\rangle^{1/2} L_x\langle\rho\rangle/\mu$,
      where $\langle\cdot\rangle$ denotes the spatial mean value. $Ma=\sqrt{g L_x}/\sqrt{\gamma R T_0}$ denotes the interface Mach number \citep{reckinger2016comprehensive}, where $R$ is the specific gas constant, $\gamma$ is the ratio of specific heats, and $T_0$ is the initial temperature at the interface.   In the table, the Reynolds number and the Kolmogorov scale are 
    calculated at dimensionless time $\widehat{t}=t/\sqrt{\frac{L_x}{\mathcal{A}g}}=4.0$.}
  \label{tab:parameter}
  \end{center}
\end{table}

Two- and three- dimensional Rayleigh-Taylor simulations are performed to analyze the energy budget and cascade in RT turbulence.
These simulations, summarized in Table \ref{tab:parameter} are performed at low and high Atwood numbers, with grid resolution up to $4096\times 8192$ in 2D and $1024\times 1024\times 2048$ in 3D.
The Atwood number $\mathcal{A}$ is defined by $\mathcal{A}=(\rho_h-\rho_l)/(\rho_h+\rho_l)$, where $\rho_h, \rho_l$ are the densities of the initial uniform heavy and light fluids.
The results shown below are primarily based on the largest simulations 2D4096 and 3D1024,
while other simulations shown in the table are for numerical convergence and verification purposes.

\begin{figure}
\centering 
\begin{minipage}[b]{1.0\textwidth}  
\subfigure[\footnotesize{2D4096 snapshot}]
{\includegraphics[height=2.8in]{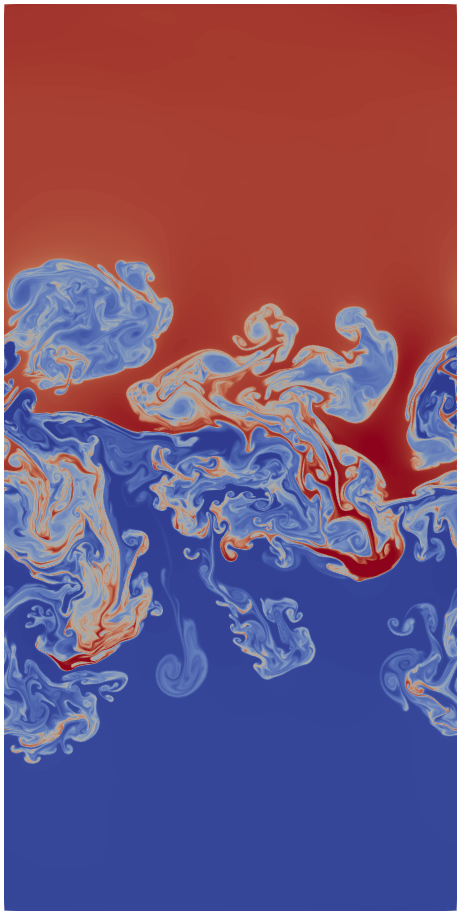} \label{fig:visual_2D}} 
\subfigure[\footnotesize{3D1024 snapshot}]
{\includegraphics[height=2.8in]{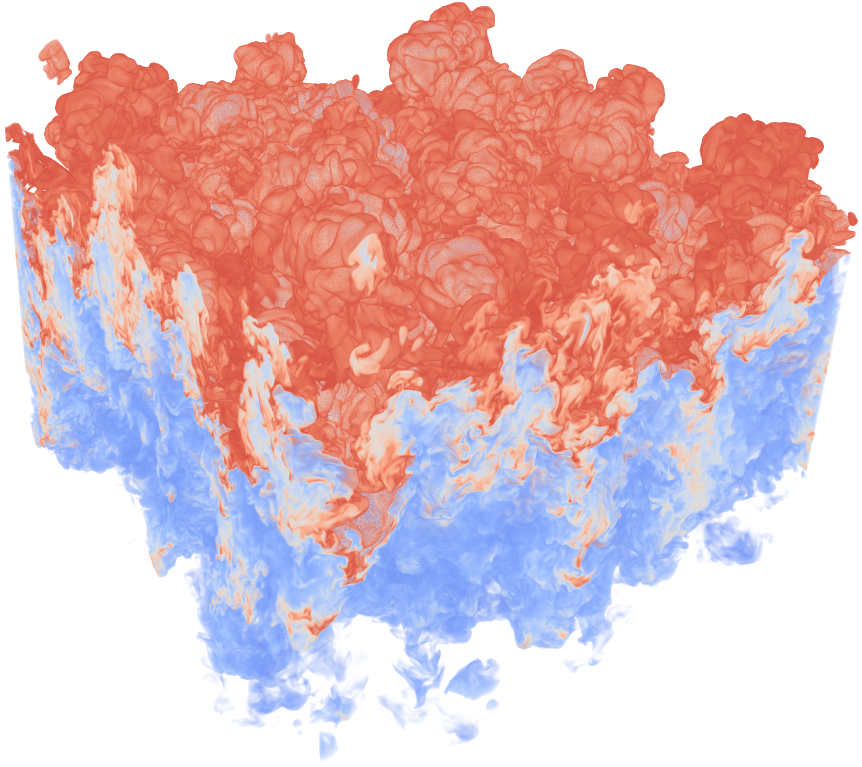}\label{fig:visual_3D}} 
\caption{\footnotesize{Density visualizations from the simulations 2D4096 and 3D1024 in Table \ref{tab:parameter}. The dimensionless time for both is
    $\widehat{t} = t/\sqrt{\frac{L_x}{\mathcal{A}g}}= 4.0$. The mixing width, which is defined and analyzed in section \ref{sec:mixing_width},
    is $h\approx 1.69$ for 2D-RT, and $h\approx 1.1$ for 3D-RT. The density fields are turbulent with the presence of a wide range of scales.} \label{fig:visual_2D3D}}
\end{minipage}
\end{figure}

The initial conditions for simulations shown in Table \ref{tab:parameter} are as follows. Initial density field is composed of uniform heavy fluid on top with $\rho_h=1$, and uniform light fluid with density $\rho_l$ at bottom. The initial pressure field satisfies hydrostatic balance $dP / dz = -\rho g$, with pressure at bottom set to be 10 for all simulations in Table \ref{tab:parameter}, except 2D1024lowM, where pressure is set to 100 at the bottom to yield lower compressibility levels. In Appendices \ref{sec:append_2DRT_Ma}, we show that our results do not depend on the compressibility level of RT flows. The velocity fields are initially zero with perturbations added to the interface of vertical velocity. The interfacial velocity perturbations are imposed in wavenumber space, and are nonzero only within a band of high wavenumbers. In 3D, the velocity amplitude is perturbed in a shell $k\in [32, 128]$, and the magnitude is set to be proportional to $e^{-\frac{1}{c}|k_x^2+k_y^2-80^2|}$, where $c = 22.63$ is a normalization constant to further limit the range of effectively perturbed wavenumber. While in 2D, the velocity amplitude is perturbed in an annulus $k\in [72, 128]$, with magnitude proportional to $e^{-\frac{1}{c}|k_x^2-100^2|}$, where $c$ is the same constant as in 3D. Since the perturbation wavenumber band is slightly different between 2D4096 and 3D1024, for verification purpose, we also perform additional 2D simulations 2D1024 and 2D2048 (see Table \ref{tab:parameter}) with initial perturbations identical to those in 3D1024. Analysis results based on these 2D simulations are  similar to the 2D4096 case, and are shown in the Supplementary Material, section \ref{suppsec:append_2DRT}.

Visualizations of density snapshots are shown in figure \ref{fig:visual_2D} for the 2D4096 case and figure \ref{fig:visual_3D} for the 3D1024 case,
both at dimensionless time $\widehat{t} = t/\sqrt{\frac{L_x}{\mathcal{A}g}}= 4.0$, where $\sqrt{\frac{L_x}{\mathcal{A}g}}$ is the characteristic time scale, and $g$ is the gravitational
acceleration. Both figures show complex flow patterns and structures
with a wide spectrum of sizes.
It is also easy to see that the 3D field develops much finer scale structures than the 2D field, a property which we will quantify using spectra below. We now present some basic properties of our simulations.

\subsection{Mixing width} \label{sec:mixing_width}
The mixing width $h(t)$ is an important quantity characterizing the total width of the mixing layer between bubble and spike fronts in RT flows.
It is used to quantify the penetration depth of RT bubbles and spikes into the unperturbed flow.
When there are sufficient multimode interactions, turbulent RT flows become self-similar at later times \citep{Birkhoff1955,Ristorcelli04,Alphagroup}. During the self-similar stage, the mixing width is well-described by $h(t)=\alpha \mathcal{A} g t^2$,
a quadratic growth in time which is proportional to a constant coefficient $\alpha$. 

The value for $\alpha$ has been the subject of vigorous research and discussion within the community. It is believed to depend on whether initial conditions render the RT flow in the bubble competition regime or in the bubble merger regime \citep[e.g.][]{Ramaprabhu05}. Numerical simulations of 3D-RT in the bubble merger regime generally report values $\alpha\approx 0.02$~to~$0.03$ (see Table 6.1 in \citet{Zhou17-1} for a summary), although values larger than $0.05$ have been reported in 3D simulations using front tracking \citep[e.g.][]{Glimmetal2013}. Experiments on the other hand generally report much larger values $\alpha\approx 0.05$~to~$0.06$ \citep{Zhou17-1}, which is thought to be a result of unintended large-scale perturbations \citep{Alphagroup}.

While we report $\alpha$ values from our simulations here, its precise value is far from being the focus of our paper. Our interest in $\alpha$ here is rather more limited to comparing its values from 2D simulations relative to 3D simulation within the same modeling framework.
For reasons that remain unclear and subject to various speculations, the parameter $\alpha$ in 2D simulations has often been found larger than in 3D simulations  \citep[e.g.][]{Youngs91POF,Youngs94,Alphagroup,Cabot06POF}. 
Such enhanced RT growth in 2D simulations has been somewhat of a puzzle given that both analytical models and simulations of single-mode RT predict that it is growth in 3D that should be faster \citep{Layzer55,Goncharov02,Bian20}. A goal of this paper is to shed light on the puzzle.

We calculate the mixing width using the formula \citep{Cabot06,Huasen19}:
$$h(t)=\frac{2}{\rho_h-\rho_l}\int_{-\infty}^\infty \text{min}\left(\langle \rho\rangle_{xy}(z)-\rho_l, \rho_h-\langle \rho\rangle_{xy}(z)\right) dz$$
where $\langle \rho \rangle_{xy}(z)$ denotes a horizontal average of the density field, which is a function of $z$ and $t$, and $\rho_h, \rho_l$ denote the original heavy and light fluid densities.
\begin{figure}
\centering 
\begin{minipage}[b]{1.0\textwidth}  
\centering
\subfigure[\footnotesize{2D mixing width}]
{\includegraphics[height=1.8in]{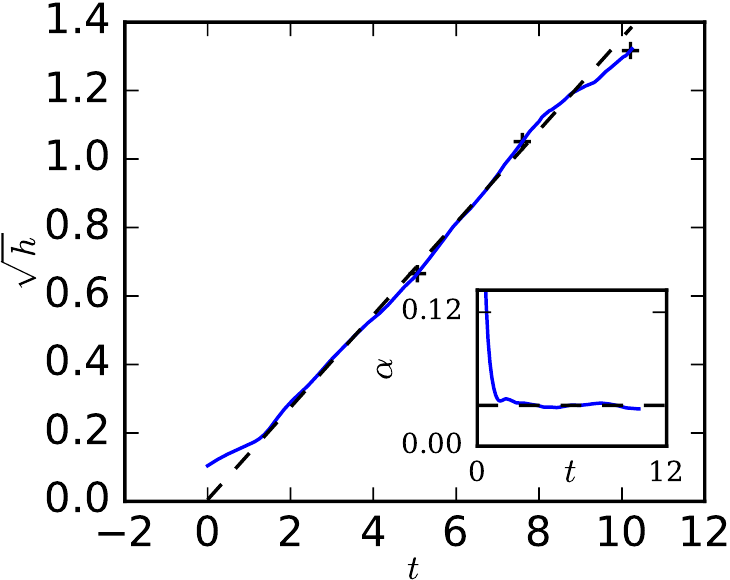} \label{fig:mixingwidth2d}} 
\phantom{a}
\subfigure[\footnotesize{3D mixing width}]
{\includegraphics[height=1.8in]{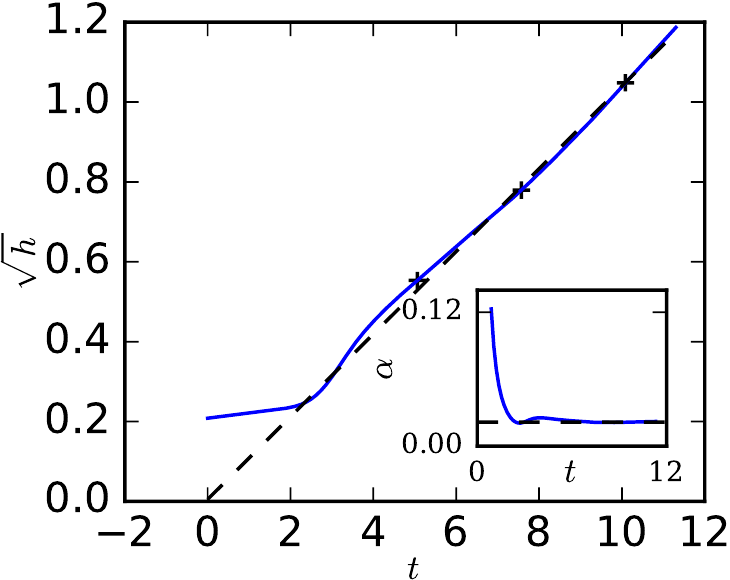} \label{fig:mixingwidth3d}} 
\caption{\footnotesize{Evolution of the square root of mixing width $\sqrt{h(t)}$ vs. time in 2D and 3D, for which self-similarity analysis predicts a linear relation 
    at late time \citep{Olson09}. The slope of the straight line is 0.135 in (a)  and 0.104 in (b),
    which corresponds to $\alpha=0.036$  and $\alpha=0.021$, respectively. The `+' markers in both figures correspond to dimensionless time $\widehat{t}=t/\sqrt{\frac{L_x}{\mathcal{A}g}}= 2, 3, 4$. Inset: the compensated plot $\alpha=h(t)/(\mathcal{A}gt^2)$ versus time,
    in which the horizontal lines correspond to the $\alpha$ values obtained by the linear fit.} \label{fig:mixingwidth}}
\end{minipage}
\end{figure}
After obtaining $h(t)$, $\alpha$ can be calculated using one of two methods. The first is by directly calculating $h(t)/\mathcal{A}gt^2$, while the second is by carrying a best linear 
fit between $\sqrt{h(t)}$ and $\sqrt{\mathcal{A}g}t$ in the self-similar regime \citep{Olson09}.
Figure \ref{fig:mixingwidth} plots $\sqrt{h(t)}$ versus time from the 
2D4096 and 3D1024 data, and a linear fit in the self-similar regime is marked with a dashed line. The inset of the figures corresponds to the first approach: $\alpha=h/\mathcal{A}gt^2$ is plotted against time, while the $\alpha$ value obtained by the linear fit is marked with a horizontal dashed line. We obtain the same results from
these two methods.  Figure \ref{fig:mixingwidth} shows that there is a substantial time range of self-similarity in both 2D and 3D,
in which there is an excellent linear fit between $\sqrt{h(t)}$ and $t$. 
The values of $\alpha$ we obtain (0.036 in 2D and 0.021 in 3D) fall within the range reported from simulations in the literature \citep{Cabot06,Alphagroup}. Note that $\alpha$ from the 2D-RT is almost twice as large as that from 3D-RT.

\subsection{Overall energy balance}

\begin{figure}
\centering 
\begin{minipage}[b]{1.0\textwidth}  
\centering
\subfigure[\footnotesize{2D instantaneous budget}]
{\includegraphics[height=1.8in]{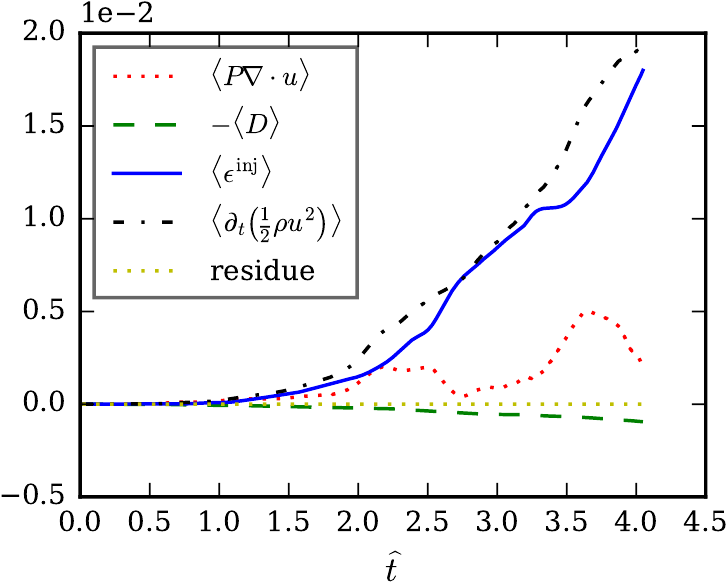} \label{fig:ke_detailed_balance_2d}} 
\subfigure[\footnotesize{2D overall budget}]
{\includegraphics[height=1.8in]{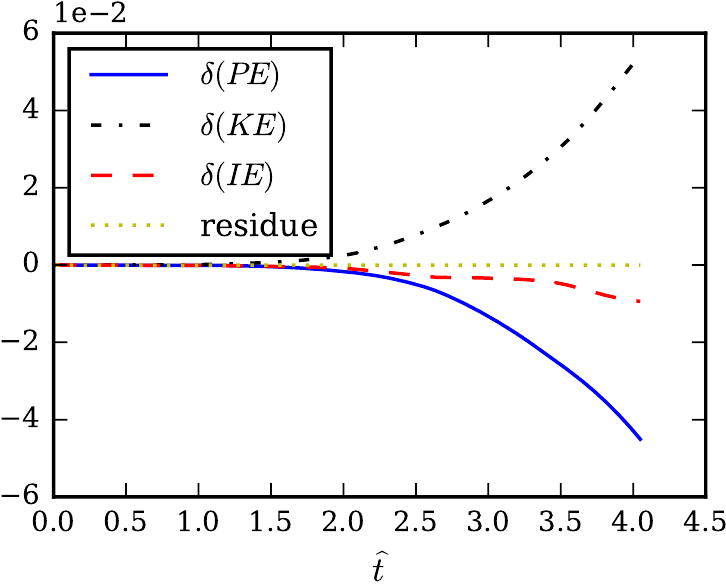} \label{fig:ke_budge_2d}}  \\
\subfigure[\footnotesize{3D instantaneous budget}]
{\includegraphics[height=1.8in]{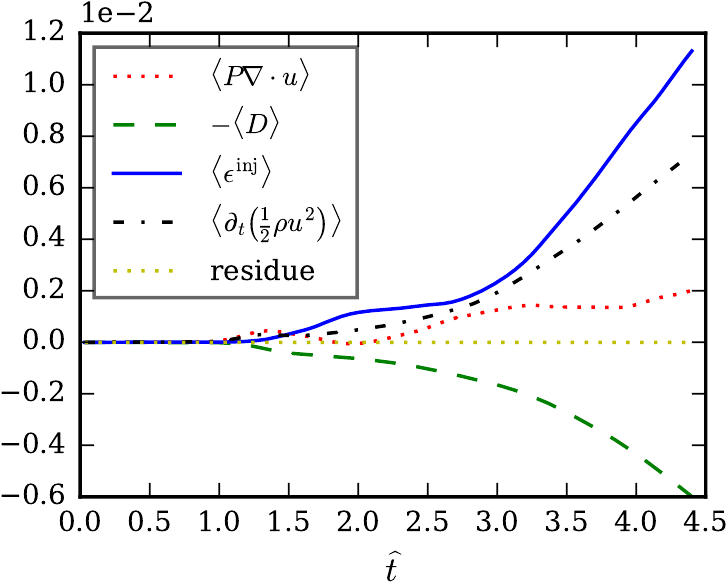} \label{fig:ke_detailed_balance}} 
\subfigure[\footnotesize{3D overall budget}]
{\includegraphics[height=1.8in]{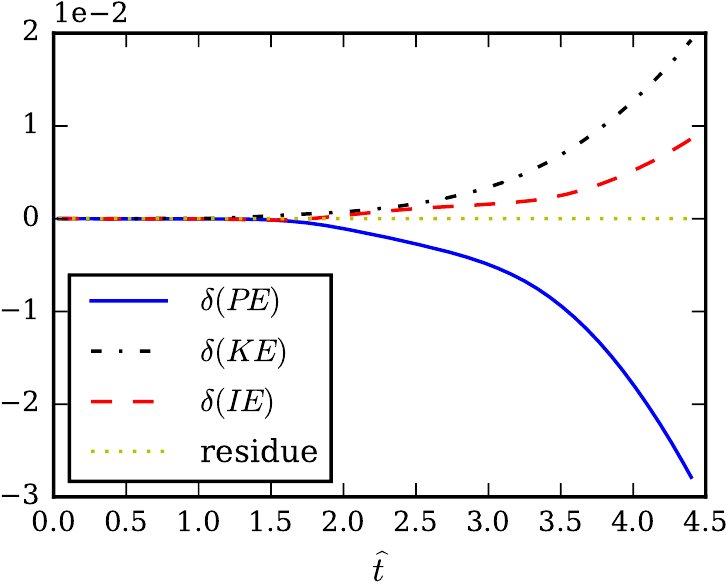} \label{fig:ke_budget}} 
\caption{\footnotesize{Temporal evolution of terms in the kinetic energy budget, and overall energy balance in 2D (a),(b) and 3D (c),(d). $D=-\sigma_{ij} S_{ij}$
    is the kinetic energy dissipation rate, $\epsilon^{\textrm{inj}}=\rho u_i g_i$ is the energy injection rate. Time $\widehat{t}= t/\sqrt{\frac{L_x}{\mathcal{A}g}}$ in the figures is dimensionless.} \label{fig:full_ke_budget}}
\end{minipage}
\end{figure}

We shall now discuss the bulk balance between potential energy, kinetic energy, and internal energy in our simulated flows. Temporal evolution of the spatial mean
of each of the budget terms in the kinetic energy equation (\ref{eq:kinetic}) is shown in figures \ref{fig:ke_detailed_balance_2d} and \ref{fig:ke_detailed_balance} for
2D and 3D, where $D=S_{ij} \left( 2\mu S_{ij}-\frac{2}{d} \mu S_{kk} \delta_{ij}\right)$ is the dissipation term,
and $\epsilon^{\textrm{inj}}=-\rho g u_z$ is the energy injection.
Correspondingly, figures \ref{fig:ke_budge_2d} and \ref{fig:ke_budget} show the temporal evolution of bulk (over volume $V$ of the domain)
potential energy $\delta \text{PE}$,  bulk internal energy $\delta \text{IE}$, and  bulk kinetic energy $\delta \text{KE}$, in 2D and 3D.
The bulk energy is measured relative to its initial value:
\begin{eqnarray}
&\delta \text{PE}(t)&=\int [\rho(\bx,t)-\rho(\bx,0)]gzdV= -\int_0^t dt \int dV \epsilon^{\textrm{inj}} \lb{eq:PE_change} \\
&\delta \text{IE}(t) &= \int [\rho e(\bx,t)-\rho e(\bx,0)] dV=\int_0^t dt \int dV (D-P\div \bu) \lb{eq:IE_change} \\
&\delta \text{KE}(t) &=\int [\frac{1}{2}\rho \bu^2(\bx,t)-\frac{1}{2}\rho \bu^2(\bx,0)]dV=\int_0^t dt\int dV \partial_t (\frac{1}{2}\rho |\bu|^2) \lb{eq:KE_change}
\end{eqnarray}
The second equality in equation (\ref{eq:PE_change}) is due to the relation $\int_V \rho g u_z dV = \frac{d}{dt}\int_V \rho g z dV$ \citep{Cabot06}, while equation (\ref{eq:IE_change})
can be obtained by time integrating the internal energy equation (\ref{eq:internal}).

Plot of 2D instantaneous energy budget terms in figure \ref{fig:ke_detailed_balance_2d} shows the time derivative of kinetic energy and 
the energy injection rate keep increasing in time, while the average value of $P\nabla\cdot \bu$ is positive so that  
internal energy is converted to kinetic energy via this term.  The overall viscous dissipation term $-\langle D\rangle$ is negative and converts kinetic energy to internal, but the
amount of conversion is smaller in magnitude relative to $\langle P \nabla\cdot \bu\rangle$ in 2D. Thus, in figure \ref{fig:ke_budge_2d} both the overall internal energy and potential energy decrease and are converted to kinetic energy. The overall bulk energy balance satisfies $\delta \text{PE}+\delta \text{KE} + \delta \text{IE} = 0$, as is marked
by the `residue' term in the figure.

The corresponding 3D results are shown in figures \ref{fig:ke_detailed_balance} and \ref{fig:ke_budget}.
The injection rate and time derivative of KE keeps increasing, while the dissipation and pressure dilatation terms convert between
kinetic and internal energy. One difference from 2D is that in 3D, the dissipation term contributes more than $P\nabla\cdot \bu$, and the overall energy conversion is from kinetic to internal,
opposite to the 2D case. Total energy conservation, $\delta \text{PE}+\delta \text{KE} + \delta \text{IE} = 0$, still holds in 3D. This conservation holds in all simulations in Table \ref{tab:parameter}. In section \ref{sec:detailed_balance}, we will present a more refined balance 
relation applied to the coarse-grained quantities, where the balance of energy at different length scales is established.

The rate-governing quantities in our figure \ref{fig:full_ke_budget} seem more sensitive to fluctuations around the self-similar trajectory than the mixing width $h(t)$ in figure \ref{fig:mixingwidth}, which has often been the main focus of previous RT studies. The fluctuations may be either due to intrinsic variability or due to a transient response to initial conditions, the removal of which has been recently discussed in \citet{HorneLawrie2020}. That figure \ref{fig:appendix_ke_detailed_balance_2d_lowMa} in the Appendix shows similar fluctuations under different initial and compressibility conditions suggests it may be the former, yet a careful investigation of this observation is beyond the scope of our paper. Nevertheless, we shall show below that energy transfer across scales exhibit  clear self-similar evolution despite the presence of these fluctuations in the rate-governing quantities.

One may notice that mean injection, $\langle\epsilon^{\textrm{inj}}\rangle \equiv -\langle\rho g u_z\rangle$, shown in figures \ref{fig:ke_detailed_balance_2d} and \ref{fig:ke_detailed_balance},
grows more rapidly in 2D than in 3D. There are two (related) ways to understand this: (i) the mixing width growth rate $\alpha$
in 2D is larger than in 3D and, thus, the mixing layer width should be larger in 2D than in 3D at any physical time.
This implies that relatively more potential energy is released in 2D due to the rising light fluid and sinking heavy fluid, leading to a higher mean injection $\langle\epsilon^{\textrm{inj}}\rangle$.
(ii) From the balance relation $\delta \text{PE}+\delta \text{KE} + \delta \text{IE} = 0$, we have $\delta \text{IE}<0$ in 2D, which is converted to kinetic energy, while $\delta \text{IE}>0$ in 3D.
This implies a larger mean kinetic energy $\delta \text{KE}$ is expected in 2D, and thus $u_z$ is expected to be larger in 2D than in 3D.
This leads to a larger $\langle\epsilon^{\textrm{inj}}\rangle = -\langle\rho g u_z\rangle$ in 2D.
We shall show below that there is a deeper physical explanation behind these differences between 2D and 3D RT arising from differences in the energy pathways across scales.
Given the difference in $\langle\epsilon^{\textrm{inj}}\rangle$ between the RT flows in 2D and 3D, we shall normalize by it to allow for a fair comparison of the budget terms between 2D and 3D.

\subsection{`Filtering' spectra} \label{sec:rt_spectra}
For completeness, we now present the spectra of density, velocity, and kinetic energy using our RT data. In homogeneous constant-density flows, the energy spectrum is obtained by taking the
Fourier transform of the velocity field and summing over shells in wavenumber space: $E(k) = \frac{1}{2}\Sigma_{k-0.5 < |\bk| \leq k+0.5}|\hat{u}(\bk)|^2$.
Obtaining the spectra for variable density RT flows is difficult because i) the domain is not periodic in the vertical direction so the Fourier transform is not straightforward, and ii) the non-quadratic nonlinearity of kinetic energy poses another problem.

Previous works in RT \citep{CookZhou02,Cabot04} obtained the spectra for density and velocity by 
Fourier transforming in the horizontal periodic directions and then taking an average in the vertical direction. By taking Fourier transforms only in the horizontal directions, spectral information in the vertical direction is completely ignored. 

Unlike in constant density turbulence, where the spectrum of energy is the same as that of velocity, in variable density the two quantities are different. Moreover, when kinetic energy is non-quadratic, there are many possible ways to construct a ``spectrum'' in $k$-space having units of energy per wavenumber. These different ``energy spectra'' correspond to the Fourier transform (or scale decomposition) of different combinations of variables and are not equivalent. To avoid the difficulty pertaining to its non-quadratic nature, previous studies constructed a kinetic energy spectrum by introducing a new variable $w_i=\sqrt{\rho}u_i$ and analyzing it in Fourier space, $\wh{w}_i(k_x,k_y)$. However, as emphasized in \citet{Zhao18}, the evolution of $w_i$ is fundamentally different from the evolution of momentum, $(\rho u_i)$, especially when each of these variables is being decomposed in scale. A length-scale $\sim 1/k$ does not exists on its own but is associated with the variable being decomposed.
\citet{Zhao18} showed that this new variable $w_i=\sqrt{\rho}u_i$ fails to satisfy the `inviscid criterion' in 
variable density flows, which means that dissipation can contaminate its large scale dynamics in the presence of significant density variations.
Thus, $\sqrt{\rho}u_i$ is not an appropriate quantity to analyze because it is inconsistent with inertial-range dynamics. Moreover, as was observed in \citet{Zhao18}, a putative power-law scaling of a quantity alone, such as $|\widehat{\sqrt{\rho}\bu}|^2(k)$, is not indicative of an inertial cascade (see Supplementary Material, section \ref{suppsec:append_spectra}).

Here we adopt a new approach to obtain the spectra using  
spatial coarse-graining. For any field $\bf{u}$, the \emph{filtering spectrum} is defined as \citep{Sadek18}:
\begin{align} \label{eq:nonperiodic_spectra}
E_{\bf{u}}(k_\ell)\equiv \frac{d}{dk_\ell}\langle |\OL{\bu}_\ell(\bx)|^2\rangle /2
\end{align}
where $k_\ell=L/\ell$, $L$ is the domain size of interest, $\ell$ is the scale we are probing, and $\langle\cdot\rangle$ stands for spatial averaging. 
The filtering spectrum of kinetic energy is
\begin{align}E_{\text{KE}}(k_{\ell})\equiv \frac{d}{dk_\ell}\langle \OL\rho_\ell|\wt{\bu}_\ell(\bx)|^2\rangle /2
\end{align}
\citet{Sadek18} showed that this filtering spectrum guarantees energy conservation, similar to the Plancherel theorem when working with Fourier spectra. The filtering spectrum allows us to measure the energy content
at scale $\ell$ by probing the change in $\langle \OL\rho_\ell|\wt{\bu}_\ell(\bx)|^2\rangle /2$ as $\ell$ is varied.
The calculation can be done in physical space without any Fourier transforms. The obvious advantage is that definition \eqref{eq:nonperiodic_spectra} allows us to extract the spectrum in inhomogeneous flows such as RT.
Moreover, the filtering spectrum generalizes easily to non-quadratic quantities, 
such as kinetic energy, $\frac{1}{2}\rho \bu^2$, in variable density flows. The Fourier spectrum, on the other hand, is limited to quadratic quantities and requires treating kinetic energy as such, even when this is inconsistent with the dynamics.


\begin{figure}
\centering 
\begin{minipage}[b]{1.0\textwidth}  
\subfigure[\footnotesize{Density spectrum 2D}]
{\includegraphics[height=1.3in]{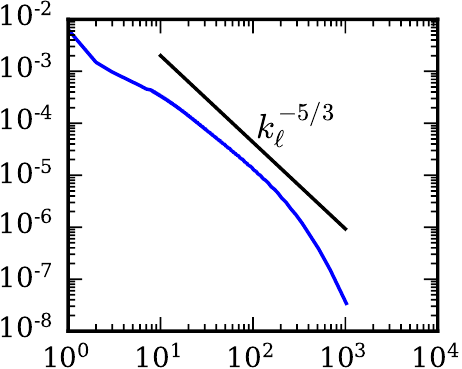} } 
\subfigure[\footnotesize{Velocity spectrum 2D}]
{\includegraphics[height=1.3in]{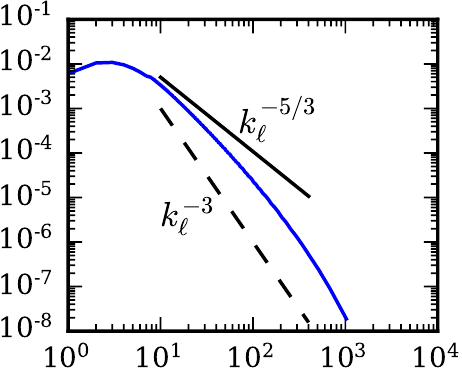} } 
\subfigure[\footnotesize{KE spectrum 2D}]
{\includegraphics[height=1.3in]{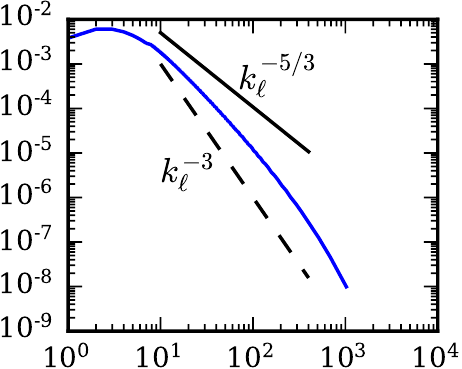} } \\
\subfigure[\footnotesize{Density spectrum 3D}]
{\includegraphics[height=1.3in]{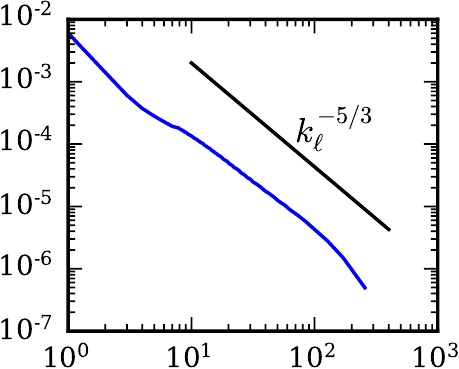} } 
\subfigure[\footnotesize{Velocity spectrum 3D}]
{\includegraphics[height=1.3in]{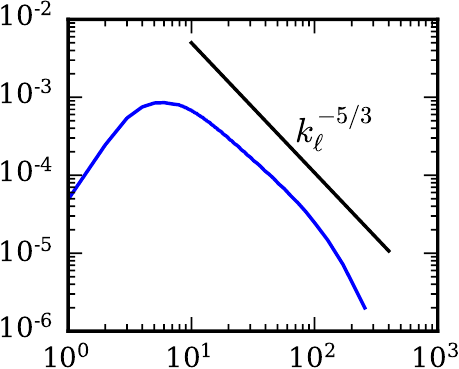} } 
\subfigure[\footnotesize{KE spectrum 3D}]
{\includegraphics[height=1.3in]{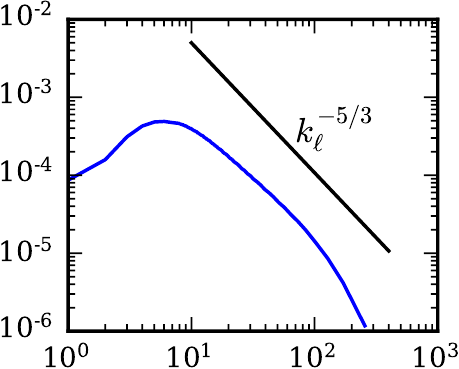} }
    \caption{\footnotesize{Density, velocity and kinetic energy `filtering' spectra versus filtering wavenumber $k_\ell=L_z/\ell$ (horizontal axes) for 2D and 3D RT turbulence at dimensionless time $\widehat{t}=4$, visualized
    in figure \ref{fig:visual_2D3D}. The spectra are calculated using Gaussian filter. A $k_\ell^{-5/3}$ scaling is shown for reference, and for 2D velocity and KE spectra an additional $k_\ell^{-3}$ scaling is shown. Note that $k_\ell$ is not a Fourier wavenumber, just a proxy for scale.} \label{fig:spectra}}
\end{minipage}
\end{figure}

Figure \ref{fig:spectra} shows the spectra of density, velocity, and kinetic energy for 2D and 3D RT flows calculated with the above definition. The density spectra follow
a putative $k_\ell^{-5/3}$ scaling over an intermediate wavenumber range in both 2D and 3D. The velocity spectrum in 2D is steeper than $k_\ell^{-5/3}$ but shallower than $k_\ell^{-3}$ as would have been expected from 2D homogeneous constant-density turbulence \citep[e.g.][]{BoffettaEcke12}. In 3D the filtering spectrum is slightly shallower than a $k_\ell^{-5/3}$ scaling over a decade. The kinetic energy spectra follow similar scalings as the velocity field.
Both velocity and kinetic energy spectra attain a peak value at low wavenumbers, where most of the energy content resides. 
In contrast, the density spectra show the importance of much larger scales due to density variation in the vertical. Indeed, the density jumps between heavy and light fluids is expected to yield a spectrum $\sim k_\ell^{-2}$, which is consistent with the scaling at the largest scales. Plots in figure \ref{fig:spectra} demonstrate the utility of the coarse-graining approach to probe the spectral content in all directions, including scales along the inhomogeneous (vertical) direction, as discussed in \cite{Sadek18}. For completeness, we show in the Supplementary Material section \ref{suppsec:append_spectra} the filtering spectra for three different scale decompositions of kinetic energy.

\section{Multi-scale energy balance in RT flows} \label{sec:detailed_balance}
We now investigate the energy scale pathways of Rayleigh-Taylor flows. We employ the Favre decomposition
as discussed in sections \ref{sec:Coarse-graining} and \ref{sec:rt_spectra} to chart the channels for potential energy, internal energy, and kinetic energy at large and small scales. Energy conversion between different forms and across different length scales is investigated by analyzing the budget terms from
numerical data. 

\subsection{Coarse-grained energy budget equations}\label{sec:coarse-grainedBudgets}

Within the Favre decomposition, the budgets for large and small scale kinetic energy, internal energy, and potential energy are:
\begin{align}
\partial_t\bar{\rho}_\ell \frac{|\widetilde{\boldsymbol{u}}_\ell |^2}{2}+\nabla\cdot \boldsymbol{J}_\ell &=
    -\Pi_\ell-\Lambda_\ell+\bar{P}_\ell \nabla\cdot\bar{\boldsymbol{u}}_\ell-D_\ell+\epsilon_\ell^{\textrm{inj}} \label{eq:KE_budget} \\
\partial_t \frac{\bar{\rho}_\ell \widetilde{\tau}_\ell(u_i, u_i)}{2} + \nabla \cdot \boldsymbol{J}_\ell^{\text{small}} &=
    \Pi_\ell +\Lambda_\ell +\bar{\tau}_\ell(P,\nabla\cdot \bu)-D_\ell^{\text{small}}+\epsilon_\ell^{\text{small}}\label{eq:small_KE_budget} \\
\partial_t \overline{\rho e}_\ell + \nabla \cdot [\overline{\rho e \bu} -\overline{\kappa \nabla T}] &=
    -\bar{P}_\ell \nabla \cdot \bar{\bu}_\ell - \bar{\tau}_\ell(P, \nabla\cdot \bu)+D_\ell^{\text{int}} \label{eq:IE_budget} \\
    \partial_t \OL{\mathcal{P}}_\ell + \nabla\cdot \OL{\mathcal{P} \bu} &= -\epsilon_\ell^{\textrm{inj}} \label{eq:PE_budget}
\end{align}
where $e$ is specific internal energy, $\mathcal{P}=\rho g z$ is potential energy density, $\boldsymbol{J}_\ell$ is 
spatial transport of kinetic energy, $\Pi_\ell$ and $\Lambda_\ell$ are kinetic energy fluxes across scale $\ell$,
$-\bar{P}_\ell\nabla \cdot \bar{\bu}_\ell$ is large scale pressure dilatation, $D_\ell$ is viscous dissipation,
and $\epsilon_\ell^{\textrm{inj}}$ is KE injection (hence, the `inj' superscript) due to gravity at scales larger than $\ell$. These terms are defined as:
\begin{align} 
\begin{split}
& \Pi_\ell(\boldsymbol{x})=-\bar{\rho} \partial_j \widetilde{u}_i \widetilde{\tau}(u_i,u_j); \  
    \Lambda_\ell(\boldsymbol{x})=\frac{1}{\bar{\rho}}\partial_j \bar{P} \bar{\tau}(\rho,u_j);\ 
    \epsilon_\ell^{\textrm{inj}}(\boldsymbol{x})=\widetilde{u}_i\bar{\rho}\widetilde{g}_i; \\
& D_\ell(\boldsymbol{x})=\partial_j \widetilde{u}_i \left[2\overline{\mu S_{ij}}-\frac{2}{d}\overline{\mu S_{kk}}\delta_{ij}\right];
    \   J_j(\boldsymbol{x})=\bar{\rho} \frac{|\widetilde{\boldsymbol{u}}|^2}{2}\widetilde{u}_j+\bar{P}\bar{u}_j+
    \widetilde{u}_i \bar{\rho} \widetilde{\tau}(u_i,u_j)-\widetilde{u}_i\bar{\sigma}_{ij} \\
& D_\ell^{\text{small}}(\bx)=\overline{\partial_j u_i \sigma_{ij}}-\partial_j \widetilde{u}_i \overline{\sigma}_{ij};  \ 
    \epsilon_\ell^{\text{small}}=\bar{\rho}\widetilde{\tau}(u_i,g_i); \  
    D_\ell^{\text{int}}=\overline{\partial_j u_i \sigma_{ij}} = D_\ell + D_\ell^{\text{small}} \\  
&J_j^{\text{small}}(\bx)=\frac{\bar{\rho}\widetilde{\tau}(u_i,u_i)}{2} \widetilde{u}_j+\frac{1}{2} 
    \bar{\rho}\widetilde{\tau}(u_i,u_i,u_j)+\bar{\tau}(P,u_j)-(\overline{u_i\sigma_{ij}}-\widetilde{u}_i\bar{\sigma}_{ij})\\
\end{split} \label{eq:budget_defs}
\end{align}
in which $\tau(\cdot,\cdot)$ is the generalized 2nd order moment \citep{Germano92}: $\wt{\tau}(u_i,u_j)=\wt{u_i u_j}-\wt{u}_i\wt{u}_j$ and 
$\OL{\tau}(\rho,u_j)=\overline{\rho u_j}-\OL{\rho}\ \OL{u}_j$.
Note that our small scale kinetic energy, defined as $\frac{1}{2}\OL\rho(\wt{|\bu|^2} - |\wt{\bu}|^2)$, is sometimes referred to as `subscale' or `subgrid' kinetic energy in the LES literature.

Coarse-grained potential energy density equation (\ref{eq:PE_budget}) can be derived by noting that
\begin{align*}
    \frac{D \mathcal{P}}{Dt} &= \frac{\partial \mathcal{P}}{\partial t} + \bu \cdot \nabla\mathcal{P} \\
    &= g z \frac{\partial\rho}{\partial t} + \rho g u_z + g z \bu\cdot \nabla\rho \\
    &= -\mathcal{P} \nabla\cdot \bu + \rho g u_z
\end{align*}

We analyze the energy pathways by inspecting the budget equations (\ref{eq:KE_budget})-(\ref{eq:PE_budget}) using our simulations in Table \ref{tab:parameter}.
In particular, we shall demonstrate the existence of an inertial range in such flows. Note that we do not formulate budgets for small scale internal energy and potential energy,
since it is unclear if these two quantities cascade \citep{EyinkDrivas17a}.

\subsection{Energy pathways in RT flows}\label{sec:EnergyPathways}
This section presents some of the main results of our paper, which we summarize in Section \ref{sec:EnergyPathwaysSummary} for the reader's convenience. Here, we refine our analysis of the global energy balances in figure \ref{fig:full_ke_budget} at different scales via coarse-grained budgets. The types of energy
involved in the scale-transfer, as shown in equations (\ref{eq:KE_budget})-(\ref{eq:PE_budget}), are  potential energy,  internal energy, large scale kinetic energy, and also small scale kinetic energy. Differences in the energy pathways between 2D and 3D will be emphasized during our presentation.

\subsubsection{Conversion between different forms of energy}

\begin{figure}
\centerline{\includegraphics[height=2.2in]{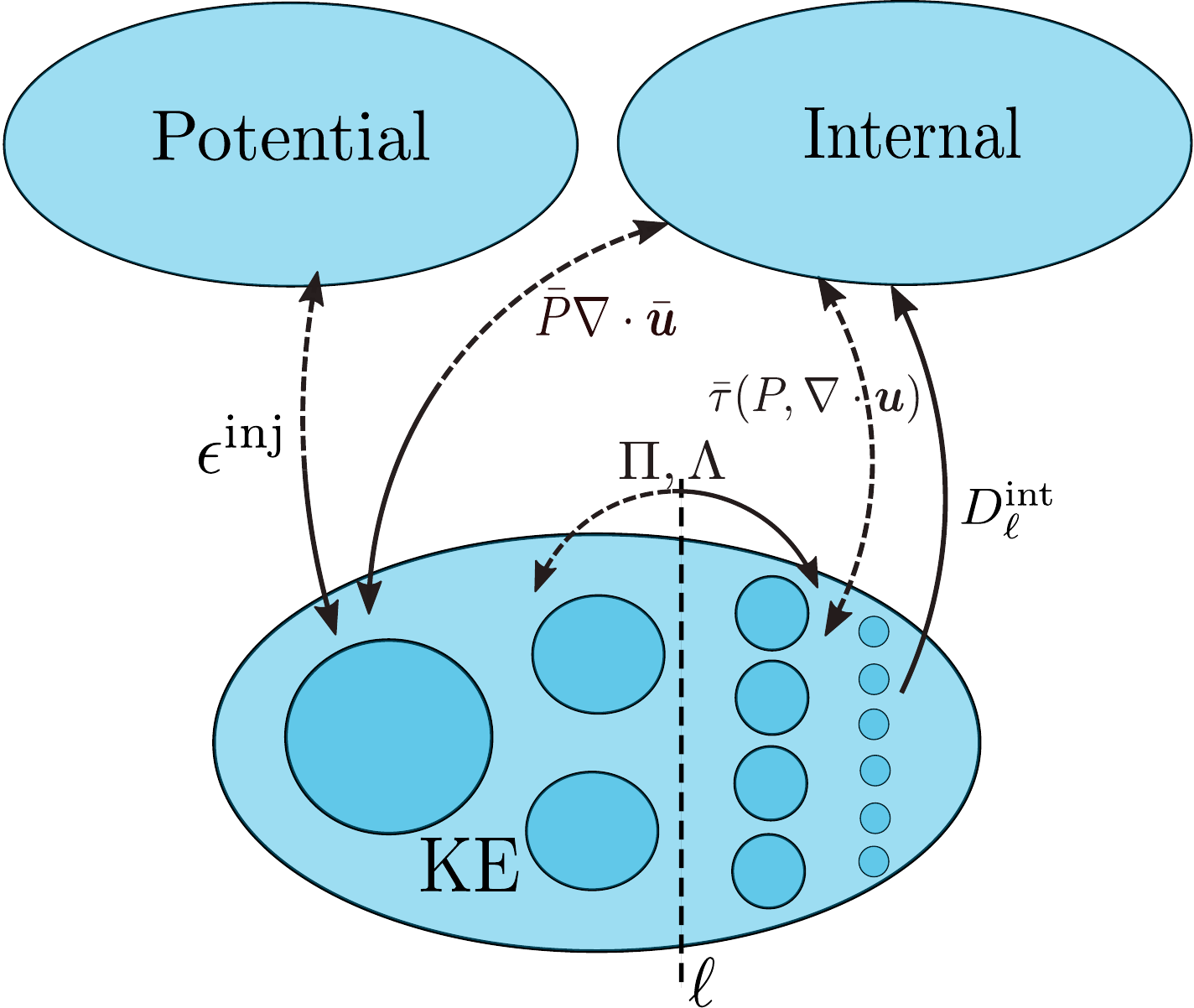} } 
\caption{\footnotesize{Schematic of energy pathways in RT flows between reserviors of kinetic energy (KE), potential energy (PE), and internal energy. 
The flow is further decomposed into its constituent scales within the KE reservoir, where a vertical dashed line denotes an arbitrary inertial scale $\ell$ we are probing. \emph{Kinematically possible} pathways are depicted with dashed arrows, while those that are \emph{dynamically manifested} in 3D-RT are depicted with solid arrows (see sections \ref{sec:ke_inertial}-\ref{sec:ke_fluxes}).  Gravitational PE is converted into KE at the largest scale by $\epsilon^{\textrm{inj}}$. This energy is then transferred to inertial scales by fluxes $\Lambda$ and $\Pi$.
On average, $\Lambda$ acts as a conduit for PE, transferring KE to smaller inertial scales in both 2D and 3D. On average, $\Pi$ transfers energy downscale in 3D and upscale in 2D. KE is coupled to internal energy at largest scales via pressure dilatation, and at the smallest viscous scales via dissipation.}\label{fig:combined_energy_pathway}}
\end{figure}

Figure \ref{fig:combined_energy_pathway} summarizes the RT energy pathways we discuss in this section. In RT flows, the ultimate energy source is the release of potential 
energy from the initial unstable density stratification.  From equation (\ref{eq:PE_budget}), large scale
potential energy is a source for large scale kinetic energy via $\epsilon^{\textrm{inj}}$ in equation (\ref{eq:KE_budget}).
Small scale kinetic energy injection $\epsilon^{\text{small}}_\ell = \OL{\rho}\wt{\tau}(u_i, g_i)$ is identically zero for RT flows 
since the gravitational acceleration $\boldsymbol{g}$ is constant in space.
Thus, potential energy is available directly to large scale kinetic energy. Meanwhile, kinetic energy is also linked to internal energy via pressure dilatation and viscous
dissipation. The term $\OL{P}_\ell \nabla\cdot \OL{\bu}_\ell$ appears as a source  in large scale KE equation (\ref{eq:KE_budget})
and a corresponding term $\OL{\tau}(P, \nabla\cdot \bu)$ is a source in small scale KE budget equation (\ref{eq:small_KE_budget}), while both appear as sinks 
in the internal energy equation (\ref{eq:IE_budget}). Similarly, dissipation terms $D_\ell$ and $D_\ell^{\text{small}}$ are sinks in large and small scale KE budgets, respectively,
and sources in the internal energy budget: $D_\ell^{\text{int}} = D_\ell + D_\ell^{\text{small}}$. These channels, summarized in figure \ref{fig:combined_energy_pathway},
are responsible for converting energy between different forms. In addition to these conversion channels, the transfer between large and small scale kinetic energy is now presented.


\subsubsection{Kinetic energy scale pathways} \label{sec:ke_pathways}
There are two kinetic energy fluxes across scales, $\Pi$ and $\Lambda$. Deformation work $\Pi$ is solely due to the turbulent velocity field, which is closely related to
vortex stretching in constant-density 3D turbulence \citep[e.g.][]{Eyink06JFM}. Baropycnal work $\Lambda$ was shown in \citet{Aarne19} to
transfer energy by strain and vorticity generation, in which density, pressure,
and velocity fields all play a role. These fluxes provide a pathway for kinetic energy across scales, and operate in the `inertial range', as we shall see shortly.

Figure  \ref{fig:combined_energy_pathway} sketches the \emph{mean} or \emph{bulk} pathways based on the energy budgets. Some arrows in the schematic are dashed, representing energy transfer directions which are not realized dynamically in our RT flows. For example, gravity converts potential energy to large scale KE despite the kinematic possibility for the transfer to be in reverse. Pressure dilatation converts internal energy to large scale KE. Baropycnal work, $\Lambda$, transfers energy from large-scale KE to smaller scales, consistent with its role as as an energy flux. Deformation work, $\Pi$, also transfers KE energy downscale, but only in 3D. We find that $\Pi$ transfers KE in the reverse direction in 2D, from small to large scales. However, the total flux $\Pi+\Lambda$ is always downscale in both 2D and 3D. Irreversible viscous dissipation converts only small-scale (but not large-scale) KE to internal energy  due to our scale-decomposition satisfying the inviscid criterion.
Last, although mean $\OL{\tau}(P, \nabla\cdot\bu)$ is a kinematically possible pathway between small scale KE and internal energy, we observe it is close to zero in our RT flows. This is consistent with previous studies showing that mean pressure dilatation is a large-scale process and does not couple kinetic and internal energy at small scales, on average \citep{Aluie11c,Aluie12,KritsukWagner13,Wang13}.

Note that in the LES modeling literature, another formulation of the kinetic energy budget is often used in which baropycnal work, $\Lambda$, is missing by lumping it with pressure dilatation. The traditional LES formulation is mathematically equivalent to our equation (\ref{eq:KE_budget}), but the physical interpretation of energy pathways is different. Stated briefly, our KE budget allows for an inertial range over which kinetic energy can cascade \citep{Aluie11c, Aluie13}, unlike the more common LES formulation, which misses the distinct physics of baropycnal work $\Lambda$ that is essential in variable density flows such as RT. A discussion is presented in Supplementary Material, section \ref{suppsec:pathway_comparison}.

\subsubsection{The inertial range in turbulent RT flows} \label{sec:ke_inertial}
Having identified the pathways, we now present a quantitative analysis. We calculate all terms in the large scale kinetic energy budget (\ref{eq:KE_budget}) 
as a function of scale $\ell$, which allows us to test and validate ideas on energy conversion and the cascade. 

\begin{figure}
\centerline{\includegraphics[height=2.5in]{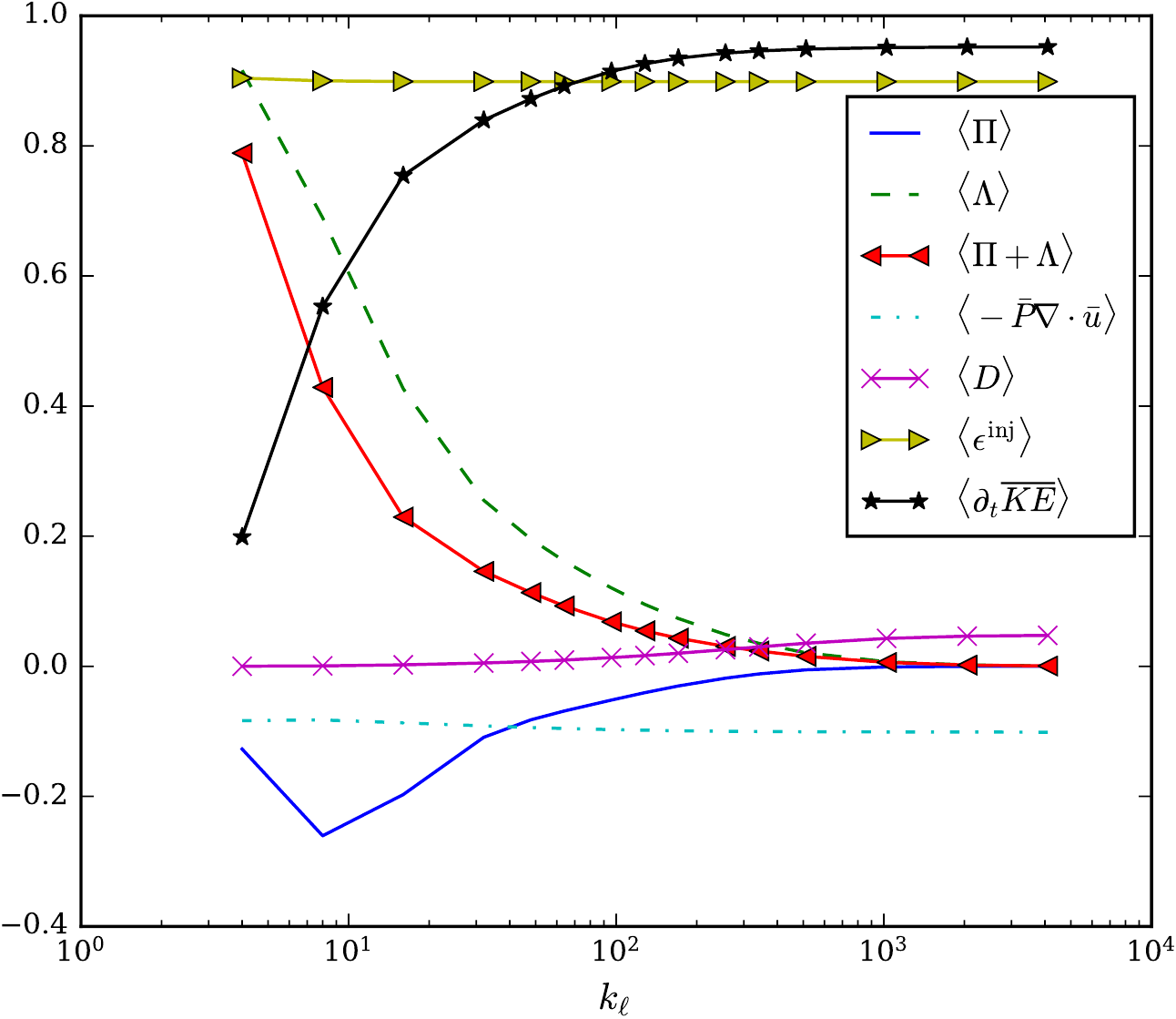} } 
\caption{\footnotesize{Kinetic energy processes as a function of scale in the 2D4096 simulation at dimensionless time $\widehat{t}=4.0$. Filtering wavenumber $k_\ell=L_z/\ell$ is a proxy for length-scale.
    Plots are normalized by $\langle \epsilon^{\textrm{inj}} + P\nabla\cdot \bu\rangle$, the (unfiltered) available mean  source of kinetic energy.
    All mean budget terms in the filtered large scale kinetic energy equation (\ref{eq:KE_budget}) are plotted as a function of $k$. The time derivative of filtered
    kinetic energy $\partial_t \OL{\text{KE}}\equiv \partial_t \left(\OL{\rho}_\ell \frac{|\wt{\bu}_\ell|^2}{2}\right)$ is also shown for this unsteady RT flow.} 
\label{fig:2D4096_cascade}}\end{figure}

\begin{figure}
\centerline{\includegraphics[height=2.5in]{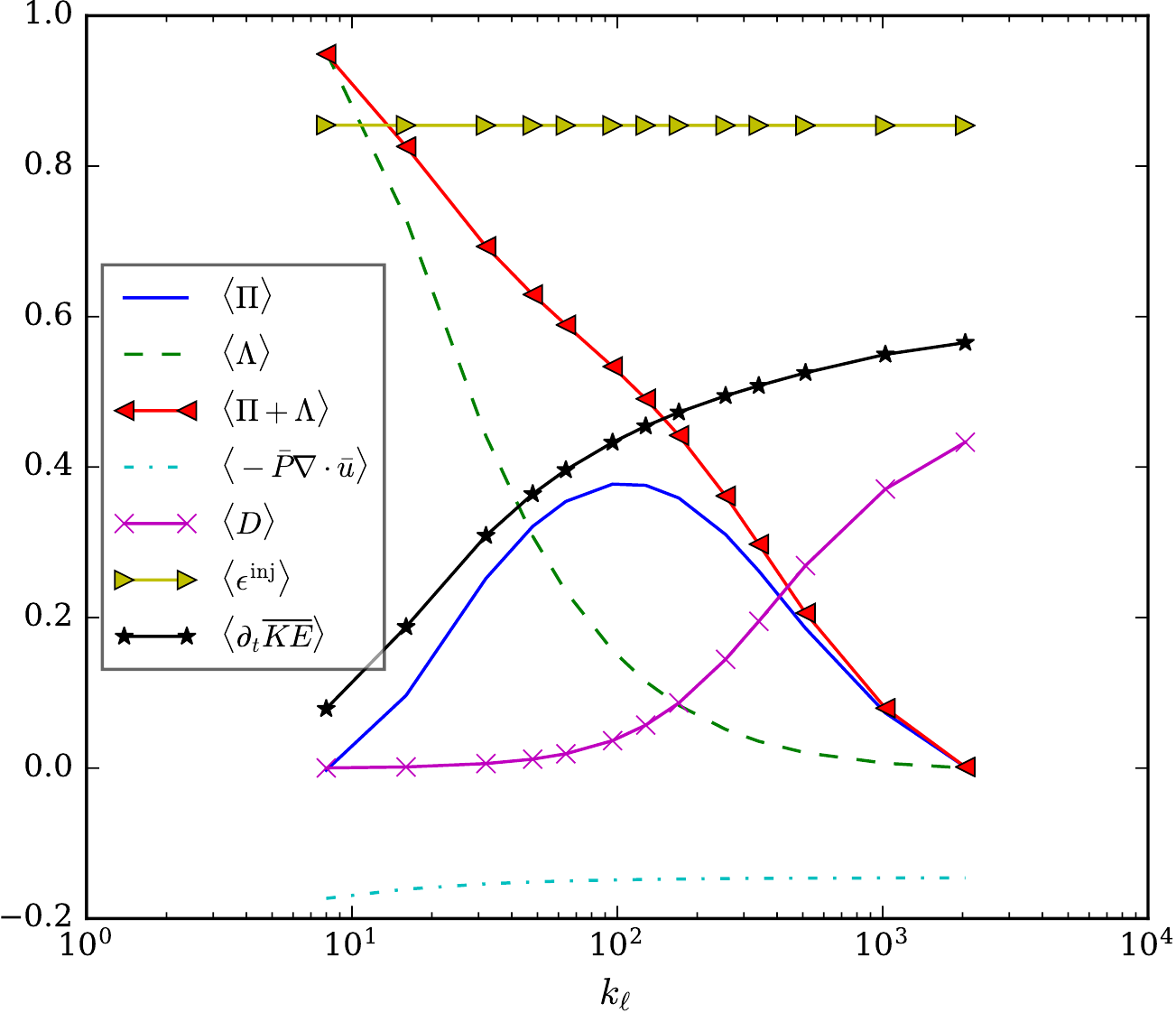} } 
\caption{\footnotesize{Same as in figure \ref{fig:2D4096_cascade} but applied to the 3D1024 data at the same dimensionless time $\widehat{t}=4.0$. Plots are normalized by $\langle \epsilon^{\textrm{inj}} + P\nabla\cdot \bu\rangle$ at that time. Main differences from 2D are that here $\langle\Pi\rangle > 0$ (blue), transferring energy to smaller scales and resulting in much higher dissipation (magneta). Also, small scales at $k_\ell>100$ are still growing in time (black) unlike in figure \ref{fig:2D4096_cascade} where $\langle\partial_t \OL{\text{KE}}\rangle$ (a cumulative quantity with contributions from \emph{all} scales $<k_\ell$) becomes $k_\ell$-independent at large $k_\ell$, indicating that there is negligible growth at $k_\ell>100$ with most of the temporal KE increase happening at very larger scales in 2D (see figure \ref{fig:2D4096_cascade}, jump in black curve over $k\in[4,30]$).}
\label{fig:3D1024_cascade}}\end{figure}

 The results from the 2D4096 simulation (see Table \ref{tab:parameter}) are shown in figure \ref{fig:2D4096_cascade}, while the 3D1024 results are in
figure \ref{fig:3D1024_cascade}. These figures plot the domain averaged terms in equation (\ref{eq:KE_budget}) as a function of filtering wavenumber (inverse of scale) $k_\ell=L_z/\ell$. Note that our ``filtering wavenumber'' $k_\ell$ does not involve any Fourier transform, and we only use it to refer to the corresponding scales, as is conventional in the turbulence literature.
For large $k_\ell$ (small scales), these terms approach their unfiltered values. Therefore, they represent cumulative processes acting at \emph{all} scales larger than $\ell$ and not just the contribution from scale $\ell$ itself. For example,
if $\langle\partial_t \OL{\text{KE}}\rangle\equiv \langle\partial_t \left(\OL{\rho}_\ell \frac{|\wt{\bu}_\ell|^2}{2}\right)\rangle$ increases in value over the range $[k_1,k_2]$ ($k_1<k_2$), then the KE content of scales within the band $[L_z/k_2,L_z/k_1]$ is growing. The only exceptions are the flux terms, $\Lambda_\ell$ and $\Pi_\ell$, which are hybrid quantities involving coarse-grained fields ($\grad\OL{P}_\ell$ or $\wt{\bS}_\ell$) acting against subscale contributions ($\OL\tau_\ell(\rho,\bu)$ or $\wt\tau_\ell(\bu,\bu)$) and, therefore, represent the energy transfer \emph{across} the partitioning scale $\ell$ \citep{Aluie11c,Aluie13}.

Figures \ref{fig:2D4096_cascade}, \ref{fig:3D1024_cascade} show results from the 2D4096 and 3D1024 simulations, respectively. 
Spatial mean injection $\langle \epsilon^{\textrm{inj}}\rangle$ shown in the figures is constant in $\ell$, which
is due to the spatially uniform gravitational acceleration used here \citep{Aluie13}.  The reason becomes clear when considering that $\langle \epsilon^{\textrm{inj}}\rangle = \langle \OL{\rho} \wt{u}_i \wt{g}_i\rangle=
-\langle \OL{\rho u_z} \rangle g=-\langle \rho u_z \rangle g$ is independent of  $\ell$. This implies that mean injection from potential energy is released directly only at the largest scale \citep{Aluie13}. To elaborate, consider two scales, one with length $L$, the largest in the system, and another with a smaller length scale $\ell$.
Since mean injection at the two distinct scales are the same, injection within the band of scales $(\ell, L)$ is zero.
Taking the limit $\ell\rightarrow 0$, it is obvious that no mean injection occurs at any scale below $L$. Thus, mean injection occurs only at the largest length scale, that of the domain size. Subsequent transfer to smaller scales occurs via the fluxes $\Lambda$ and $\Pi$, as we shall discuss below.

Similar to injection, mean pressure dilatation $\langle -\OL{P}\nabla\cdot\OL{\bu}\rangle$ is almost flat as a function of $\ell$ in 
figures \ref{fig:2D4096_cascade}, \ref{fig:3D1024_cascade}. This is consistent with predictions in \citet{Aluie11c} and simulations by \citet{Aluie12}, where the pressure dilatation was shown to operate over a transitional  scale range of limited extent. 
Hence, mean pressure dilatation only affects the large scale dynamics. Dissipation $\langle D_\ell\rangle$ within the Favre scale decomposition has already been proved to be negligible at large scales \citep{Aluie13, Zhao18}, which is again confirmed here in figures \ref{fig:2D4096_cascade}, \ref{fig:3D1024_cascade}. Thus, in well developed RT turbulence, there is
a range of length scales which is immune from injection and pressure dilatation at large scales, and also from dissipation at small scales. 
This range of length scales is the `inertial range' in the spirit of Kolmogorov's turbulence theory. Within this inertial range, kinetic energy dynamics is governed by the two fluxes $\Pi$ and $\Lambda$.

\subsubsection{Energy fluxes across scales} \label{sec:ke_fluxes}
 The narrative that will emerge throughout this paper indicates that baropycnal work, $\Lambda$, in RT flows acts as a conduit for gravitational potential energy, transferring (non-locally in scale) the energy deposited by $\epsilon^{\textrm{inj}}$ at scale $L$ to smaller scales. Figures \ref{fig:2D4096_cascade},\ref{fig:3D1024_cascade} show that $\langle\Lambda\rangle$ is positive, acting as a KE source for intermediate to small scales. Mean $\Lambda$ decays monotonically with $k_\ell$ in both 2D and 3D, indicating that $\Lambda$ does not persist to arbitrarily small scales. As we discuss in Supplementary Material, section \ref{suppsec:flux_efficiency}, this decay is consistent with $\Lambda$ being predominantly a non-local transfer in scale.
As shown in figures \ref{fig:2D4096_cascade},\ref{fig:3D1024_cascade}, the decrease in $\Lambda$ is accompanied by an increase in $\partial_t \OL{\text{KE}}$ at those scales. This implies that as RT turbulence is evolving, $\Lambda$ acts as a KE source for unsteady scales within the inertial range, increasing their KE content. The largest scales increase their KE content the most, leaving less energy to be transferred by $\Lambda$ to smaller inertial scales, which explains why $\Lambda$ is not persistent (\textit{i.e.} constant) but decays with smaller $\ell$. Further analysis of $\Lambda$'s scale-locality and decay in $k_\ell$ is undertaken in Supplementary Material, section \ref{suppsec:flux_efficiency}.

As these inertial scales become energetic enough and vortex stretching starts operating in 3D (figure \ref{fig:3D1024_cascade}), deformation work 
 $\Pi$ takes over and transfers energy to smaller inertial scales similar to constant-density homogeneous 3D turbulence. Transfer by $\Pi$ is strongest (peak in figure \ref{fig:3D1024_cascade}) at an inertial scale that grows in time and is $O(10)$ smaller than the mixing width. In the absence of vortex stretching in 2D (figure \ref{fig:2D4096_cascade}), $\langle\Pi\rangle$ is negative as in constant-density homogeneous 2D turbulence, transferring energy upscale. Hence, the only mechanism for growth of small scale kinetic energy in 2D is baropycnal work $\Lambda$. 
 
 The downscale cascade of energy by $\langle\Pi\rangle$ in 3D delivers energy to viscous scales which is manifested by large values of viscous dissipation in figure \ref{fig:3D1024_cascade}. In contrast, dissipation levels in 2D are miniscule since $\langle\Pi\rangle <0$, cascading energy upscale. This fundamental difference also explains the difference in spectra in figure \ref{fig:spectra}, where there is less energy at small scales ($k_\ell\approx100$) in 2D relative to 3D, despite there being more overall KE (integrated across all scales) in 2D relative to 3D. This is because $\Pi$ in 3D acts as a source term for these small inertial scales, in addition to $\Lambda$, delaying the saturation of KE at these scales in 3D relative to 2D. 
 
 Indeed, note that $\langle \partial_t \OL{\text{KE}} \rangle$ as a function of scale in 2D in figure \ref{fig:2D4096_cascade} 
saturates to a constant value at $k_\ell=L_z/\ell\approx 100$, with almost zero KE growth at higher $k_\ell$. This indicates that at the
time being analyzed, these small scales have reached a steady state and are not increasing on average. However, in 2D, unlike in 3D,  their energy is subsequently transferred to large scales by $\Pi$ (which is negative), resulting in a low level of dissipation of kinetic energy. In contrast, $\langle\Pi\rangle$ and  $\langle\Lambda\rangle$ are both positive in 3D, transferring a larger amount of energy to small scales compared to 2D. This leads to higher KE levels at small scales in 3D, as evidenced by plots of spectra in figure \ref{fig:spectra} and also of 
$\langle \partial_t \OL{\text{KE}} \rangle$ in figure \ref{fig:3D1024_cascade}, which keeps increasing as a function of $k_\ell$, indicating that those scales are still growing and have not yet reached a steady state. Since viscous dissipation is most effective at the smallest scales (within the Favre decomposition), this substantially higher energy input at high $k_\ell$ in 3D (versus 2D) leads to the higher levels of energy dissipation in figure \ref{fig:3D1024_cascade} (versus figure \ref{fig:2D4096_cascade}).

Our analysis points to fundamental differences in physics between 2D and 3D RT instabilities. It is often observed that the turbulent RT instability mixing layer grows faster in 2D simulations relative to 3D simulations\footnote{It is worth noting that \cite{HorneLawrie2020}  have reported smaller $\alpha=0.012$ in quasi-2D RT experiments. However, Ekman friction is prominent as discussed lucidly in that work and it would be expected to reduce the $\alpha$ value since it acts preferentially on large-scales.}. This is captured by the $\alpha$ parameter appearing in the mixing width evolution, $h=\alpha \mathcal{A} g t^2 $, which has often been found to be larger 
in 2D than in 3D \citep{Youngs91POF,Youngs94,Alphagroup,Cabot06POF}, consistent with the values we found in figure \ref{fig:mixingwidth}. The enhanced growth in 2D simulations has been somewhat of a puzzle since single-mode (non-turbulent) RT instability models and simulations predict that it is growth in 3D that should be faster \citep{Layzer55,Goncharov02,Bian20}. The difference in $\alpha$ values in turbulent RT instability had been attributed to the smaller dissipation levels in 2D \citep{Youngs91POF,Youngs94}, although we have seen above that viscous dissipation levels are merely a symptom (or a consequence) of the energy scale-fluxes and do not directly influence inertial-range dynamics governing the mixing layer. 

The fundamental difference in inertial-range physics is that deformation work $\langle\Pi\rangle$ channels energy upscale in 2D and downscale in 3D. In 2D-RT, $\langle\Pi\rangle$ redirects energy delivered by $\Lambda$ to  inertial scales back to larger scales, driving a positive feedback loop that is absent in 3D-RT. In this feedback loop, the upscale cascade leads to an enhanced growth of the mixing layer, which in turn leads to enhanced potential energy release (as seen in figure \ref{fig:full_ke_budget}) by raising the light fluid and lowering the heavy fluid. The excess release of potential energy  results in enhanced baropycnal energy transfer by $\Lambda$ to smaller scales which is re-channeled back upscale by $\Pi_\ell$, closing the feedback loop. 
This narrative is consistent with plots of filtering spectra in figure \ref{fig:spectra}, where we see that the KE (and velocity) spectral peak in 2D is at larger scales and has a higher magnitude compared to 3D. \citet{Cabot06POF} had speculated that the difference in $\alpha$ between 2D and 3D may be due to an inverse energy cascade in 2D-RT, although that study did not conduct an analysis of energy transfer across scales. Our results here provide thorough evidence in support of that hypothesis, and highlight the potentially misleading 2D-RT flow physics when  modeling real-world 3D-RT flows.

Another important conclusion we can draw from our analysis is the absence of a net upscale energy transfer in 3D-RT. In many studies discussing RT flows in 3D, motivated by ICF (e.g. \cite{Shvartsetal1995,Alonetal1995,Oferetal1996,Oronetal2001,Zhou03}), or astrophysics (e.g. \cite{Joggerstetal2010,Porthetal2014}), or fundamental fluid dynamics (e.g. \cite{Abarzhi1998,Chengetal2002,Cabot06POF}), it is often claimed that the generation of successively larger bubbles following the presence of smaller ones is due to an inverse cascade of energy which causes small bubbles to merge into larger ones. According to such a narrative, the small-scales are the source of energy feeding and sustaining the large-scales, even in 3D.

Such claims are often based on qualitative observations of the flow development, including spectral energy content, but without clarifying the causality of energy spectral broadening and do not analyze the energy transfer across scales directly (e.g. scale fluxes $\Pi$ and $\Lambda$) as we do here. While   energy at small scales saturates before energy at larger scales, this does not necessarily imply that energy is being continuously channelled to the small-scales and back up to larger scales. Indeed,
our findings indicate the absence of a net upscale energy transfer in 3D-RT, and that growth of large-scale bubbles and spikes is solely due to baropycnal work $\Lambda_\ell$ in 3D, which deposits energy directly from gravitational potential energy at the largest scales in the system. No \emph{persistent} upscale transfer of energy is required\footnote{Beating of modes to seed larger scales from smaller ones can still transfer minuscule amounts of energy upscale but this is fundamentally different from the notion of a cascade, which requires \emph{persistent} transfer in scale and/or time. Consider, for example, 3D constant-density homogeneous isotropic turbulence (HIT) forced at an intermediate wavenumber $k_f$ in a triply periodic domain and having slightly smaller wavenumbers $k\lesssim k_f$ gain finite energy. Such \emph{transient} upscale energy transfer is not sustained (or persistent) in wavenumber, which is why it is widely agreed that an inverse cascade in 3D-HIT is absent. Another example is the downscale cascade of enstrophy in 2D constant-density flows: there is invariably non-zero energy transferred downscale along with enstrophy, but such energy transfer is not a cascade since it is not persistent in scale, vanishing rapidly $\sim \ell^{2}$ as $\ell \to 0$ within the inertial range \citep{Kraichnan67}.} to explain the evolution of spectral energy content. The assumed existence of an inverse cascade had been used in some instances \citep[e.g.][]{Abarzhi1998} to explain the dependence of RT flows on initial perturbations \citep{Dimonte04,Ramaprabhu05}. If this were the case, it would have created significant difficulties in the subgrid modeling of RT flows. Our results here provide compelling evidence that the inverse cascade narrative in 3D-RT is unjustified.

\subsubsection{Summary of RT energy pathways} \label{sec:EnergyPathwaysSummary}
Section \ref{sec:EnergyPathways} presented some of the main results of this paper, which we now summarize. 
We showed that Baropycnal work, $\Lambda$, in RT flows acts as a conduit for gravitational potential energy, transferring the energy deposited by $\epsilon^{\textrm{inj}}$ at the largest domain scale $L$ to smaller scales. As these inertial scales become energetic enough and vortex stretching starts operating in 3D (figure \ref{fig:3D1024_cascade}), deformation work  $\Pi$ takes over and transfers energy to smaller inertial scales. The net downscale cascade of energy by $\langle\Pi\rangle$ in 3D delivers energy to viscous scales which is manifested by large values of viscous dissipation. In contrast, in 2D-RT, $\langle\Pi\rangle<0$ redirects energy delivered by $\Lambda$ back to larger scales via an upscale cascade. The upscale cascade in 2D leads to an earlier saturation of small inertial scales compared to 3D, which explains why small inertial scales are weaker than in 3D (see spectrum in figure \ref{fig:spectra}).  This is despite the greater overall KE (integrated across all scales) in 2D relative to 3D, which is explained by the positive feedback loop in 2D-RT.

The upscale cascade in 2D-RT drives a positive feedback loop that is absent in 3D-RT. In this feedback loop, the upscale cascade leads to an enhanced growth of the mixing layer, which in turn leads to enhanced potential energy release (as seen in figure \ref{fig:full_ke_budget}) by raising the light fluid and lowering the heavy fluid. The excess PE released results in enhanced baropycnal energy transfer by $\Lambda$ to smaller scales which is re-channeled back upscale by $\Pi$, closing the feedback loop.

\subsection{Spatial distribution of fluxes}

\begin{figure}
\centering 
\begin{minipage}[b]{1.0\textwidth}  
\centering
\subfigure[\footnotesize{PDF of $\Pi$ in 2D}]
{\includegraphics[height=1.6in]{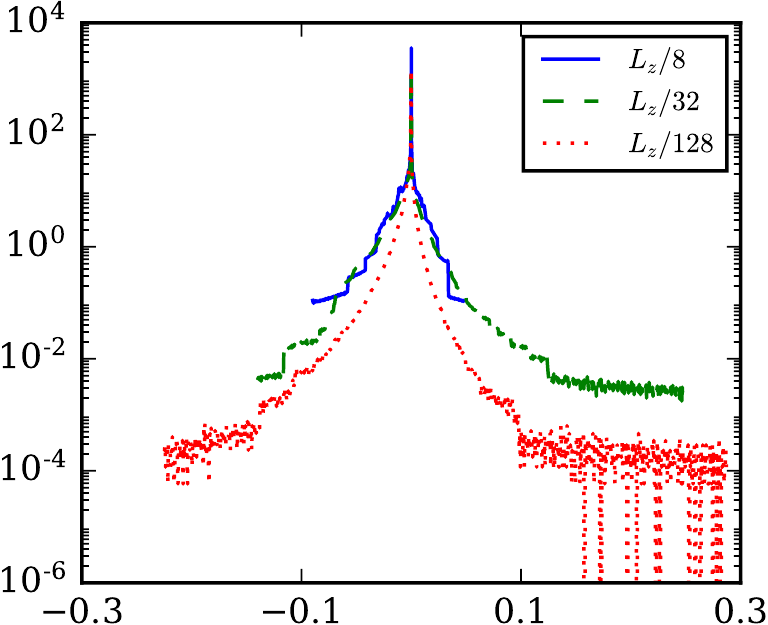} \label{fig:flux_pdfs_2d_pi}} 
\phantom{a}
\subfigure[\footnotesize{PDF of $\Lambda$ in 2D}]
{\includegraphics[height=1.6in]{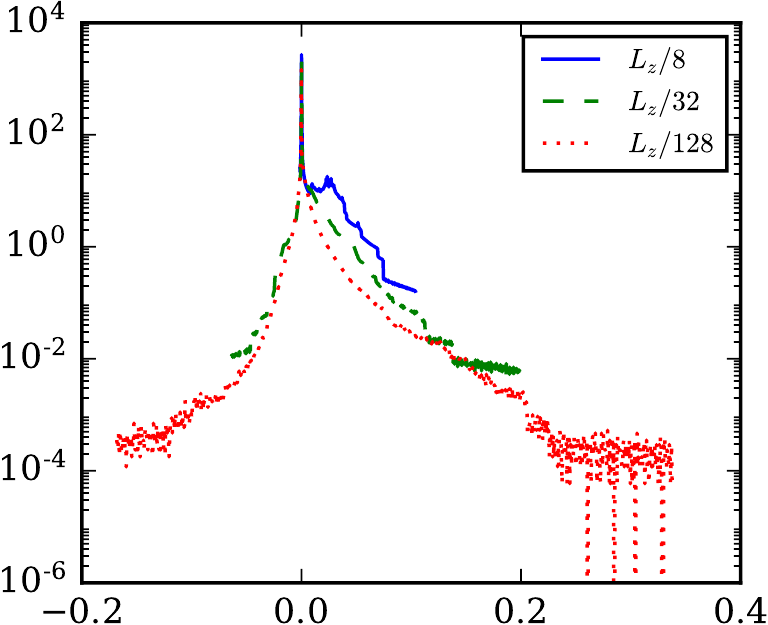} \label{fig:flux_pdfs_2d_lambda}}\\
\subfigure[\footnotesize{PDF of $\Pi$ in 3D}]
{\includegraphics[height=1.6in]{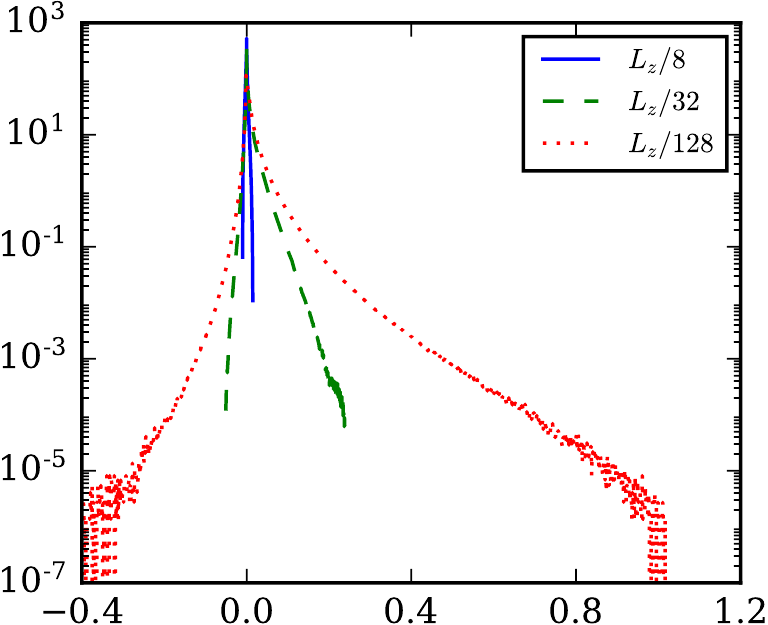} \label{fig:flux_pdfs_3d_pi}} 
\phantom{a}
\subfigure[\footnotesize{PDF of $\Lambda$ in 3D}]
{\includegraphics[height=1.6in]{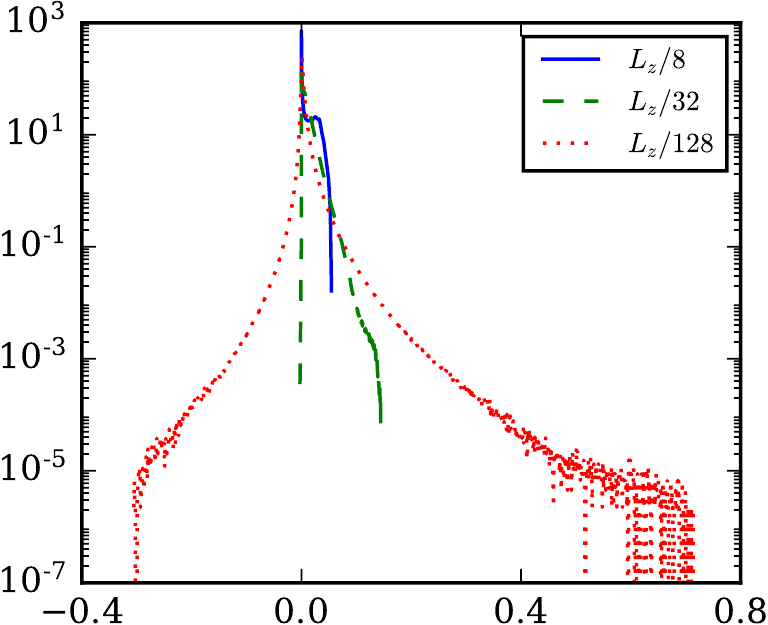} \label{fig:flux_pdfs_3d_lambda}} 
\caption{\footnotesize{Probability density functions (PDFs) of $\Pi$ and $\Lambda$ from simulations 2D4096 and 3D1024 at dimensionless time $\widehat{t}=4$.
    Results for three different scales $\ell=L_z/8, L_z/32, L_z/128$ are shown.} \label{fig:flux_pdfs}}
\end{minipage}
\end{figure}

The spatial distribution of values for fluxes $\Lambda$ and $\Pi$ is shown by
their PDFs in figure \ref{fig:flux_pdfs}, for 2D and 3D RT.
Results at three scales $\ell=L_z/8, L_z/32, L_z/128$ are shown.
All PDFs are non-Gaussian, with heavy tails and a stronger departure from Gaussianity at smaller scales. In 2D, $\Pi$ is skewed towards negative 
values, but in 3D it is positively skewed, reflecting the inverse versus forward net energy cascade by $\Pi$ in 2D versus 3D. The PDF of $\Lambda$ 
is always skewed towards positive values.
In particular, at large scales (e.g., $\ell=L_z/8$), $\Lambda_\ell$ is  positive everywhere in the domain as is seen in figures \ref{fig:flux_pdfs_2d_lambda} and 
\ref{fig:flux_pdfs_3d_lambda}. As we shall discuss in \cite{ZhaoAluie20}, negative values
of $\Lambda_\ell$ are correlated with spiraling regions of the flow, which are part of the mushroom structure that appears in late time single-mode RT.
Note that the PDF of $\Lambda$ in 2D (figure \ref{fig:flux_pdfs_2d_lambda})
contains both positive and negative values when $\ell$ is below or equal to $L_z/32$, while the PDF of $\Lambda$ in 3D
(figure \ref{fig:flux_pdfs_3d_lambda}) is almost always positive even at $\ell=L_z/32$. 
This observation is in accord with visualizations of density fields in figures \ref{fig:visual_2D} and \ref{fig:visual_3D}
where 3D-RT exhibits much finer structures than in 2D-RT, such that the spiraling structures are on average smaller in 3D.


\subsection{Temporal self-similarity of flux terms} \label{sec:temporal_self_similarity}
\begin{figure}
\centering 
\begin{minipage}[b]{1.0\textwidth}  
\centering
\subfigure[\footnotesize{Normalized $\langle\Lambda\rangle$ evolution in 2D}]
{\includegraphics[height=1.6in]{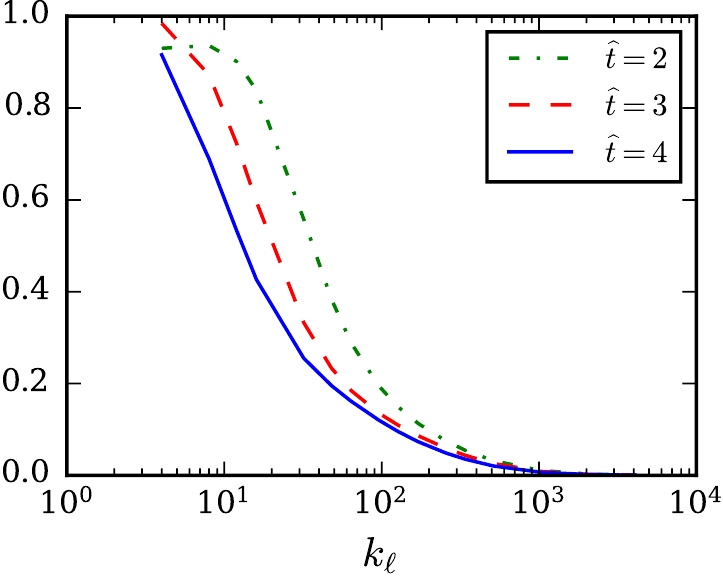} } 
\phantom{a}
\subfigure[\footnotesize{Normalized $\langle\Pi\rangle$ evolution in 2D}]
{\includegraphics[height=1.6in]{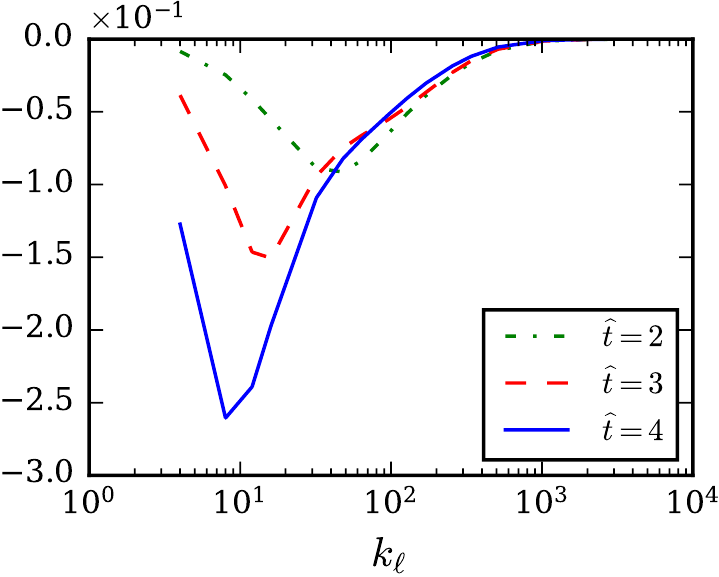} } \\
\subfigure[\footnotesize{Normalized $\langle\Lambda\rangle$ evolution in 3D}]
{\includegraphics[height=1.6in]{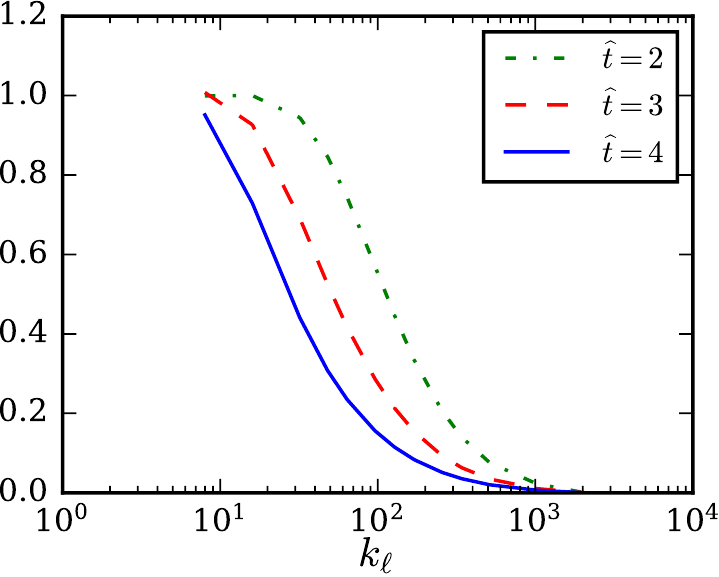} } 
\phantom{a}
\subfigure[\footnotesize{Normalized $\langle\Pi\rangle$ evolution in 3D}]
{\includegraphics[height=1.6in]{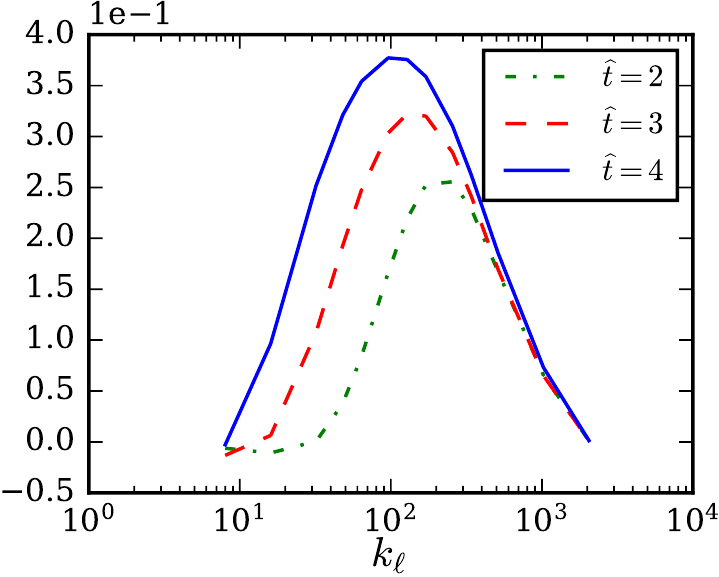} }
\caption{\footnotesize{Temporal evolution of fluxes in 2D and 3D at dimensionless time $\widehat{t}=2,3,4$. Filtering wavenumber is $k_\ell=L_z/\ell$.
    Plots are normalized by $\langle \epsilon^{\textrm{inj}} + P\nabla\cdot \bu\rangle$ at the corresponding time.} \label{fig:flux_evolution}}
\end{minipage}
\end{figure}

So far, we have discussed the fluxes as a function of scale at one time instant. In this section we investigate their temporal evolution. 
Figure \ref{fig:flux_evolution} shows  mean $\Pi$ and $\Lambda$ normalized by $\langle \epsilon^{\textrm{inj}} + P\nabla\cdot \bu\rangle$ at dimensionless times $\widehat{t}=2,3,4$ using the 2D4096 and 3D1024 RT data (see Table \ref{tab:parameter}).
As RT turbulence develops, the magnitude of \emph{normalized} $\Lambda$ decreases, whereas the normalized $\Pi$ increases in magnitude. Note, however, that the \emph{absolute} values of both fluxes keep increasing (not shown). A decrease in normalized $\Lambda$ indicates that as KE at scale $\ell$ saturates in time, the relative importance of $\Lambda$, which is responsible for redistributing injected energy from the potential release, also decreases. In contrast, turbulence levels in RT flows increase in time,
leading to an increasing relative magnitude of $\langle\Pi\rangle$.
The scale corresponding to the peak magnitude of $\Pi$, indicated by $k_{\text{peak}}$, shifts to smaller $k_\ell$ (or larger scales) over time.
For example, in 2D, $k_{\text{peak}}$ shifts from $\approx64$ at $\widehat{t}=2$, to $\approx10$ at $\widehat{t}=4$, while in 3D, $k_{\text{peak}}$ shifts from about $\approx256$ at $\widehat{t}=2$ to $\approx100$ at $\widehat{t}=4$. Thus, as expected, turbulence encroaches onto larger scales as RT evolves, so that vortex stretching (in 3D) or mergers (in 2D) which drive the $\Pi$ flux are more pronounced at these scales.

\begin{figure}
\centering 
\begin{minipage}[b]{1.0\textwidth}  
\subfigure
{\includegraphics[height=1.31in]{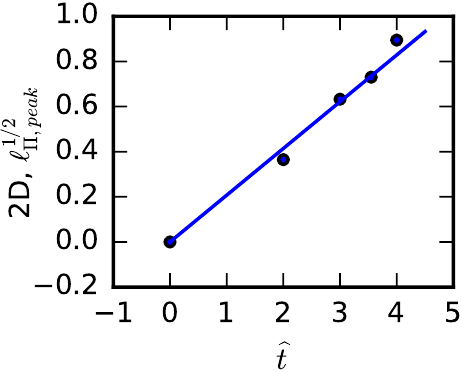} 
\llap{\parbox[b]{1.1in}{{\large (a)}\\\rule{0ex}{0.9in}}}
    \label{fig:self_similar_flux_1}} 
\subfigure
{\includegraphics[height=1.34in]{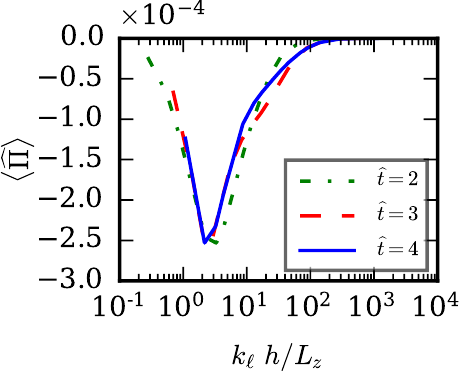} 
\llap{\parbox[b]{1.25in}{{\large (b)}\\\rule{0ex}{0.9in}}}
    \label{fig:self_similar_flux_2}} 
\subfigure
{\includegraphics[height=1.34in]{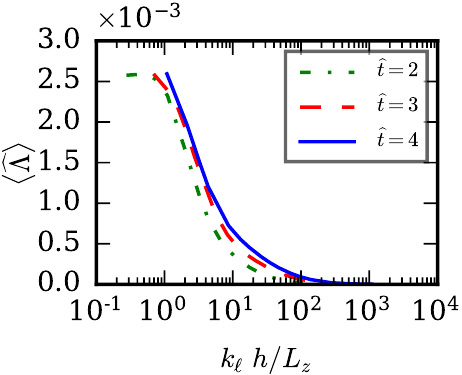} 
\llap{\parbox[b]{1.3in}{{\large (c)}\\\rule{0ex}{0.9in}}}
    \label{fig:self_similar_flux_3}}  \\
\subfigure
{\includegraphics[height=1.31in]{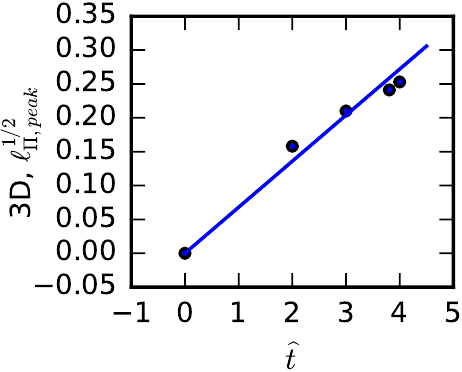} 
\llap{\parbox[b]{1.05in}{{\large (d)}\\\rule{0ex}{0.9in}}}
    \label{fig:self_similar_flux_4}} 
\subfigure
{\includegraphics[height=1.34in]{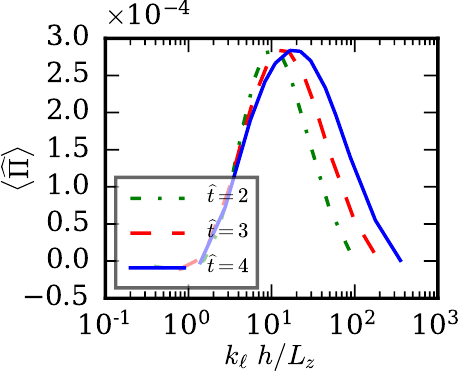} 
\llap{\parbox[b]{1.3in}{{\large (e)}\\\rule{0ex}{0.9in}}}
    \label{fig:self_similar_flux_5}} 
\subfigure
{\includegraphics[height=1.34in]{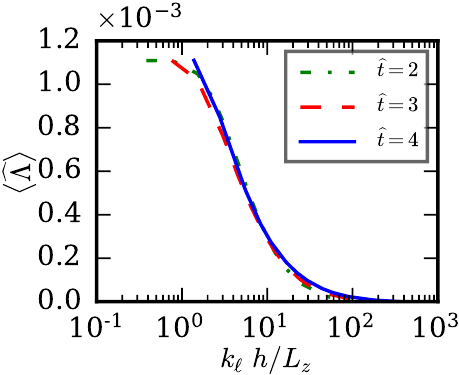} 
\llap{\parbox[b]{1.3in}{{\large (f)}\\\rule{0ex}{0.9in}}}
    \label{fig:self_similar_flux_6}}  \\
\caption{\footnotesize{Temporal self-similarity of turbulent RT fluxes. Top row shows results for 2D-RT, bottom row shows 3D-RT.
    Panels (a),(d) plot length scale $\ell_{\Pi, \text{peak}}$ associated with the peak of $\Pi$ in 2D and 3D, versus dimensionless time $\widehat{t}$. Panels (b),(e) show rescaled $\langle\widehat{\Pi}\rangle$ in
    equation (\ref{eq:self_similar_flux}). Panels (c),(f) show rescaled $\langle\widehat{\Lambda}\rangle$.} \label{fig:self_similar_flux}}
\end{minipage}
\end{figure}

The above observations indicate possible self-similar (in time) behavior of turbulent RT fluxes. If we plot the scales corresponding to the peak magnitude of 
$\langle\Pi\rangle$, which we denote as $\ell_{\Pi,\text{peak}}$, against dimensionless time $\widehat{t}$, we obtain a quadratic scaling in time as shown in figures \ref{fig:self_similar_flux}(a),(d),
similar to the mixing width $h(t)$ in figure \ref{fig:mixingwidth}. Figures \ref{fig:self_similar_flux}(a), (d) indicate that,
in both 2D and 3D, $\sqrt{\ell_{\Pi, \text{peak}}} \propto \widehat{t} \propto t$, with the coefficients
\begin{align}
    \label{eq:ell_max_pi_scaling}
    \ell_{\Pi, \text{peak}}^{2D} = \alpha_\pi^{2D} \mathcal{A} g t^2 \approx 0.37 h^{2D}(t), \quad \ell_{\Pi, \text{peak}}^{3D} = \alpha_\pi^{3D} \mathcal{A} g t^2 \approx 0.067 h^{3D}(t)
\end{align}
where $h^{2D}(t), h^{3D}(t)$ are the mixing widths from the 2D4096 and 3D1024 RT data. Note from figure \ref{fig:flux_evolution} (also figures \ref{fig:self_similar_flux}(b),(e)) how $\langle\Pi\rangle$ in 2D is skewed to larger scales, unlike in 3D, peaking at approximately half the size of the mixing width, $\ell_{\Pi, \text{peak}}^{2D}\approx \frac{1}{2}h^{2D}$ (equation \eqref{eq:ell_max_pi_scaling}). This is consistent with the net upscale transfer by $\Pi$ driving the mixing layer growth as discussed in section \ref{sec:ke_fluxes}. In comparison, $\langle\Pi\rangle$ in 3D is symmetric, peaking at a scale $O(10)$ smaller than the mixing width, $\ell_{\Pi, \text{peak}}^{3D}\approx \frac{1}{10}h^{3D}$.

The temporal scaling of $\ell_{\Pi, \text{peak}}$ in figures \ref{fig:self_similar_flux}(a),(d) is evidence that the evolution of mean fluxes $\langle\Pi\rangle, \langle\Lambda\rangle$ is self-similar.
Recall that self-similarity implies that functions of length and time, such as $\langle \Pi\rangle$ and $\langle\Lambda\rangle$, should collapse to the same curve when properly rescaled \citep{Ristorcelli04}.  Specifically, the rescaled fluxes should satisfy
\begin{align}
    \label{eq:self_similar_flux}
    \begin{split}
    \langle\widehat{\Pi}\rangle(\hat{k}_\ell)\equiv f_\pi(k_\ell\,t^2) &= F_\pi^{-1}(t)~\langle\Pi\rangle\left(\hat{k}_\ell\frac{L_z}{h},t\right),\\ 
    \langle\widehat{\Lambda}\rangle(\hat{k}_\ell)\equiv f_\lambda(k_\ell\,t^2) &=F_\lambda^{-1}(t)~\langle\Lambda\rangle\left(\hat{k}_\ell\frac{L_z}{h},t\right). 
    \end{split}
 \end{align}
where $\hat{k}_\ell$ is just filtering wavenumber $k_\ell=L_z/\ell$, rescaled by the mixing width $h(t)$, $\hat{k}_\ell\equiv k_\ell \,h/L_z$.  $F_\pi(t)$, $F_\lambda(t)$ are temporal scaling functions, and
$f_\pi(\cdot), f_\lambda(\cdot)$ are the similarity functions. If mean fluxes $\langle\Pi\rangle, \langle\Lambda\rangle$ are temporally self-similar, then
$\langle\widehat{\Pi}\rangle, \langle\widehat{\Lambda}\rangle$ should collapse onto the same curve. 

Figures \ref{fig:self_similar_flux}(b),(e) for $\langle\widehat{\Pi}\rangle$ in 2D and 3D,
and figures \ref{fig:self_similar_flux}(c),(f) for $ \langle\widehat{\Lambda}\rangle$ in 2D and 3D indicate that they indeed collapse onto each other. In figures \ref{fig:self_similar_flux}(b),(e),
the horizontal axis is rescaled by the mixing width $h(t)/L_z$. To account for the temporal function $F_\pi(t)$, we rescaled the plots at $\widehat{t}=3, 4$ so that their maxima match with that at $\widehat{t}=2$. The same rescaling amplitudes for $\langle\widehat{\Pi}\rangle$ are also applied to $\langle\widehat{\Lambda}\rangle$, as is shown in
figures \ref{fig:self_similar_flux}(c),(f), which indicates that $F_\pi(t)\approx F_\lambda(t)$. We have found that $F_\pi(t)$ and $F_\lambda(t)$ follow a scaling close to $t^3$ both in 2D and 3D (not shown). This $t^3$ growth of the fluxes can be understood from the cumulative amount of potential energy released, which follows $\delta P\sim h^2\sim t^4$ \citep{Alphagroup}, implying that the rate of kinetic energy injection should scale as $\langle\epsilon^{\textrm{inj}}\rangle\sim t^3$. This energy is eventually channeled by the fluxes $\Lambda$ and $\Pi$, which grow in magnitude accordingly.

While plots of the fluxes show good collapse in figure \ref{fig:self_similar_flux}, it is worth discussing the relatively poor collapse of $\langle\widehat{\Pi}\rangle$ in 3D (figure \ref{fig:self_similar_flux}(e)) at scales smaller than that of the peak. The reason $\langle\widehat{\Pi}\rangle$ decays at those scales is due to viscous dissipation removing the kinetic energy available to cascade. As the RT flow evolves and there is more kinetic energy cascading, the dissipation scale $\ell_d$ should become smaller as $\ell_d\sim t^{-1/4}$ as shown in \citet{Ristorcelli04} by invoking Kolmogorov's turbulence phenomenology. This is reflected in $\langle\widehat{\Pi}\rangle$ extending to smaller scales at later times in figure \ref{fig:self_similar_flux}(e). Therefore, the dissipation dynamics is not expected to follow the temporal self-similarity of inertial scales captured in equation \eqref{eq:self_similar_flux}. Note that the $\langle\widehat{\Pi}\rangle$ cascade in 2D, being upscale, is not arrested by viscous dissipation at the smallest scales, which explains the much better collapse of $\langle\widehat{\Pi}\rangle$ in figure \ref{fig:self_similar_flux}(b).


\section{Directionally split analysis of RT anisotropy} \label{sec:split_direction}
In RT flows the kinetic energy injection acts only in the vertical direction due to the downward gravitational acceleration. In addition,
the initial and boundary conditions are different in the horizontal and vertical directions.  Therefore, anisotropy both in the bulk flow and as a function of scale is expected in RT flows. Using DNS data, \citet{CabotZhou13} examined RT anisotropy mostly based on bulk flow quantities such as the Taylor microscale, Kolmogorov scale, and velocity derivatives, and using other statistical measures such as correlation functions. These quantities do not rely on a proper scale decomposition. They also analyzed anisotropy as a function of scale but only in the homogeneous directions using Fourier transforms to measure spectra. They found that large scale dynamics (based on the bulk flow and Fourier spectra) in RT is highly anisotropic, but that the flow becomes more isotropic at small scales, where velocity derivatives appear more isotropic than the velocity field itself. In a follow-up work, \citet{ZhouCabot19POF} studied the time evolution of anisotropy, mixing, and scaling in RT instability, extending their previous one-snapshot analysis to different times and Atwood numbers. Other works on anisotropy of RT relied on analyzing the normalized deviatoric part of the Reynolds stress tensor, $b_{ij}=\langle u_i u_j\rangle / \langle u_k u_k\rangle-1/3\delta_{ij}$
to measure the anisotropy in RT flows \citep{Livescu09,Livescu10,Banerjee10,Zhou17-2}, whose values are bounded between $-1/3$ and $2/3$. Nonzero values of $b_{ij}$ indicate deviation from isotropy, and typical values of $b_{33}\approx 0.3$ or $0.35$ are observed with DNS RT data \citep[e.g.][]{Livescu09}. 
The above works relied primarily on statistical approaches, such as Reynolds averaging, to analyze RT anisotropy, which are not capable of identifying anisotropy as a function of scale. On the other hand, a Fourier scale decomposition is limited to homogeneous (horizontal) directions. A more refined analysis is possible with our coarse-graining approach, which analyzes the scale-by-scale anisotropy in RT flows, as we shall discuss now and also in Supplementary Material, section \ref{suppsec:scaleanisotropy}.

\subsection{Directionally split anisotropy at scales}
Here, we investigate anisotropy of RT flows at different scales in physical space. To quantify the contribution to the kinetic energy budget from different directions, we first formulate the budgets along each direction separately, and then diagnose the constituent terms from numerical data. We call this `directionally split' analysis of anisotropy, which is different from the analysis of \emph{scale anisotropy} done in Supplementary Material, section \ref{suppsec:scaleanisotropy}.
Similar to the filtered total KE equation (\ref{eq:KE_budget}), we can get the filtered KE equations in separate directions:
\begin{align}
    \partial_t\bar{\rho}_\ell \frac{|\widetilde{\boldsymbol{u}}_\xi |^2}{2}+\nabla\cdot \boldsymbol{J}_\ell^\xi =
    -\Pi^\xi_\ell-\Lambda^\xi_\ell+\bar{P}_\ell \partial_\xi\bar{u}_\xi-D^\xi_\ell+\epsilon_\ell^{inj,\xi} \label{eq:KE_budget_split}
\end{align}
where $\xi$ is a symbol denoting a single direction $x$, $y$, or $z$. Note that since we are not concerned with scale anisotropy here, the filtering kernel is still isotropic as before,
$G(\br) = G(|\br|)$. The budget terms in the above equation are defined by:
\begin{equation} \label{eq:KE_budget_xyz_terms}
\begin{aligned}
& \Pi^\xi_\ell(\boldsymbol{x})=-\bar{\rho} \partial_j \tilde{u}_\xi \tilde{\tau}(u_\xi,u_j), \quad 
    \Lambda^\xi_\ell(\boldsymbol{x})=\frac{1}{\bar{\rho}}\partial_\xi \bar{P} \bar{\tau}(\rho,u_\xi), \quad
    D^\xi_\ell(\boldsymbol{x})=\partial_j \tilde{u}_\xi \bar{\sigma}_{\xi j} \\
    & J^\xi_j(\boldsymbol{x})=\bar{\rho} \frac{\tilde{u}_\xi^2}{2}\tilde{u}_j+\bar{P}\bar{u}_\xi\delta_{\xi j}+\tilde{u}_\xi \bar{\rho} \tilde{\tau}(u_\xi,u_j)-\tilde{u}_\xi\bar{\sigma}_{\xi j}, \quad 
    \epsilon^{inj,\xi}_\ell(\boldsymbol{x})=\tilde{u}_\xi\bar{\rho}\tilde{F}_\xi
\end{aligned}
\end{equation}
whose forms are similar to their counterparts in the total KE budget equation (\ref{eq:KE_budget}). The sum of these equations in all directions reduces to the 
full budget equation (\ref{eq:KE_budget}). Note that in the above definitions, the subscript $\xi$ represents only one direction, and $\xi$ itself does not follow the Einstein summation convention.

\begin{figure}
\centering 
\begin{minipage}[b]{1.0\textwidth}  
\centering
\subfigure[\footnotesize{2D result}]
{\includegraphics[height=1.6in]{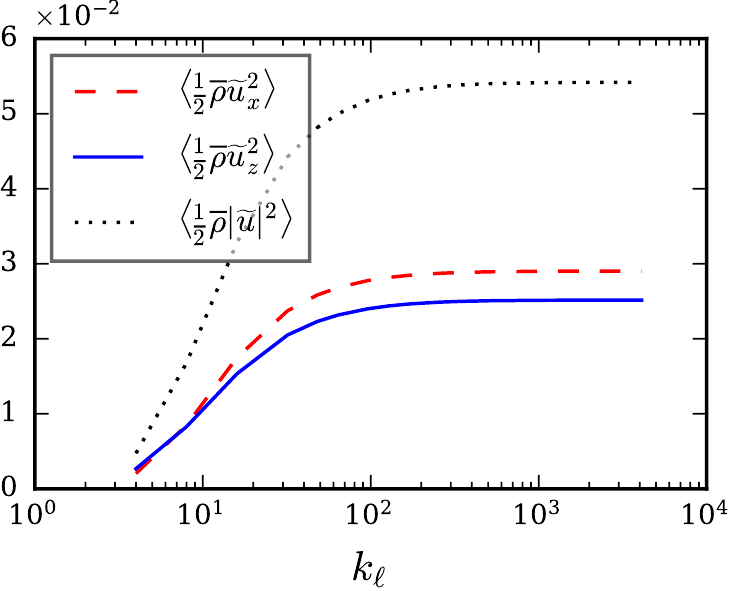} \label{fig:Favre_KE_aniso_2D}} 
\phantom{}
\subfigure[\footnotesize{3D result}]
{\includegraphics[height=1.6in]{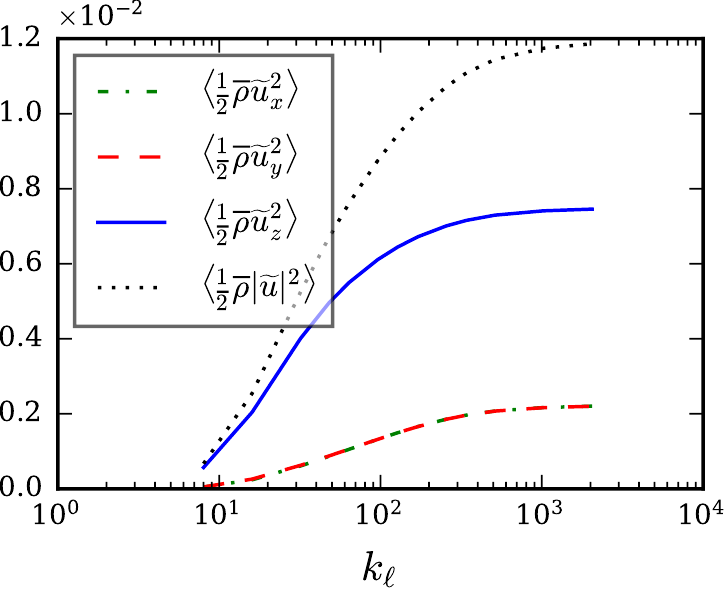} \label{fig:Favre_KE_aniso_3D}}
\caption{\footnotesize{Directional anisotropy of RT flows. Coarse-grained kinetic energy of the flow in different directions as a function of scale in 2D and 3D. Their sum over all directions is also shown for reference. Data is at dimensionless time $\widehat{t}=4.0$. These are cumulative quantities, showing energy at all scales larger than $\ell=L_z/k_\ell$. Energy content at any $k_\ell$ (\textit{i.e.} the spectrum) is obtained from the $k_\ell$-derivative of these plots.} \label{fig:Favre_KE_aniso}}
\end{minipage}
\end{figure}

Before delving into details of the budgets in separate directions, we shall first check the coarse-grained kinetic energy in each direction.
This is shown as a function of filtering wavenumber $k_\ell=L_z/\ell$ in figure \ref{fig:Favre_KE_aniso} for 2D and 3D RT data.
In 2D, mean KE is approximately isotropic at all scales, with the horizontal component slightly greater than the vertical at large scales. This may seem counterintuitive since
kinetic energy is injected in the vertical direction. 
In contrast, we see that in 3D, horizontal kinetic energy in $x$ and $y$ directions are almost equal, but are more than three times smaller than KE in the vertical at large scales. This indicates that in the horizontal plane, 3D-RT turbulent flow is isotropic as expected, but the vertical flow is much stronger than the horizontal at large scales, consistent with results of \citet{CabotZhou13} based on Fourier spectra in the homogeneous direction. 
Thus, we have different behaviors for 2D and 3D RT. In 2D-RT, the flow seems to be close to isotropic, while it is highly anisotropic in 3D-RT. Note that in figure \ref{fig:Favre_KE_aniso}, derivatives of the plots with respect to $k_\ell$ give rise to the filtering spectra defined in equation (\ref{eq:nonperiodic_spectra}). At small scales, the $k_\ell$-derivatives of both 2D and 3D filtered KE are small and comparable in all directions. At large scales, the filtering spectrum indicates that the flow is approximately isotropic in 2D (figure \ref{fig:Favre_KE_aniso_2D}) and highly anisotropic in 3D (figure \ref{fig:Favre_KE_aniso_3D}), where the vertical large-scale flow contains more energy than the horizontal large-scale flow. We will now investigate this difference carefully by analyzing the energy budgets in separate directions.

\begin{figure}
\centering 
\begin{minipage}[b]{1.0\textwidth}  
\centering
\subfigure
{\includegraphics[height=1.6in]{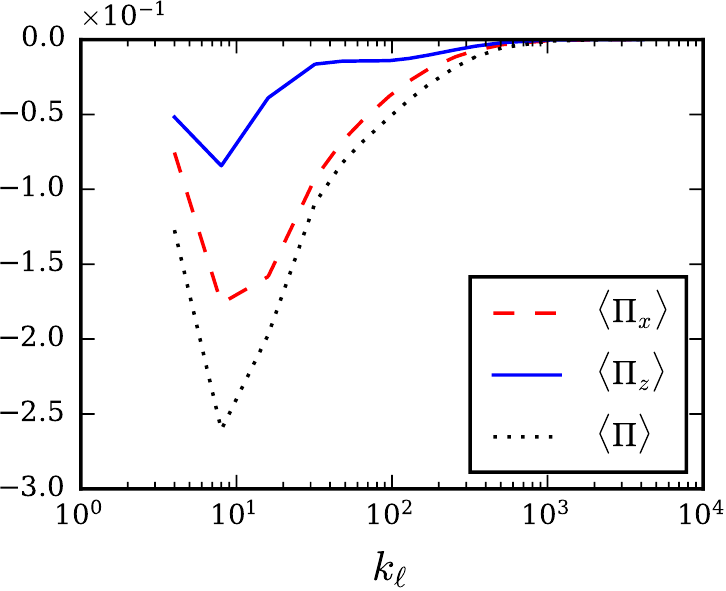} 
\llap{\parbox[b]{1.8in}{{\large (a)}\\\rule{0ex}{1.2in}}}
    \label{fig:2D_4096_aniso1}} 
\phantom{}
\subfigure
{\includegraphics[height=1.6in]{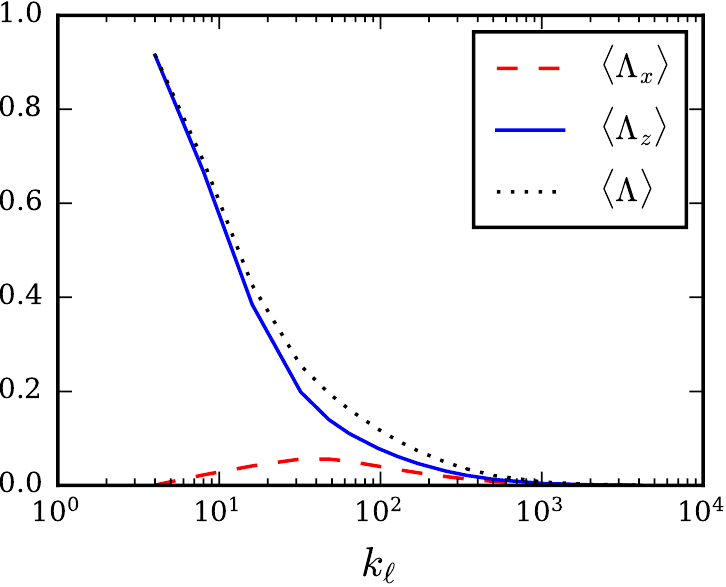}
\llap{\parbox[b]{1.8in}{{\large (b)}\\\rule{0ex}{1.2in}}}
    \label{fig:2D_4096_aniso2}} \\
\subfigure
{\includegraphics[height=1.6in]{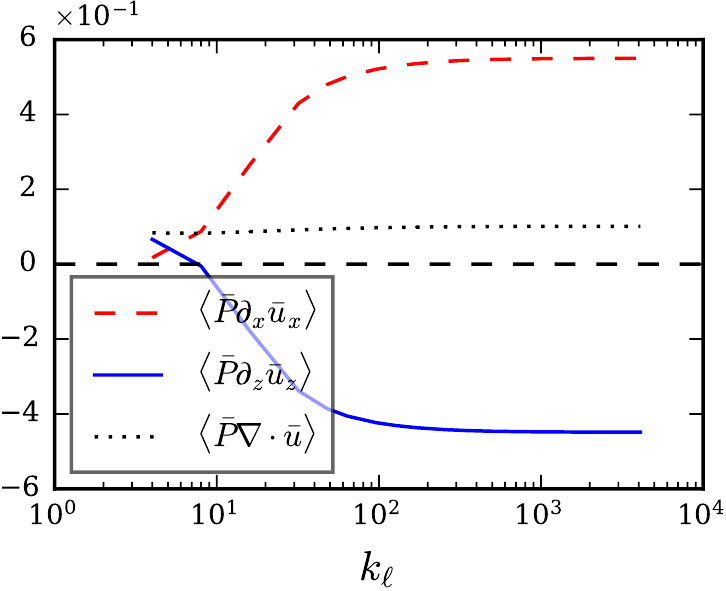}
\llap{\parbox[b]{1.8in}{{\large (c)}\\\rule{0ex}{1.2in}}}
    \label{fig:2D_4096_aniso3}} 
\phantom{}
\subfigure
{\includegraphics[height=1.6in]{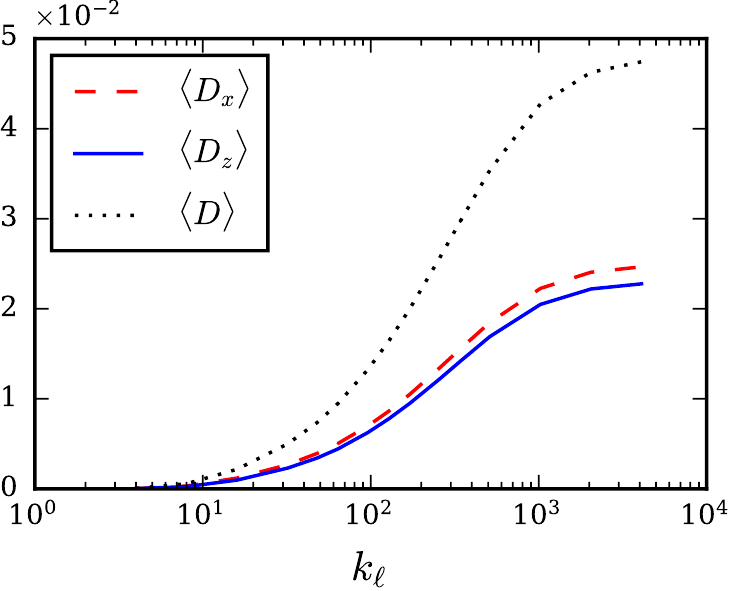}
\llap{\parbox[b]{1.2in}{{\large (d)}\\\rule{0ex}{1.2in}}}
    \label{fig:2D_4096_aniso4}} 
\caption{\footnotesize{2D-RT directionally split budget terms for filtered kinetic energy using the 2D4096 data at $\widehat{t}=4$. Filtering wavenumber
    $k_\ell=L_z/\ell$. Plots are normalized by $\langle \epsilon^{\textrm{inj}} + P\nabla\cdot \bu\rangle$. Panels show deformation work, baropycnal work, pressure-dilatation, and dissipation.} \label{fig:2D_4096_aniso}}
\end{minipage}
\end{figure}

We show in figure \ref{fig:2D_4096_aniso} mean values of the normalized budget terms in equation (\ref{eq:KE_budget_xyz_terms}) for the 2D4096 simulation at time $\widehat{t}=4.0$. Budget terms corresponding
to the evolution of horizontal KE, vertical KE, and the full KE are shown. In figure \ref{fig:2D_4096_aniso1}, $\Pi_x$ is larger in magnitude than $\Pi_z$, in accordance with the larger magnitude of kinetic energy in the horizontal 
direction in 2D-RT. The $\Lambda$ term resides mostly in the vertical component $\Lambda_z$, while $\Lambda_x$ is
almost zero. This indicates that $\Lambda$ is mainly responsible for transferring injected energy to vertical scales, and does not redistribute kinetic energy among different directions. 

The next term in figure \ref{fig:2D_4096_aniso} is the (negative) pressure dilatation, which is a cumulative quantity, similar to filtered KE in figure \ref{fig:Favre_KE_aniso}, meaning that its contribution at scale $\ell$
is determined by its derivative with respect to filtering wavenumber $k_\ell=L_z/\ell$. That is, its contribution at scale $\ell$ equals $\frac{\partial}{\partial k}\langle \OL{P}_\ell\nabla\cdot\OL{\bu}_\ell\rangle\bigg\rvert_{k=L_z/\ell}$.
Therefore, as indicated by figure \ref{fig:2D_4096_aniso3}, pressure dilatation acts as a source for mean horizontal kinetic energy,
since $\langle \OL{P} \partial_x \OL{u}_x\rangle$ increases with respect to scale,
so that its contribution to horizontal KE at scale $\ell=L_z/k_\ell$ measured by the derivative is positive.
For the mean vertical kinetic energy, the vertical component $\langle \OL{P} \partial_z \OL{u}_z\rangle$ at the largest scale is positive and acts as a source,
but at all subsequent scales its mean value decreases with increasing filtering wavenumber, $\frac{d}{dk}\langle \OL{P} \partial_z \OL{u}_z\rangle < 0$,
indicating it is a sink term at these scales. 
The combined value in all directions gives rise to a pressure dilatation that is almost constant on average with respect to filtering wavenumber.
Therefore, pressure dilatation acts to mix up horizontal and vertical velocities and isotropize the kinetic energy in different directions, similar to the role of pressure-strain in constant-density
turbulence \citep{Pope2001}. 
Considering dissipation in figure \ref{fig:2D_4096_aniso}, the two components have comparable magnitude, but since there is little kinetic energy 
residing at small scales due to the fast decay of spectrum and a net upscale cascade by $\Pi$, the magnitude of dissipation in 2D is far smaller than other budget terms.

\begin{figure}
\centering 
\begin{minipage}[b]{1.0\textwidth}  
\centering
\subfigure
{\includegraphics[height=1.6in]{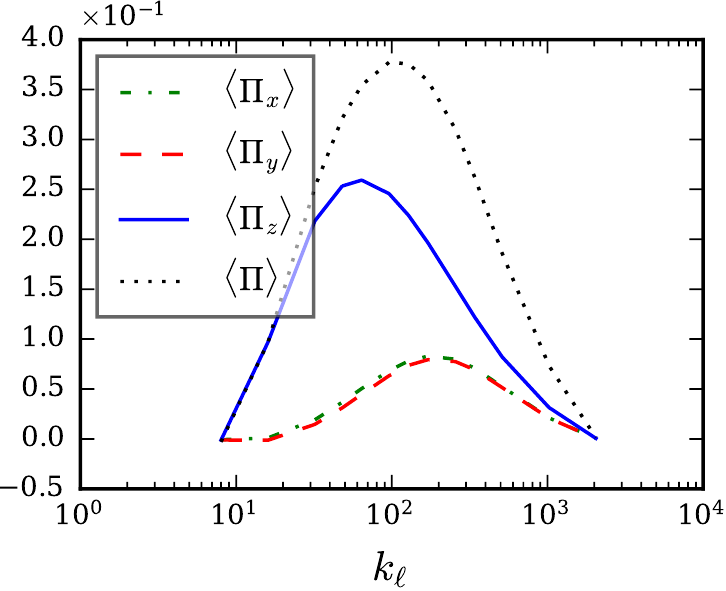} \llap{\parbox[b]{0.7in}{{\large (a)}\\\rule{0ex}{1.2in}}} \label{fig:3D_1024_aniso1}} 
\phantom{}
\subfigure
{\includegraphics[height=1.6in]{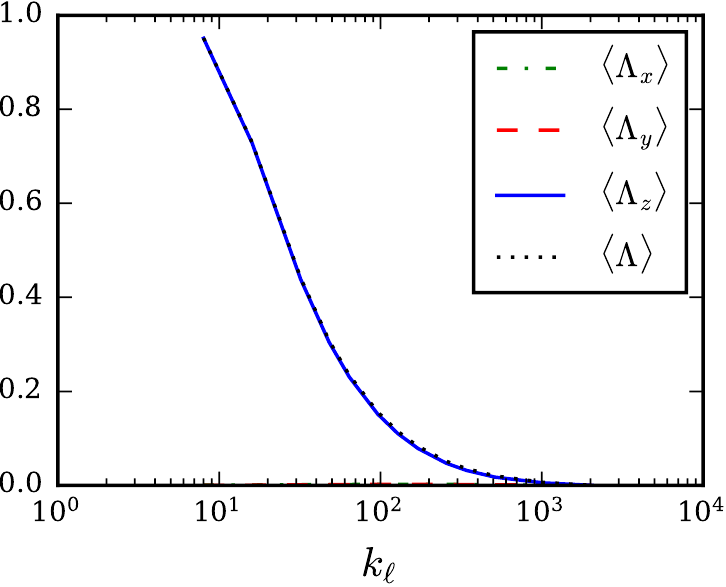} \llap{\parbox[b]{1.8in}{{\large (b)}\\\rule{0ex}{1.2in}}} \label{fig:3D_1024_aniso2}} \\
\subfigure
{\includegraphics[height=1.6in]{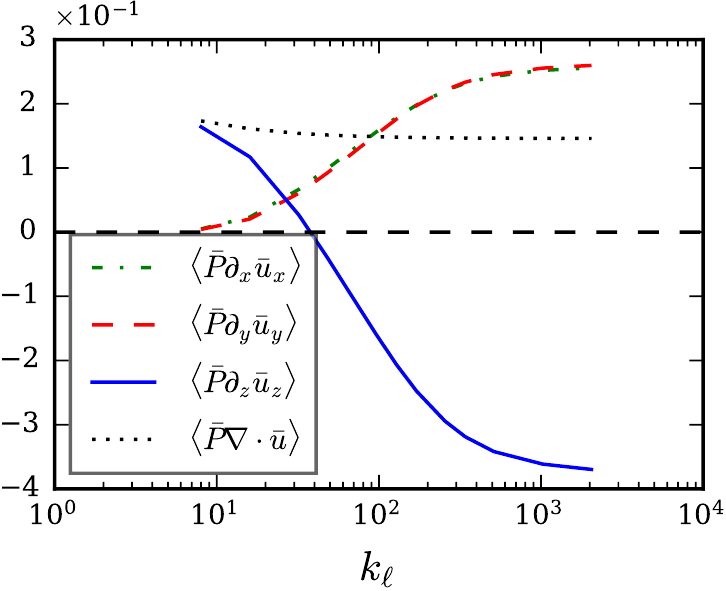} \llap{\parbox[b]{1.8in}{{\large (c)}\\\rule{0ex}{1.2in}}} \label{fig:3D_1024_aniso3}} 
\phantom{}
\subfigure
{\includegraphics[height=1.6in]{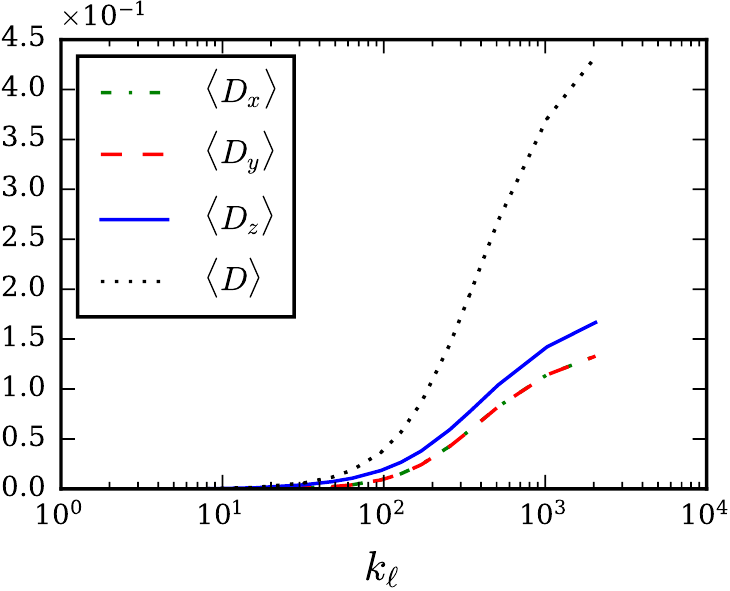} \llap{\parbox[b]{1.2in}{{\large (d)}\\\rule{0ex}{1.2in}}} \label{fig:3D_1024_aniso4}} 
\caption{\footnotesize{Same as in figure \ref{fig:2D_4096_aniso} but for 3D-RT using the 3D1024 data at $\widehat{t}=4$.} \label{fig:3D_1024_aniso}}
\end{minipage}
\end{figure}

The directionally split budget terms for 3D-RT are shown in figure \ref{fig:3D_1024_aniso}. In all four subplots, the $x$ and $y$ components are equal, indicating isotropy in the horizontal plane. However, anisotropy in the vertical direction is more pronounced than in 2D. $\langle\Pi_z\rangle$ in the vertical direction
is much larger than the horizontal components, contrary to the 2D case where the horizontal $\langle\Pi_x\rangle$ is larger in magnitude. Also in 3D, $\langle \Pi_x\rangle$ and $\langle \Pi_y \rangle$ peak at smaller scales compared to $\langle \Pi_z\rangle$, indicating a strengthening of the cascade in the horizontal direction at smaller scales, where it becomes more comparable to $\langle \Pi_z \rangle$. Mean $\Lambda$ is only significant
in the vertical direction, with almost zero values in the horizontal directions, similar to 2D. Also, pressure dilatation is the source of kinetic energy input in the horizontal directions, similar to 2D.  Dissipation is more isotropic in all directions, consistent with a trend towards isotropy of $\langle \Pi\rangle$ at these small scales, although the vertical component attains a slightly higher value.

In summary, the directionally split energy budgets in 2D and 3D RT indicates that mean injection and its conduit $\Lambda$ contribute only to the vertical component of kinetic energy.
Horizontal kinetic energy is forced by the horizontal component of pressure dilatation in both 2D and 3D. Dissipation, which is at the smallest scales, is approximately isotropic. However, significant differences between 2D and 3D appear in the fluxes $\langle \Pi_x\rangle$ and $\langle \Pi_z\rangle$ at inertial scales.

  In the earlier discussion of this section, we observed a sharp contrast between 2D and 3D in figure \ref{fig:Favre_KE_aniso}, which indicated that the 2D flow is surprisingly isotropic at large-scales, while it is highly anisotropic in 3D due to the expected stronger vertical velocity.
  The reason can be gleaned from the foregoing results, especially those in figures \ref{fig:2D_4096_aniso} and \ref{fig:3D_1024_aniso}. The primary difference between the directionally split budgets in 2D and 3D is due to the cascade $\Pi$. In 2D, figure \ref{fig:2D_4096_aniso}(a) shows that $\Pi_x$ is an energy \emph{source} for mean horizontal large-scale KE, and similarly, $\Pi_z$ is a source for vertical large-scale KE. However, the magnitude of $\langle \Pi_x\rangle$ is approximately two to three times greater than $\langle \Pi_z\rangle$, indicating that large-scale KE gains more energy from $\Pi$ in the horizontal direction than in the vertical, offsetting anisotropy due to $\Lambda$, which energizes the vertical flow. Thus, the inverse cascade $\Pi$ acts alongside pressure-dilatation to isotropize the large-scale flow, where most of the kinetic energy resides. In contrast, results from 3D-RT indicate that $\langle\Pi_x\rangle$ and $\langle\Pi_y\rangle$ are positive, acting as an energy \emph{sink} for horizontal large-scale KE. Therefore, the large-scale horizontal flow in 3D has only pressure-dilatation as a source of energy, unlike the vertical flow that is being driven by $\Lambda$, which ultimately leads to the large anisotropy in 3D-RT. 
  
  While the anisotropy analyzed above is due to horizontal and vertical components of the flow, another measure of anisotropy
is that of length-scales by using anisotropic filtering kernels as we do in the Supplementary Material, section \ref{suppsec:scaleanisotropy}.

\section{Conclusions \label{sec:conclude}}
In this work, we have undertaken a thorough analysis of the energy pathways across scales in RT turbulence. We coarse-grained the dynamics using the Favre decomposition to probe nonlinear processes at different scales without resorting to Fourier transforms. In particular, this allowed us to probe scales in the inhomogeneous (vertical) direction, which had not been done before in variable density (non-Boussinesq) RT flows. Our approach does not assume the \textit{a priori} presence of turbulence  and is applicable to general multi-scale flows that may not be considered as turbulent in the traditional sense.

We focused on the role of two flux terms responsible for kinetic energy transfer across scales. One of these, baropycnal work $\Lambda_\ell$, has  been often neglected in LES modeling and lumped with pressure dilatation. \citet{Aluie11c,Aluie13} had identified $\Lambda_\ell$ as being essential in the transfer of energy across scales in variable density flows such as RT, arising from work done by large-scale pressure gradients on small-scales of density and velocity. Here, we have presented a first analysis of $\Lambda_\ell$ in RT flows and found that it acts as a conduit for potential energy, transferring energy non-locally from the largest scales to smaller scales in the inertial range where the other flux term, $\Pi_\ell$, takes over the cascade process.

Unlike $\Lambda_\ell$, which is inherently due to variations in density, deformation work, $\Pi_\ell$, also arises in constant-density flows and may be thought of as representing the process of vortex stretching in 3D, or of vortex mergers in 2D. In 3D-RT, we have shown how $\Pi_\ell$ takes the energy delivered by $\Lambda_\ell$ at inertial scales $\ell$ and transfers it via a persistent cascade to smaller scales, until it is dissipated at the viscous scales. On the other hand, we have shown how 2D-RT is fundamentally different from 3D, since $\Pi_\ell$ re-channels the energy back to larger scales, driving a positive feedback loop that is absent in 3D-RT. In this feedback loop, the net upscale cascade leads to an enhanced growth of the mixing layer, which in turn leads to enhanced potential energy release by raising the light fluid and lowering the heavy fluid. The excess PE released results in enhanced baropycnal energy transfer by $\Lambda_\ell$ to smaller scales which is re-channeled back upscale by $\Pi_\ell$, closing the feedback loop. Despite higher bulk kinetic energy levels in 2D compared to 3D as a result of this feedback loop, small inertial scales are weaker than in 3D.
Our findings also indicate the absence of a net upscale energy transfer in 3D-RT as is often claimed, and that growth of bubbles and spikes at the largest scales is solely due to baropycnal work $\Lambda_\ell$ in 3D.

Our scale analysis demonstrated the presence of an inertial range of scales in RT flows over which kinetic energy fluxes $\Lambda_\ell$ and $\Pi_\ell$ are not directly influenced by either external inputs at the largest scales or by dissipation at the smallest scales. Our measurement of $\Lambda_\ell$ and $\Pi_\ell$ over this scale-range revealed that they exhibit a self-similar evolution that is quadratic in time, $\sim t^2$, similar to the RT mixing layer.

It had been shown by many prior studies that the mixing layer width $h=\alpha \mathcal{A} g t^2$ grows quadratically in time. It had also been often observed that the growth coefficient $\alpha$ is significantly larger in 2D than in 3D, despite the precise value of $\alpha$ still being an open problem. Our identification of the positive feedback loop in 2D-RT serves to explain this discrepancy as arising from the fundamentally different scale dynamics between 2D and 3D. The possibility of an upscale cascade had been suggested as a possible reason in prior work \citep{Cabot06POF} but, to our knowledge, never demonstrated by an analysis of energy scale transfer in variable density RT flows before now. 
We also showed that the net upscale cascade by $\Pi_\ell$ in 2D tends to isotropize the large-scale flow, in stark contrast to the high level of anisotropy in 3D at large-scales.  These differences in the fundamental flow processes highlight the stark disparity between the evolution of 2D-RT and 3D-RT, and pinpoint the potentially misleading physics when employing 2D-RT simulations to model real-world 3D systems.


\section*{Acknowledgement}
We thank X. Bian, E. Blackman, and D. Livescu for valuable discussions. We also thank the anonymous reviewers for their time and valuable suggestions, which helped improve our manuscript. This research was funded by DOE FES grants DE-SC0014318 and DE-SC0020229. Partial support from NNSA award DE-NA0003856 is acknowledged. HA was also supported by NASA grant 80NSSC18K0772, NSF grant PHY-2020249, DOE grant DE-SC0019329, and NNSA grant DE-NA0003914. Computing time was provided by the National Energy Research Scientific Computing Center (NERSC) under Contract No. DE-AC02-05CH11231, and by an award from the INCITE program, using resources of the Argonne Leadership Computing Facility, which is a DOE Office of Science User Facility supported under Contract DE-AC02-06CH11357. We also acknowledge using data from the JHU Turbulence database (\url{http://turbulence.pha.jhu.edu/}). 


\section*{Declaration of interests}
The authors report no conflict of interest.

\appendix

\section{Effects of Interface Mach number} \label{sec:append_2DRT_Ma}
To investigate the influence of interface Mach number \citep[e.g.][]{wieland2017effects} on our conclusions, we performed another 2D-RT simulation using $1024\times 2048$ grid points, at a lower Mach number, which we name 2D1024lowM in Table \ref{tab:parameter}. To achieve this, we set the initial pressure at the bottom boundary to be 100, instead of 10 as in simulations 2D4096 and 3D1024, in order to lower the interface Mach number. The interface Mach number, $Ma = \sqrt{\frac{g\lambda}{\gamma R T_0}}$, is based on the ratio between characteristic gravity wave speed $\sqrt{g\lambda}$, where $\lambda=L_x$ is the width of the domain, and sound speed at the interface $\sqrt{\gamma R T_0}$, where $T_0$ is initial temperature at the interface, and $R$ is the specific gas constant \citep{reckinger2016comprehensive}. The interface Mach number for simulations 2D4096 and 3D1024 in Table \ref{tab:parameter} is 0.45, while in simulation 2D1024lowM, it is 0.13.

\begin{figure}
\centering
\begin{minipage}[b]{1.0\textwidth}
\centering
    \subfigure{\includegraphics[height=1.7in]{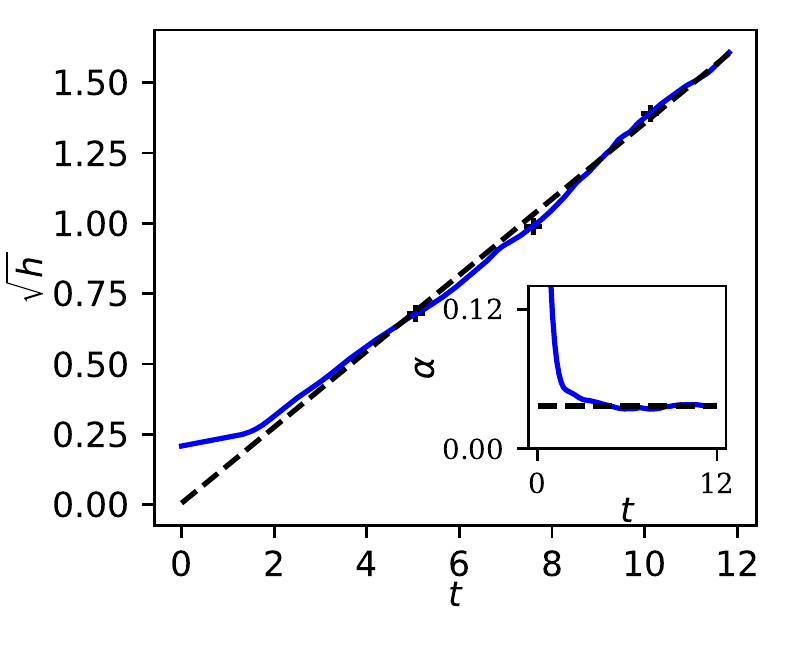}}
    \caption{\footnotesize{Similar to figure \ref{fig:mixingwidth}, the square root of mixing width $\sqrt{h(t)}$ versus time for the simulation 2D1024lowM with low Mach number, and the plot corresponds to $\alpha=0.037$. The `+' markers correspond to dimensionless time $\widehat{t}=t/\sqrt{\frac{L_x}{\mathcal{A}g}}= 2, 3, 4$. Inset: the compensated plot $\alpha = \frac{h(t)}{\mathcal{A}gt^2}$ versus time, in which the horizontal lines correspond to the $\alpha$ value obtained by linear fit. } \label{fig:appendix_2d_mixing_width_lowMa}}
\end{minipage}
\end{figure}

\begin{figure}
\centering
\begin{minipage}[b]{1.0\textwidth}
\centering
    \subfigure
{\includegraphics[height=1.5in]{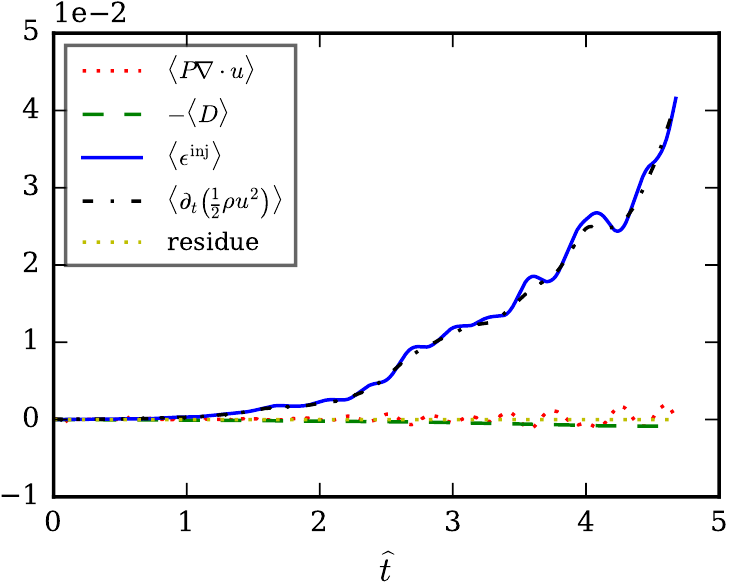}}
\phantom{}
    \subfigure
{\includegraphics[height=1.5in]{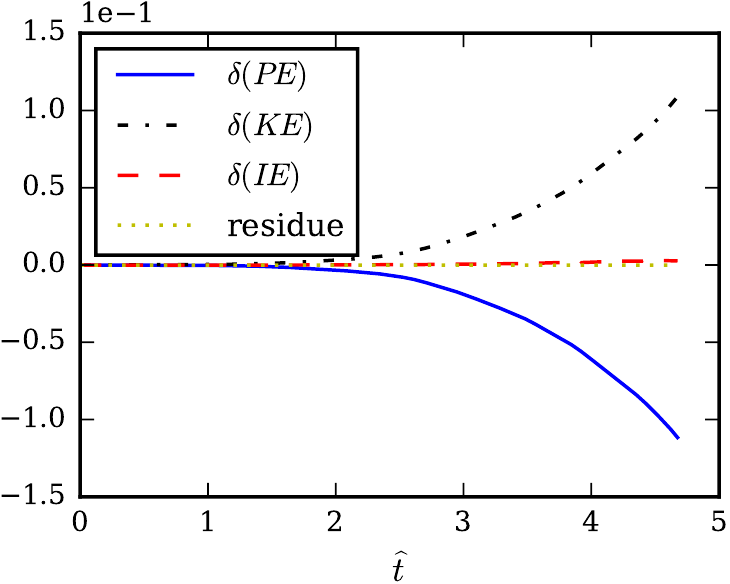}}
    \caption{\footnotesize{Left and right figures are temporal evolution of kinetic energy budget, and overall energy balance for the simulation 2D1024lowM, similar to figure \ref{fig:ke_detailed_balance_2d} and \ref{fig:ke_budge_2d}.} \label{fig:appendix_ke_detailed_balance_2d_lowMa}}
\end{minipage}
\end{figure}

\begin{figure}
\centering
\begin{minipage}[b]{0.75\textwidth}
\centering
    \subfigure{\includegraphics[height=2.5in]{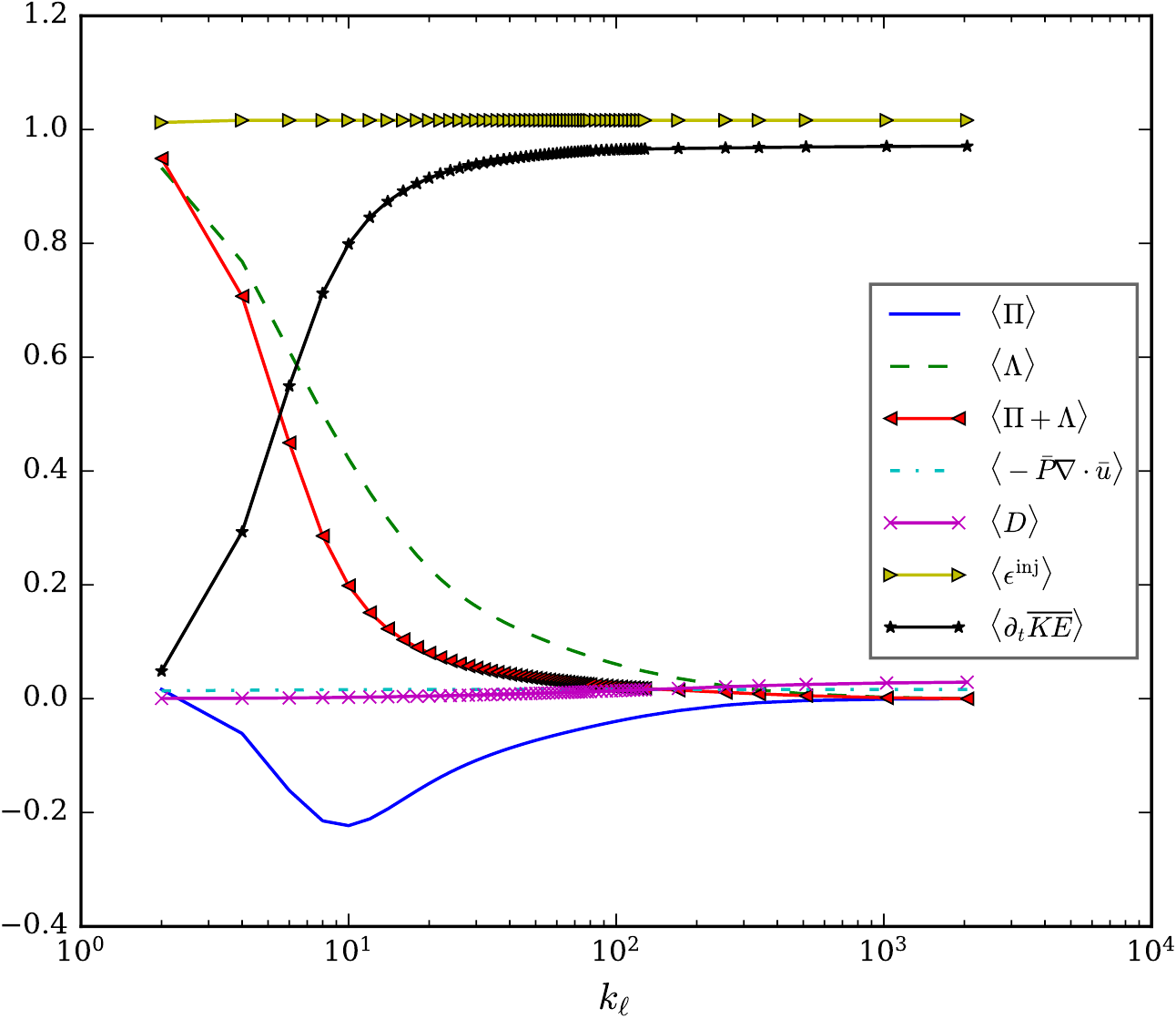}}
    \caption{\footnotesize{Similar to figure \ref{fig:2D4096_cascade}, mean kinetic energy budget as a function of scale in 2D at dimensionless time $\widehat{t}=4.0$ for the simulation 2D1024lowM.
    The plot is normalized by $\langle \epsilon^{\textrm{inj}} + P\nabla\cdot \bu\rangle$, the  available mean source of kinetic energy. } \label{fig:append_2D1024_cascade_lowMa}}
\end{minipage}
\end{figure}

\begin{figure}
\centering 
\begin{minipage}[b]{1.0\textwidth}  
\subfigure
{\includegraphics[height=1.3in]{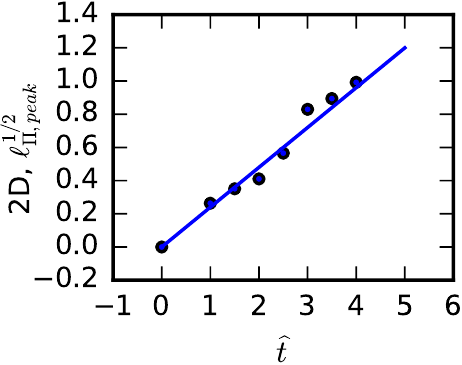} 
\llap{\parbox[b]{1.1in}{{\large (a)}\\\rule{0ex}{0.9in}}}
  } 
\subfigure
{\includegraphics[height=1.33in]{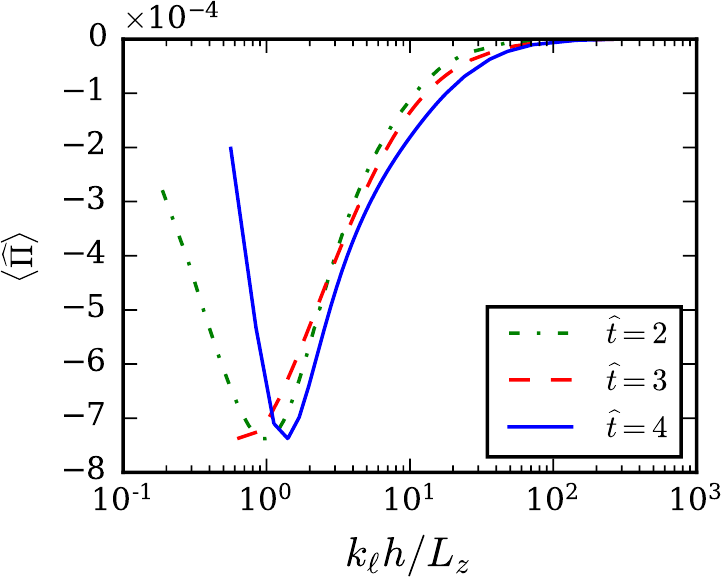} 
\llap{\parbox[b]{1.25in}{{\large (b)}\\\rule{0ex}{0.9in}}}
    } 
\subfigure
{\includegraphics[height=1.33in]{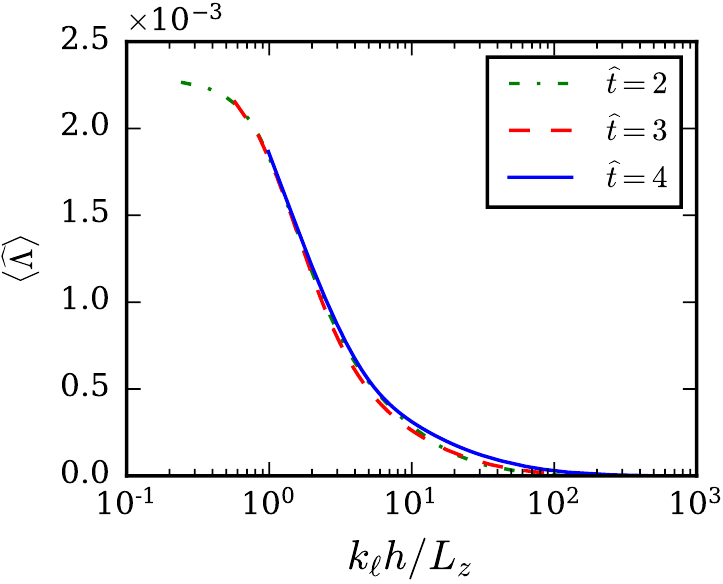} 
\llap{\parbox[b]{1.3in}{{\large (c)}\\\rule{0ex}{0.9in}}}
    }
\caption{\footnotesize{Similar to figure \ref{fig:self_similar_flux}, panels (a)-(c) show the temporal self-similarity of turbulent RT fluxes for the simulation 2D1024lowM. 
    Panel (a) plot length scale $\ell_{\Pi, \text{peak}}$ associated with the peak of $\Pi$ versus dimensionless time $\widehat{t}$. Panel (b) shows rescaled $\langle\widehat{\Pi}\rangle$ in
    equation (\ref{eq:self_similar_flux}). Panel (c) shows rescaled $\langle\widehat{\Lambda}\rangle$.} \label{fig:appendix_self_similar_flux_lowMa}}
\end{minipage}
\end{figure}

Results from run 2D1024lowM are similar to 2D4096, except for pressure-dilatation as would be expected. Figure \ref{fig:appendix_2d_mixing_width_lowMa} shows the mixing width versus time, with growth rate $\alpha=0.037$. Figure \ref{fig:appendix_ke_detailed_balance_2d_lowMa} shows the instantaneous and overall energy budget versus non-dimensional time. The major difference compared to 2D4096 is that pressure-dilatation is greatly suppressed, and the cumulative change of internal energy $\delta(IE)$ plays a minor role in 2D1024lowM compared to 2D4096. This is consistent with physical expectations when reducing the compressibility of the RT flow by lowering its Mach number. Figure \ref{fig:append_2D1024_cascade_lowMa} plots the KE budget terms as a function of scale. All results are consistent with those in figure \ref{fig:2D4096_cascade} from run 2D4096, except for the suppressed pressure-dilatation.
 Finally, figure \ref{fig:appendix_self_similar_flux_lowMa} shows that energy fluxes at low Mach number also exhibit a  self-similar behavior.  

\section{Comparison with 2-species incompressible RT}\lb{sec:2-species}
 Throughout the paper we use the single-species two-density model to describe RT flows, but there are other models commonly used in the literature.
One notable example is the two-species incompressible model \citep{sandoval1995thesis,Cook01,Cabot04,Livescu07,Livescu08}.
Here, we compare the energy budgets between the two-species model and our single-species two-density RT model to highlight similarities and differences. 

Two-species incompressible RT falls into the category of variable density flows, similar to the compressible model. Governing equations consist of transport equations for each of the two species of mass fraction $Y_1$ and $Y_2$, and densities $\rho_1$ and $\rho_2$, respectively. Average density $\rho$ satisfies $\frac{1}{\rho}=\frac{Y_1}{\rho_1}+\frac{Y_2}{\rho_2}$, where $Y_1+Y_2=1$. The dynamics reduces to \citep{sandoval1995thesis,Cook01,Livescu07}:
\begin{align} 
    \partial_t\rho + \partial_j (\rho u_j) &= 0, \label{eq:two_species_mass}\\
    \partial_t(\rho u_i) + \partial_j (\rho u_i u_j) &= -\partial_i P + \partial_j \sigma_{ij} + \rho g_i \label{eq:two_species_momentum}\\
    \partial_j u_j &= -\partial_j(\mathcal{D}\partial_j \ln{\rho}). \label{eq:two_species_diffusion}
\end{align}
Equation (\ref{eq:two_species_diffusion}) represents the non-zero velocity divergence due to the local change in volume when mixing incompressible fluids of different densities \citep[e.g.][]{BaltzerLivescu2020}.
$\sigma_{ij}$ is viscous stress tensor as in the compressible equations, and $\mathcal{D}$ is the binary species diffusivity. The system described by equation (\ref{eq:two_species_mass})-(\ref{eq:two_species_diffusion}) can be obtained \citep{Livescu04} by taking the $\mathrm{Ma}\to 0$ limit of the compressible equations (\ref{eq:continuity})-(\ref{eq:kinetic}).

Compared to equations (\ref{eq:continuity})-(\ref{eq:kinetic}) for compressible dynamics, a key difference is that the two-species model here does not account for internal energy and its coupling to the flow via pressure-dilatation and viscous dissipation. In fact, the primary difference between these two models is a finite (single-species two-density model) versus a zero (two-species model) Mach number \citep{Livescu04,Livescu20}. If we decompose the pressure in equations (\ref{eq:continuity})-(\ref{eq:kinetic}) into a mean $P_0$ and fluctuation $p'$, $P=P_0+p'$, and we take the limit $P_0 \to \infty$, then we approach the $\mathrm{Ma}\to 0$ limit in the compressible model. Meanwhile, the mean of internal energy equation (\ref{eq:internal}) in this infinite $P_0$ limit reduces to $\partial_j u_j = -\partial_j(\alpha \partial_j \ln{\rho})$, which is similar to equation (\ref{eq:two_species_diffusion}), except that mass diffusivity $\mathcal{D}$ in equation (\ref{eq:two_species_diffusion}) is replaced by thermal diffusivity $\alpha$ \citep{Livescu2013}. Thermal diffusivity is defined as $\alpha = \frac{\kappa}{\rho c_p}$, where $\kappa$ is thermal conductivity and $c_p$ is the specific heat at constant pressure. Hence, in the $\mathrm{Ma}\to 0$ limit, the two models are equivalent.  Similar to the analysis we have done in section \ref{sec:coarse-grainedBudgets}, by coarse-graining the two-species dynamics, we obtain
a large-scale kinetic energy budget of the same form as the filtered KE budget (\ref{eq:KE_budget}).

For a quantitative comparison, we download and analyze the homogeneous buoyancy driven Rayleigh-Taylor data from the Johns Hopkins Turbulence Database \citep{JHU_database, Livescu07}. The database stores the output of an incompressible 
two-species RT simulation at a $1024^3$ grid resolution. A total of 1015 snapshots are stored densely in time. Unlike our 3D-RT flow presented earlier, this flow is statistically homogeneous in all directions, with a triply periodic domain of equal dimensions, $L_x=L_y=L_z=2\pi$. A spatially uniform pressure gradient is imposed externally and changes with the bulk mixing between species \citep{Livescu07}. The two species are initialized as random blobs, with a characteristic size $\approx1/5$ of the domain. Flow is driven from rest due to buoyancy, eventually leading to turbulence. However, since the two fluids are miscible, potential energy is ultimately depleted as the fluids mix, leading to turbulence decay. The Atwood number is 0.05, which is close to the Boussinesq limit, but the velocity divergence is nonzero due to the diffusion between the two fluids \citep{JHU_database, Livescu07}. We download a snapshot at time $t=6.56$ when the Reynolds number $Re$ is maximum, and another at $t=14.56$ when the dissipation rate peaks.


\begin{figure}
\centering 
\begin{minipage}[b]{1.0\textwidth}  
\subfigure[\footnotesize{At time of maximum $Re$}]
{\includegraphics[height=2.1in]{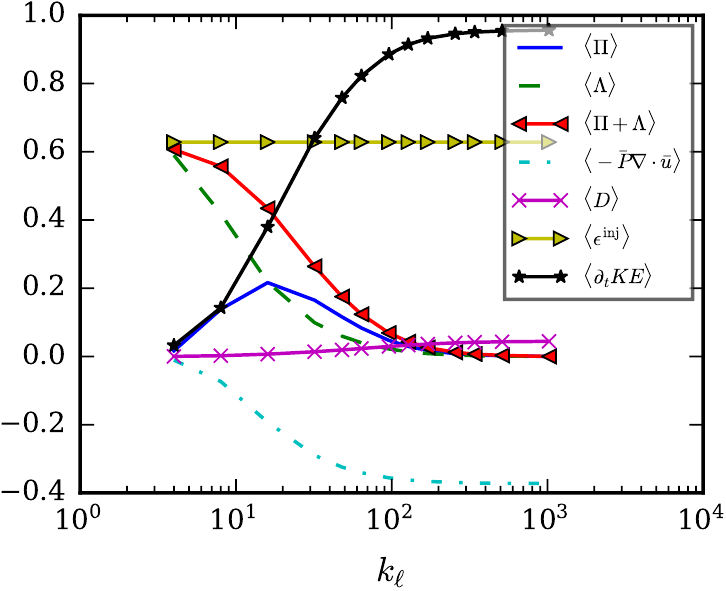} \label{fig:JHU_develop}} 
\subfigure[\footnotesize{At time of maximum dissipation}]
{\includegraphics[height=2.1in]{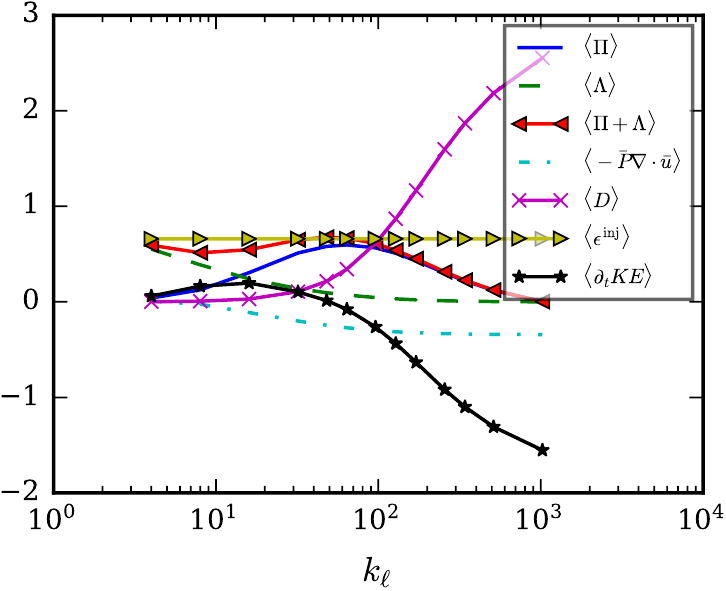} \label{fig:JHU_decay}} 
\caption{\footnotesize{Kinetic energy processes as a function of scale from the 
homogeneous incompressible two-species RT data \citep{JHU_database,Livescu07}.
Two different snapshots are analyzed, (a) at time $t=6.56$ when $Re$ is maximum, and
(b) at time $t=14.56$ when dissipation is maximum. Filtering wavenumber $k_\ell=L_z/\ell$ is a proxy for length-scale. Plots are normalized by $\langle \epsilon^{\textrm{inj}} + P\nabla\cdot \bu\rangle$ at the corresponding time. The legend is the same as in figures \ref{fig:2D4096_cascade},\ref{fig:3D1024_cascade}}. \label{fig:JHU_cascade}}
\end{minipage}
\end{figure}

We study the scale dynamics following our earlier analysis in sections \ref{sec:ke_inertial}-\ref{sec:ke_fluxes}.
Normalized kinetic energy budget terms are shown in figure \ref{fig:JHU_cascade},
with a normalization factor $\langle \epsilon^{\textrm{inj}} + P\nabla\cdot \bu\rangle$ at the corresponding time.

Figure \ref{fig:JHU_develop} shows the result at early time $t=6.56$ when the RT is still in the developing stage, which is qualitatively similar to our 3D-RT  budgets in figure \ref{fig:3D1024_cascade}. Mean injection is constant in scale, and pressure dilatation saturates at small scales but with a slower convergence rate compared to our results using the compressible Navier-Stokes equations. Fluxes $\Pi$ and $\Lambda$ have a similar scale behavior as in our 3D-RT case. Specifically, both are positive transfering energy to small scales, and $\Pi$ peaks at filtering wavenumber $k_\ell\approx20$. During this developing stage of the flow, the forward cascade via $\Pi$ is arrested not by dissipation but by growth of KE at scales smaller than $\ell \approx L_z/20$. This can be seen from the plot of $\langle  \partial_t \OL{\text{KE}}\rangle$, which keeps increasing as a function of $k_\ell$ until $k_\ell\approx200$ (positive $k_\ell$-derivative indicates those scales are experiencing a net gain of KE). Dissipation at this time of the flow (figure \ref{fig:JHU_develop}) is negligible.

At the later time when dissipation is maximum (figure \ref{fig:JHU_decay}), we observe negative values of $\langle  \partial_t \OL{\text{KE}}\rangle$, indicating decaying turbulence at that time step. Available potential energy is close to being exhausted in this decaying phase of RT, so that the mean injection and $\Lambda$ become relatively small, while dissipation dominates all other terms. This stage of the homogeneous flow is never reached in our inhomogeneous RT simulations since the effects of walls becomes important at those times. Note in figure \ref{fig:JHU_decay} that unlike the earlier time (figure \ref{fig:JHU_develop}), the forward cascade via $\Pi$ operates across a much broader scale range and is arrested by viscous dissipation.

In summary, RT using the  two-species incompressible model and  RT using the single-species two-density  model both belong to the category of variable density flows, and exhibit  scale processes in the KE evolution that are  similar in many aspects.

\bibliographystyle{jfm}
\bibliography{RT_citation,Compressible_citation,Turbulence_citation,Numerical_citation}

\clearpage
\newpage
\makeatletter 
\renewcommand{\thefigure}{S\@arabic\c@figure}
\makeatother
\setcounter{figure}{0} 
 \renewcommand{\theequation}{S-\arabic{equation}}
\setcounter{equation}{0}  
\renewcommand{\thepage}{{\it Supp. Mat. -- \arabic{page}}} \setcounter{page}{1}

\section*{Supplementary Information for ``Scale interactions and anisotropy in Rayleigh-Taylor turbulence'' by Dongxiao Zhao, Riccardo Betti, and Hussein Aluie}
\renewcommand{\thesubsection}{\Alph{subsection}}

\subsection{Efficiency of the fluxes} \label{suppsec:flux_efficiency}
We have observed (figures \ref{fig:2D4096_cascade}-\ref{fig:3D1024_cascade} in the main text) that baropycnal work $\langle\Lambda_\ell\rangle$ dominates at the largest scales. It acts as a conduit for mean injection from potential energy by transferring energy from the largest scales to smaller scales in the inertial range where deformation work $\Pi$ takes over and dominates the cascade process. 

Since this work is the first to investigate $\Lambda_\ell$ as a function of scale $\ell$, a basic question that arises is whether its decay at small $\ell$ is due to a decay in its magnitude or due to a mis-alignment of $\grad\OL{P}_\ell$ and $\OL{\tau}_\ell(\rho,\bu)$. We shall answer the question in this subsection.
We first reiterate the flux expressions in equation (\ref{eq:budget_defs}) in the main text for convenience: 
\begin{align*}
    \begin{split}
        \Pi&=-\bar{\rho}\partial_j \wt{u}_i \wt{\tau}(u_i,u_j)=-\bar{\rho} \wt{S}_{ij} \wt{\tau}(u_i,u_j) \\
        \Lambda&= \frac{1}{\OL{\rho}} \partial_i \OL{P} \OL{\tau}(\rho, u_i)
    \end{split}
\end{align*}
where $\wt{S}_{ij}=1/2(\partial_i\wt{u}_j+\partial_j\wt{u}_i)$ is the Favre filtered velocity strain tensor. Following (\citeSupp{Liao14,FangOuellette16,Ouellette18}), 
we define the cascade efficiency $\Gamma_{\Lambda}$ and $\Gamma_{\Pi}$ to be 
\begin{align} \label{eq:flux_efficiency}
    \begin{split}
    \Gamma_\Lambda &=\frac{\partial_i \bar{P}\ \OL{\tau}(\rho, u_i)}{(\partial_j \bar{P} \partial_j \bar{P})^{1/2} (\OL{\tau}(\rho, u_k) \OL{\tau}(\rho, u_k))^{1/2}} \\
    \Gamma_\Pi &= -\frac{\wt{S}_{ij}\  \wt{\tau}(u_i,u_j)}{(\wt{S}_{kl}\wt{S}_{kl})^{1/2}(\wt{\tau}(u_m,u_n)\wt{\tau}(u_m,u_n))^{1/2}}
    \end{split}
\end{align} 
which resembles the cosine of the angle between two vectors. These definitions in equation \eqref{eq:flux_efficiency} follow that in \citeSupp{Ouellette18} who studied the cascade efficiency in constant-density turbulence. By the Cauchy-Schwartz inequality, 
$-1\le\Gamma_\Lambda,\Gamma_\Pi\le1$.

Similarly, the magnitudes of these two fluxes, denoted by $M_\Lambda$ and $M_\Pi$, are defined by the denominator in the flux efficiency equation (\ref{eq:flux_efficiency}):
\begin{align}
\begin{split}
    M_\Lambda&=(\partial_j \bar{P} \partial_j \bar{P})^{1/2} (\OL{\tau}(\rho, u_k) \OL{\tau}(\rho, u_k))^{1/2}\\
M_\Pi&=(\wt{S}_{kl}\wt{S}_{kl})^{1/2}(\wt{\tau}(u_m,u_n)\wt{\tau}(u_m,u_n))^{1/2}
\end{split}
\end{align}
and can be regarded as the product of tensor magnitudes. The actual $\Pi$ and $\Lambda$ terms are proportional to the product of efficiency and magnitude
$$\Lambda \propto \Gamma_\Lambda \cdot M_\Lambda,\qquad \Pi \propto \Gamma_\Pi \cdot M_\Pi$$
with a proportionality factor involving density.

\begin{figure}
\centering 
\begin{minipage}[b]{1.0\textwidth}  
    \subfigure[\footnotesize{PDF of $\Gamma_\Lambda$ in 2D-RT. Mean efficiency equals 0.50, 0.48, 0.44 for $\ell=L_z/8$, $L_z/32$, $L_z/128$, respectively.}]
    {\includegraphics[height=2.1in]{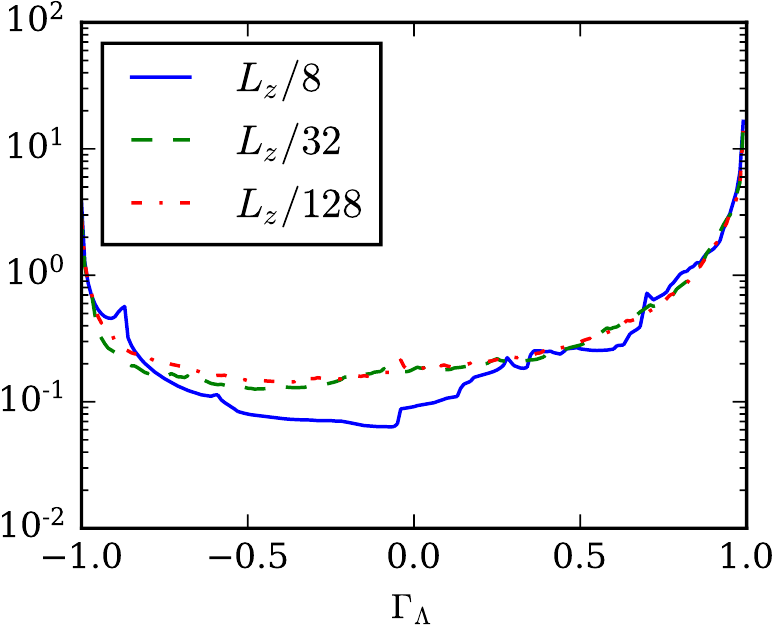} \label{fig:lambda_eff_mag_a}} \phantom{ }
\subfigure[\footnotesize{PDF of $M_\Lambda$ in 2D-RT. Mean magnitude equals 7.8E-3, 3.7E-3, 1.68E-3 for $\ell=L_z/8$, $L_z/32$, $L_z/128$, respectively.}]
{\includegraphics[height=2.1in]{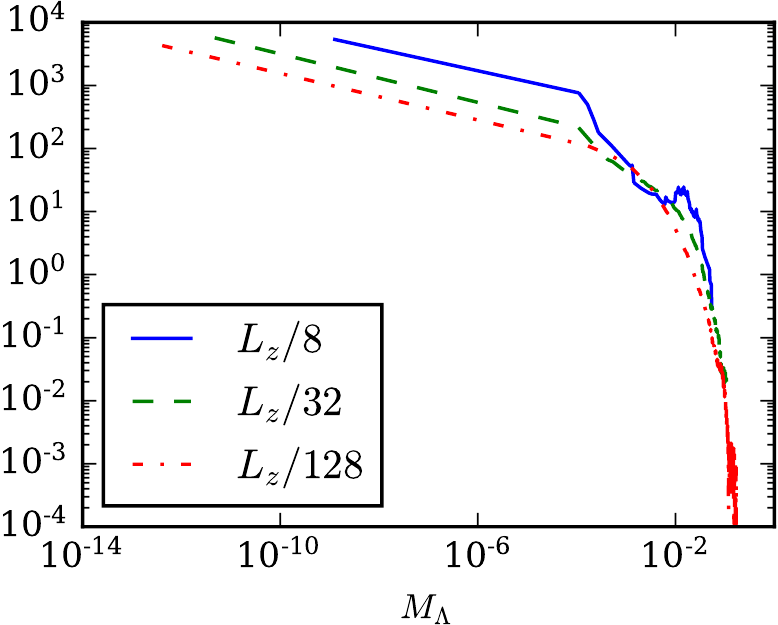} \label{fig:lambda_eff_mag_b}} \\
\subfigure[\footnotesize{PDF of $\Gamma_\Lambda$ in 3D-RT. Mean efficiency equals 0.71, 0.70, 0.54 for $\ell=L_z/8$, $L_z/32$, $L_z/128$, respectively.}]
{\includegraphics[height=2.1in]{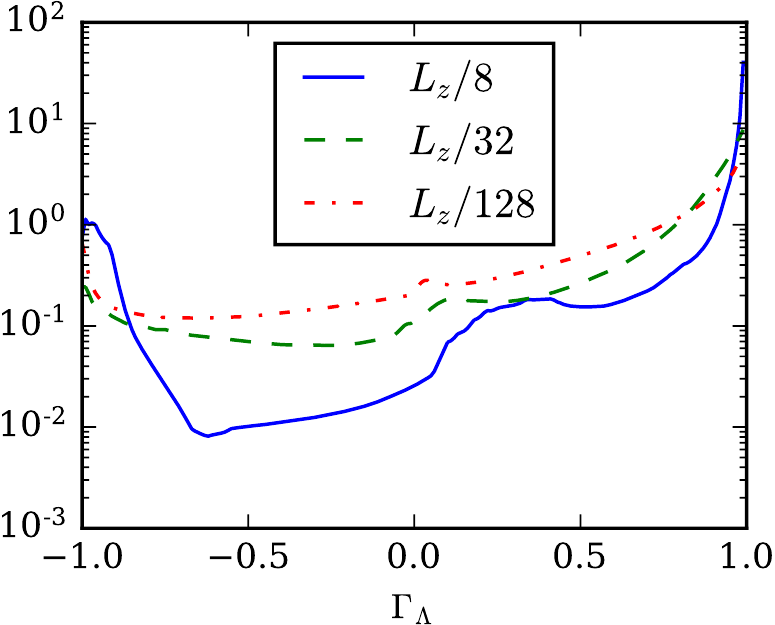} \label{fig:lambda_eff_mag_c}} \phantom{ }
\subfigure[\footnotesize{PDF of $M_\Lambda$ in 3D-RT. Mean magnitude equals 6.1E-3, 3.0E-3, 1.2E-3 for $\ell=L_z/8$, $L_z/32$, $L_z/128$, respectively.}]
{\includegraphics[height=2.1in]{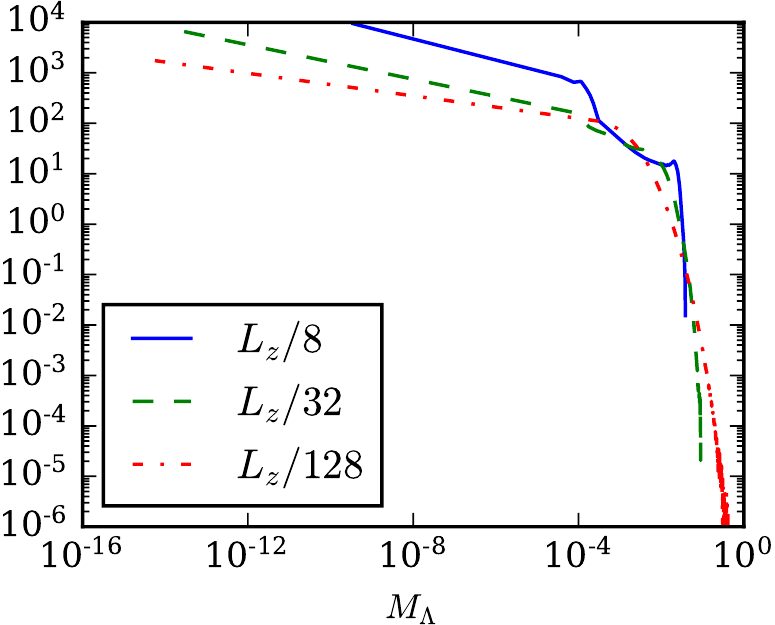} \label{fig:lambda_eff_mag_d}} 
\caption{\footnotesize{PDFs of the efficiency and magnitude of $\Lambda$ at different scales from the 2D4096 and 3D1024 data at time $\widehat{t}=4$.
    The PDFs of efficiency are on a linear-log plot, while the PDFs of magnitude are on a log-log plot for clarity.} \label{fig:lambda_eff_mag}}
\end{minipage}
\end{figure}

\begin{figure}
\centering 
\begin{minipage}[b]{1.0\textwidth}  
\centering
    \subfigure{\includegraphics[height=2in]{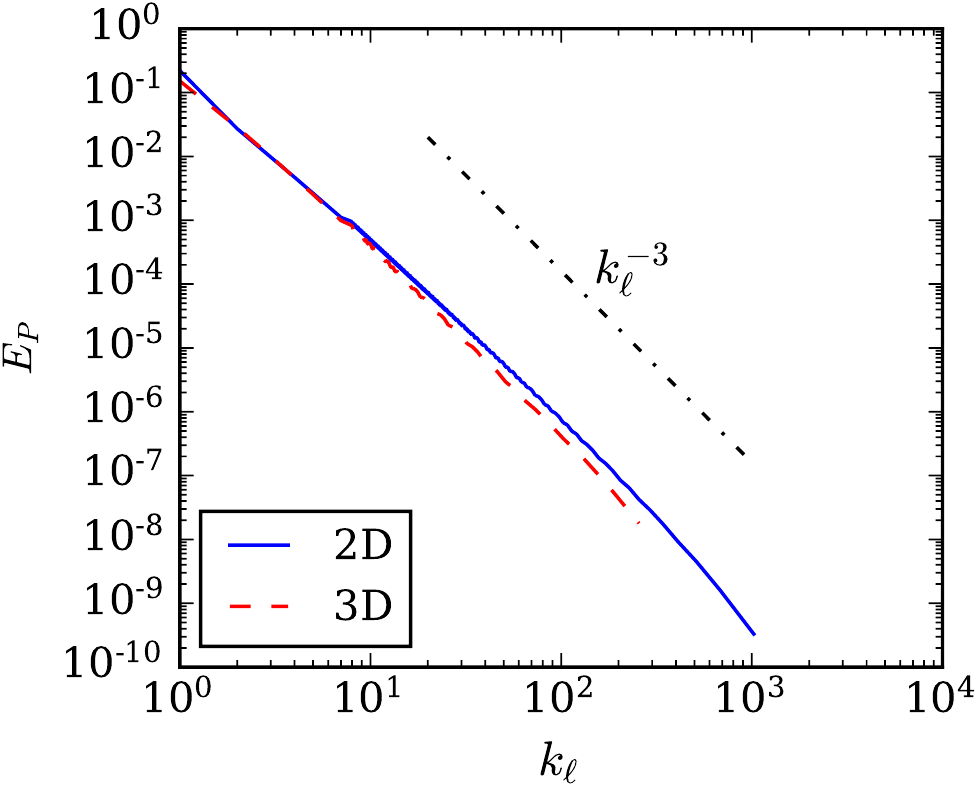} \llap{\parbox[b]{0.5in}{{\Large (a)}\\\rule{0ex}{1.6in}}} } 
    \subfigure{\includegraphics[height=2in]{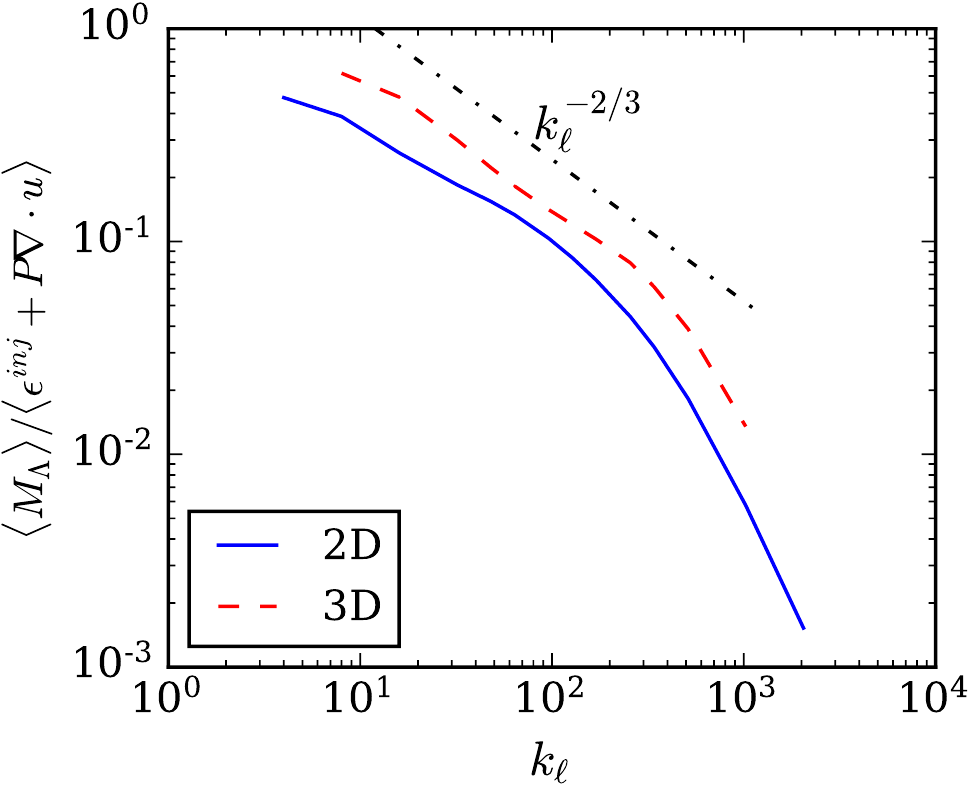} \llap{\parbox[b]{0.5in}{{\Large (b)}\\\rule{0ex}{1.6in}}} } \\
    \subfigure{\includegraphics[height=2in]{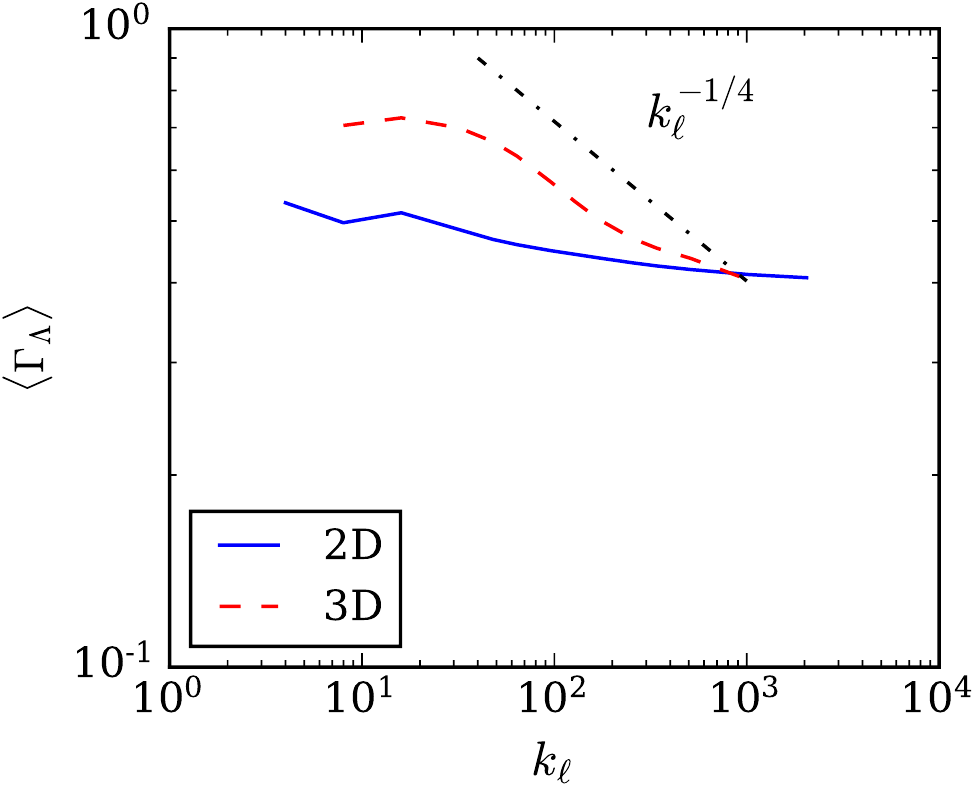} \llap{\parbox[b]{0.5in}{{\Large (c)}\\\rule{0ex}{1.6in}}} } 
    \subfigure{\includegraphics[height=2in]{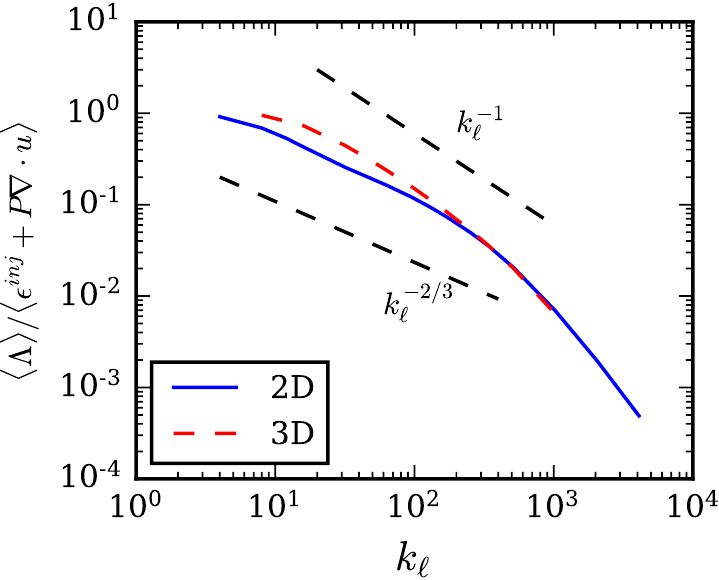} \llap{\parbox[b]{0.5in}{{\Large (d)}\\\rule{0ex}{1.6in}}} } 
    \caption{\footnotesize{Explaining the decay of $\Lambda$ at small scales. (a) Filtering spectrum, $E_P$, of pressure using the 2D4096 and 3D1024 data at $\widehat{t}=4$. $E_P\equiv \frac{d}{dk}\langle|\OL{P}_\ell|^2\rangle$, where the $k_\ell$-derivative is in the filtering wavenumber $k_\ell=L_z/\ell$ following equation \eqref{eq:nonperiodic_spectra} in the main text.
    (b) Scaling of $M_\Lambda$ in 2D and 3D.
    The normalization, $\langle \epsilon^{\textrm{inj}}+P \nabla\cdot \bu \rangle$, is calculated from the corresponding 2D and 3D data. (c) Mean efficiency of $\Lambda$ as functions of scale in 2D \& 3D, both of which show a weak dependence on $k_\ell$.  (d) Decay of $\langle\Lambda\rangle$ in 2D \& 3D as functions of scale.} \label{fig:lambda_2D_3D_scaling}}
\end{minipage}
\end{figure}

\begin{figure}
\centering 
\begin{minipage}[b]{1.0\textwidth}  
\subfigure[\footnotesize{PDF of $\Gamma_\Pi$ in 2D-RT. Mean efficiency equals -0.018, -0.012, -0.0092 for $\ell=L_z/8$, $L_z/32$, $L_z/128$, respectively.}]
{\includegraphics[height=2.1in]{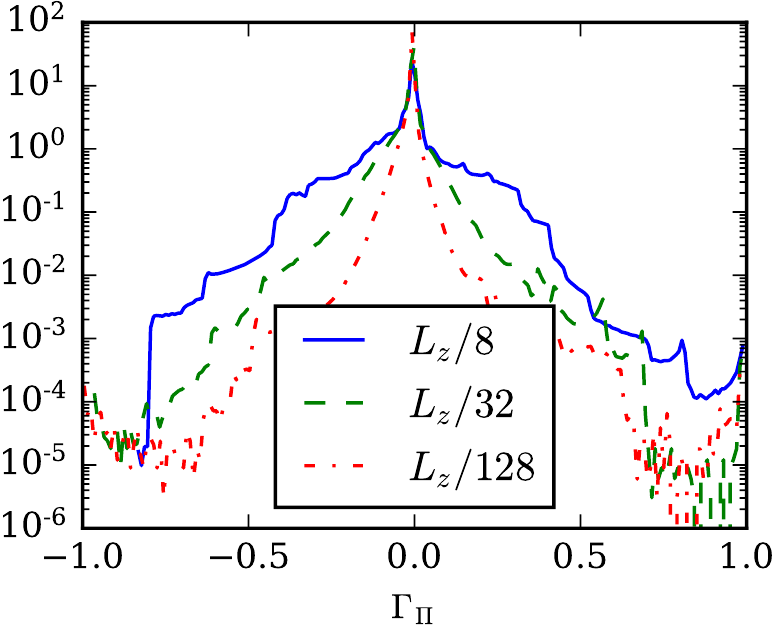} \label{fig:pi_eff_mag_a}} \phantom{ }
\subfigure[\footnotesize{PDF of $M_\Pi$ in 2D-RT. Mean magnitude equals 0.064, 0.11, 0.093 for $\ell=L_z/8$, $L_z/32$, $L_z/128$, respectively.}]
{\includegraphics[height=2.1in]{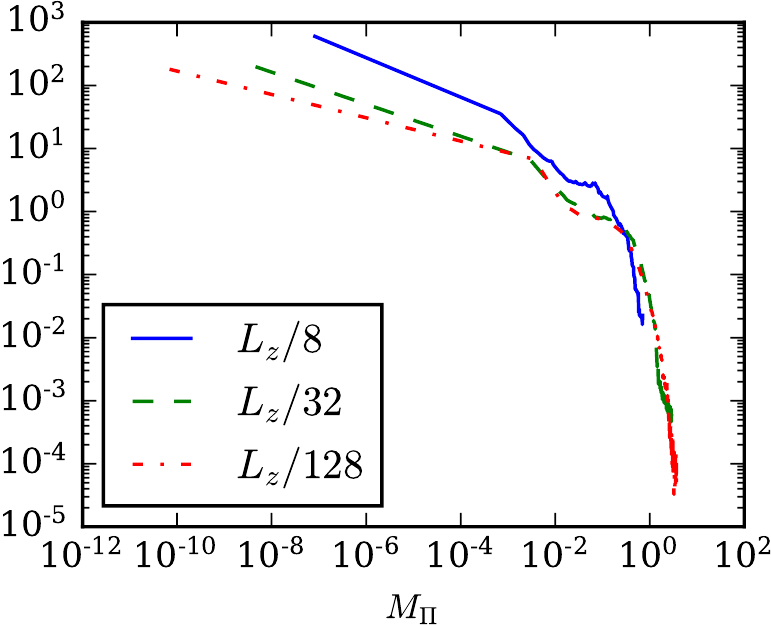} \label{fig:pi_eff_mag_b}} \\
\subfigure[\footnotesize{PDF of $\Gamma_\Pi$ in 3D-RT. Mean efficiency equals 2E-3, 5E-3, 2.2E-3 for $\ell=L_z/8$, $L_z/32$, $L_z/128$, respectively.}]
{\includegraphics[height=2.1in]{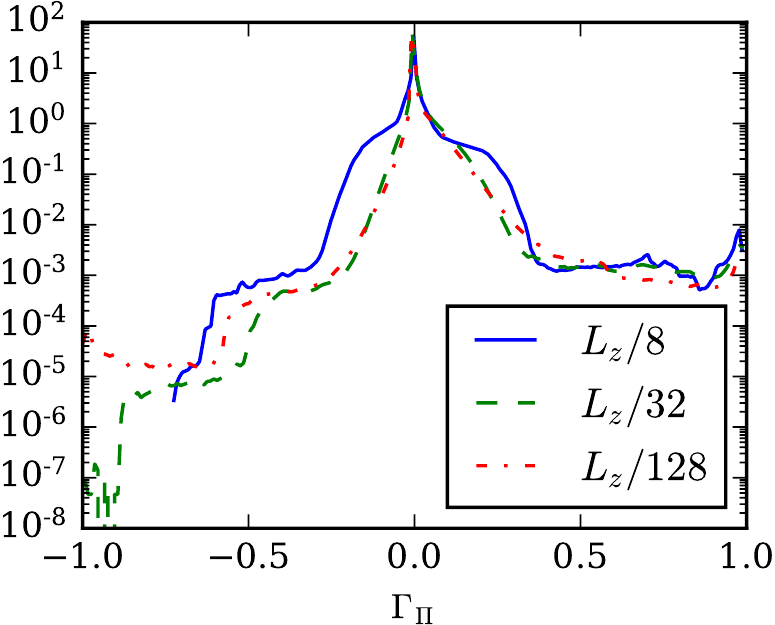} \label{fig:pi_eff_mag_c}} \phantom{ }
\subfigure[\footnotesize{PDF of $M_\Pi$ in 3D-RT. Mean magnitude equals 5.5E-2, 0.26, 0.75 for $\ell=L_z/8$, $L_z/32$, $L_z/128$, respectively.}]
{\includegraphics[height=2.1in]{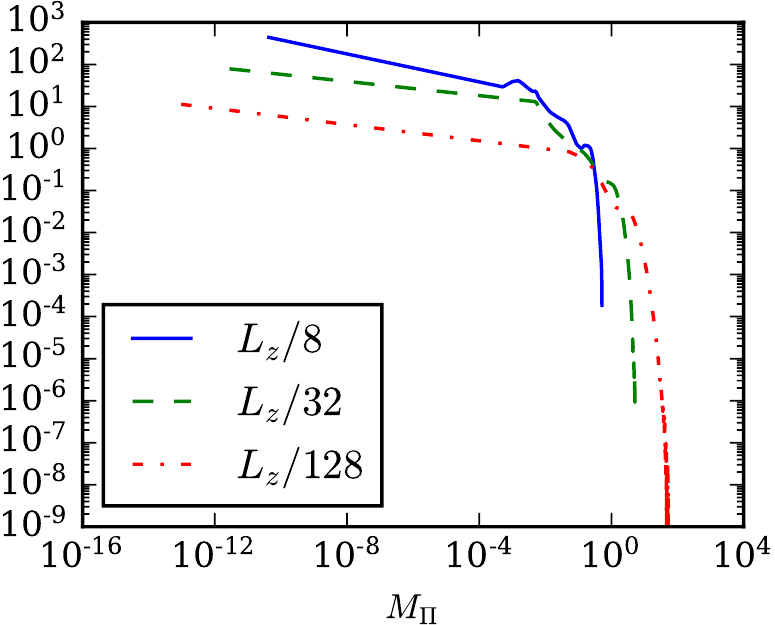} \label{fig:pi_eff_mag_d}}
\caption{\footnotesize{PDFs of the efficiency and magnitude of $\Pi$ at different scales from the 2D4096 and 3D1024 data at time $\widehat{t}=4$.
    The PDFs of efficiency are on a linear-log plot, while the PDFs of magnitude are on a log-log plot for clarity.} \label{fig:pi_eff_mag}}
\end{minipage}
\end{figure}

The cascade efficiency and magnitude of $\Lambda$ in 2D and 3D are shown in figure \ref{fig:lambda_eff_mag}.  Magnitude $M_\Lambda$ in both 2D and 3D decreases at smaller scales
as appears in figures \ref{fig:lambda_eff_mag_b} and \ref{fig:lambda_eff_mag_d}. In fact, figure \ref{fig:lambda_2D_3D_scaling}(b) shows that $M_\Lambda$ seems to scale $\sim k^{-2/3}$.
This decay can be understood from basic scaling estimates for the magnitude of $\Lambda=\frac{1}{\OL{\rho}} \nabla\OL{P}_\ell \OL{\tau}_\ell(\rho, \bu)$, following (\citeSupp{Eyink_notes,Eyink05,Aluie11c}). We first observe that in our flows, both 2D and 3D, the pressure field is mostly smooth as evidenced by the scaling of its filtering spectrum in figure \ref{fig:lambda_2D_3D_scaling}(a), decaying at least as fast\footnote{As \citeSupp{Sadek18} elaborate, the filtering spectrum for a field with actual scaling that is steeper than $k^{-3}$ will appear as $k^{-3}$ when using 1st-order filtering kernels such as a Gaussian.} as $k^{-3}$, where $k$ is filtering wavenumber (inverse of scale).
Therefore, the pressure gradient is dominated by the largest scale $L$, 
$$\nabla\OL{P}_\ell \sim \frac{\delta P(L)}{L},$$
and is (to leading order) independent of $\ell$, where an increment is defined as usual, $\delta f(r) \equiv f(x+r)-f(x)$. Here, $L$ is a characteristic large scale of RT, comparable to the domain size. On the other hand, subscale mass flux scales as (\citeSupp{Aluie11c}):
$$\OL{\tau}_\ell(\rho, \bu)\sim \delta\rho(\ell)\delta \bu(\ell) \sim \ell^{\sigma_\rho+\sigma_u} \sim k^{-2/3}$$
where $\delta \rho(\ell)\sim \ell^{\sigma_\rho}, \ \delta \bu(\ell)\sim \ell^{\sigma_u}$, and from figure \ref{fig:spectra}, the two scaling exponents $\sigma_\rho \lessapprox 1/3, \sigma_u\gtrapprox 1/3$,
yielding $\sigma_\rho+\sigma_u\approx 2/3$. This explains decay of $M_\Lambda$ as a function of filtering wavenumber $k_\ell=L_z/\ell$ in both 2D and 3D-RT. The decay is faster in the dissipation range since fluctuations giving rise to $\OL{\tau}_\ell(\rho, \bu)$ decay faster at those scales.

An important conclusion from the scaling of the pressure spectrum in figure \ref{fig:lambda_2D_3D_scaling}(a) is that energy transfer by $\Lambda$ is (infrared) non-local in scale due to dominance of the largest scale $L$ in the pressure gradient (\citeSupp{Eyink05,EyinkAluie09,Aluie11c}). This is similar to energy transfer in the Batchelor range of magnetohydrodynamic turbulence at high magnetic Prandtl numbers (\citeSupp{batchelor1959small,AluieEyink10}), where energy is transferred nonlocally from the flow at  viscous scales $\ell_\nu$ to the magnetic field at much smaller scales due to smoothness of the velocity at those small viscous scales. We note, however, that such non-local transfer by $\Lambda$ is mostly likely a hallmark of RT turbulence, where $\grad P$ is dominated by the largest scales in the system, and may not hold in general variable density turbulence in which the pressure field is not smooth (\citeSupp{Aluie11c}).

While the magnitude $M_\Lambda$ decays at smaller scales both in 2D-RT and 3D-RT, efficiency $\Gamma_\Lambda$ seems to have a much weaker decay at small scales. Figure \ref{fig:lambda_eff_mag}(a) shows that $\Gamma_\Lambda$ in 2D-RT is almost independent of scale, which is reinforced by scaling plots in figure \ref{fig:lambda_2D_3D_scaling}(c). In 3D-RT,  figure \ref{fig:lambda_2D_3D_scaling}(c) shows a very shallow scaling, with $\langle\Gamma_\Lambda\rangle \sim k_\ell^{-1/4}$.

To explain the decay of $\langle\Lambda\rangle$  at small scales we observed in 2D-RT (figure \ref{fig:2D4096_cascade} in the main text) and in 3D-RT (figure \ref{fig:3D1024_cascade} in the main text), we assume that the efficiency and magnitude of $\Lambda$ are statistically independent, $\langle\Lambda\rangle \sim \langle \Gamma_\Lambda\rangle \langle M_\Lambda\rangle$ (within the factor $\frac{1}{\OL{\rho}}$).
When this is combined the above observations on the scaling of each of $M_\Lambda$ and $\Gamma_\Lambda$, we are able to explain the decay of $\langle\Lambda\rangle$, which figure \ref{fig:lambda_2D_3D_scaling}(d) shows to be $\sim k_\ell^{-2/3}$ in 2D and $\sim k_\ell^{-1}$ in 3D. Therefore, we can conclude from the foregoing discussion that the decay in $\langle\Lambda_\ell\rangle$ at smaller $\ell$ is primarily due to a decay in its magnitude rather than due to a mis-alignment of $\grad\OL{P}_\ell$ and $\OL{\tau}_\ell(\rho,\bu)$.

Unlike $\Lambda$, which decays monotonically in the inertial range, deformation work $\Pi$ peaks in the inertial range as it takes over the energy transfer process. PDFs of the efficiency of $\Pi$ (figures \ref{fig:pi_eff_mag}(a),(c)) indicate a profound difference from $\Lambda$. Whereas the PDFs of $\Gamma_\Lambda$ in figures \ref{fig:lambda_2D_3D_scaling}(a) and (c) peak around the maximum possible values, -1 and 1, $\Gamma_\Pi$ has a highest probability around zero. This is consistent with the investigation by \citeSupp{Ouellette18}, who concluded that $\Pi$ is inefficient at transferring energy in constant-density homogeneous turbulence (see also \citeSupp{FangOuellette16}). Thus, the $\Pi$ flux, which transfers energy due to subscale stress acting against a larger-scale strain, is less efficient than $\Lambda$, which operates by the barotropic and baroclinic generation of vorticity and strain (\citeSupp{Aarne19,ZhaoAluie20}).

\subsection{$\Pi$ and $\Lambda$ with anisotropic filters\lb{suppsec:scaleanisotropy}}
In section \ref{sec:split_direction} of the main text, we analyzed how the fluxes can lead to anisotropic motion in different directions by directional splitting of the kinetic energy budget, which we coarse-grained with isotropic kernels.
 In this section, we shall quantify the \emph{anisotropy of length-scales} energized by $\Pi$ and $\Lambda$, through coarse-graining with anisotropic kernels. We use  the following `horizontal kernel' and `vertical kernel':
$$G_\ell^{x}(\bx)=\left(\frac{6}{\pi\ell^2}\right)^{1/2} e^{-\frac{6}{\ell^2}x^2}, \quad G_\ell^{z}(\bx)=\left(\frac{6}{\pi\ell^2}\right)^{1/2} e^{-\frac{6}{\ell^2}z^2}$$
which perform the filtering operation in only one direction. These filters only capture the spectral information in the corresponding direction and ignore variations in the other
directions. We denote the fluxes calculated by these anisotropic kernels using notation such as $\Pi_{G_x}$ or $\Lambda_{G_z}$. Note that unlike the `directionally split' analysis presented
earlier, here the sum of anisotropically filtered fluxes does not equal the full flux, i.e., $\Lambda_{G_x}+\Lambda_{G_y}+\Lambda_{G_z}\neq \Lambda$.

\begin{figure}
\centering 
\begin{minipage}[b]{1.0\textwidth}  
\subfigure[\footnotesize{Isotropic filtering}]
{\includegraphics[height=1.6in]{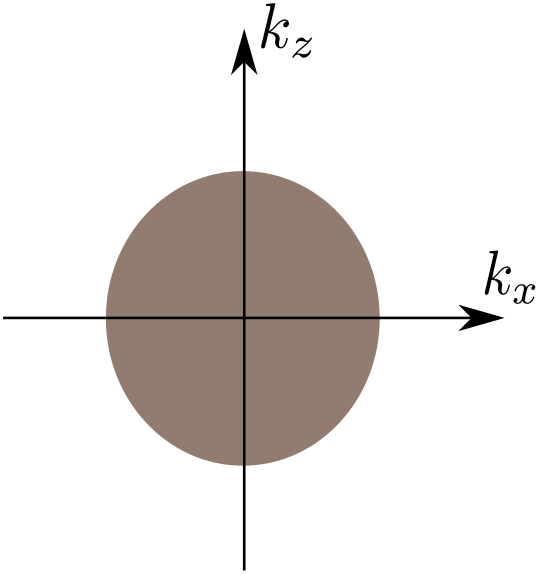} \label{fig:iso_aniso_filters_a}} 
\subfigure[\footnotesize{Anisotropic filtering with $G^x_\ell$}]
{\includegraphics[height=1.6in]{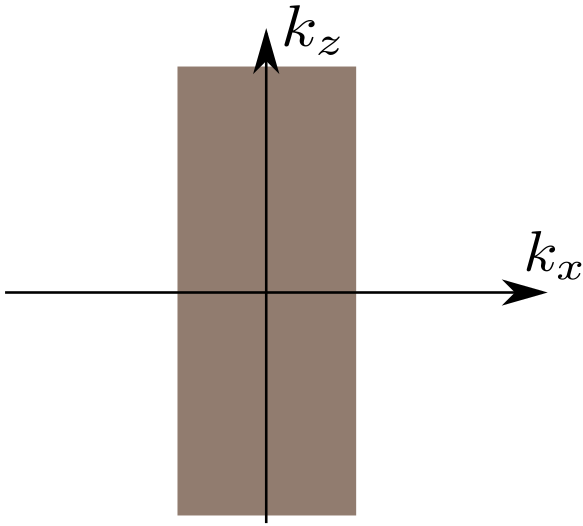} \label{fig:iso_aniso_filters_b}}
\subfigure[\footnotesize{Anisotropic filtering with $G^z_\ell$}]
{\includegraphics[height=1.6in]{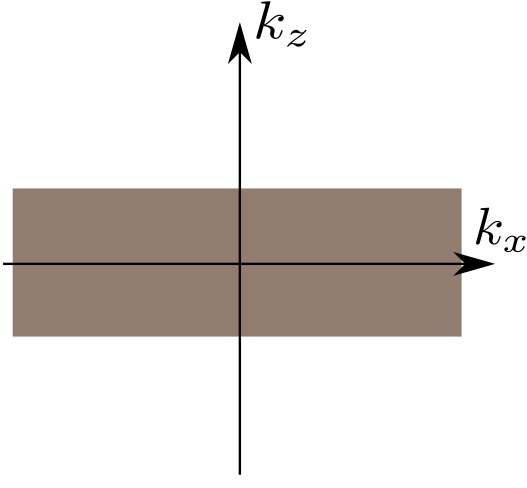} \label{fig:iso_aniso_filters_c}}
\caption{\footnotesize{Schematics of isotropic and anisotropic filtering, associated with scale $\ell$. Filtering is performed in physical space to disentangle scales, without resorting to Fourier transforms.  Filtering wavenumbers $\bk=(k_x,k_z)=(L/\ell_x,L/\ell_z)$ are only a proxy for length-scales, but may be thought of as Fourier wavenumbers conceptually.
    (a) Isotropic (low-pass) filtering with kernel $G_\ell$ retains scales within the shaded sphere in (scale) k-space, satisfying $|\bk| \le L/\ell$.
    (b) Anisotropic filtering with kernel $G^x_\ell$ retains scales within the shaded slab in (scale) k-space, satisfying $|k_x| \le L/\ell$ without decomposing scales in the z-direction.
    (c) Anisotropic filtering with kernel $G^z_\ell$ retains scales within the shaded slab in (scale) k-space, satisfying $|k_z| \le L/\ell$ without decomposing scales in the x-direction.
    } \label{fig:iso_aniso_filters}}
\end{minipage}
\end{figure}

The essence of isotropic and anisotropic filtering is illustrated in figure \ref{fig:iso_aniso_filters} in  $k$-space, where 
filtering wavenumbers $\bk=(k_x,k_z)=(L/\ell_x,L/\ell_z)$ are only a proxy for length-scales, but can be conceptually associated with Fourier wavenumbers commonly used in the community. Here, $L$ is any characteristic large-scale to be used only as a \emph{common} reference in all directions to define filtering wavenumbers. Figure \ref{fig:iso_aniso_filters} depicts a two-dimensional wavenumber space, with horizontal and vertical filtering wavenumbers corresponding to scales in the x-direction and z-direction, respectively. Isotropic filtering with kernel $G_\ell$ retains scales within the shaded sphere in $k$-space (figure \ref{fig:iso_aniso_filters_a}), satisfying $|\bk| \le L/\ell$. On the other hand, anisotropic filtering with kernel $G^x_\ell$, for example, retains scales within the shaded slab in $k$-space (figure \ref{fig:iso_aniso_filters_b}), satisfying $|k_x| \le L/\ell$ but does not allow us to probe scales in the z-direction since it does not partition those scales. The analysis generalizes to 3D in a straightforward way.

\begin{figure}
\centering 
\begin{minipage}[b]{1.0\textwidth}  
\subfigure[\footnotesize{Anisotropic $\Lambda$ in 2D}]
{\includegraphics[height=2.1in]{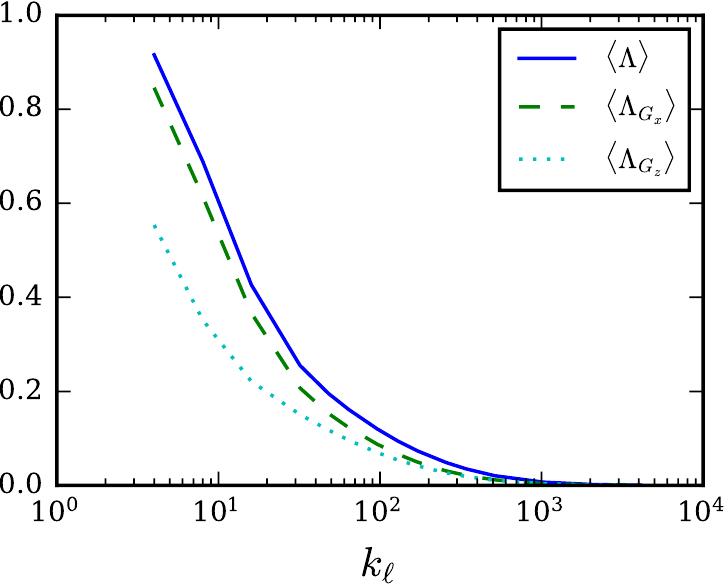} \label{fig:anisotropic_flux_a}} 
\phantom{a}
\subfigure[\footnotesize{Anisotropic $\Pi$ in 2D}]
{\includegraphics[height=2.1in]{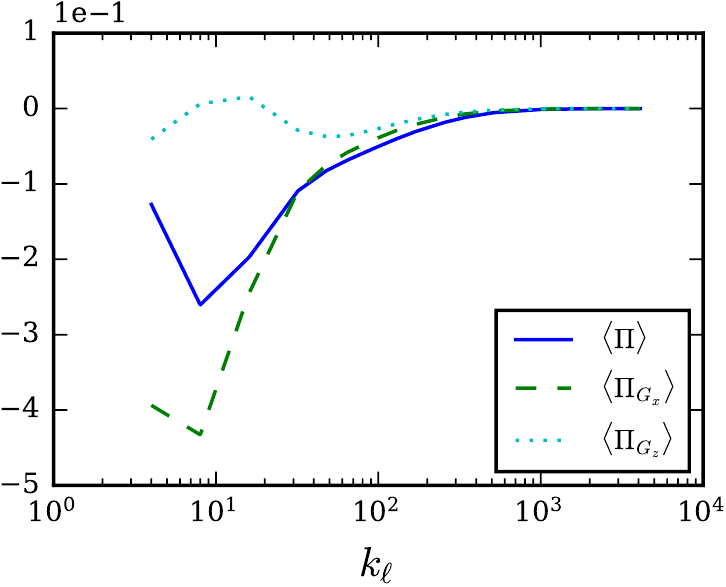} \label{fig:anisotropic_flux_b}} \\
\subfigure[\footnotesize{Anisotropic $\Lambda$ in 3D}]
{\includegraphics[height=2.1in]{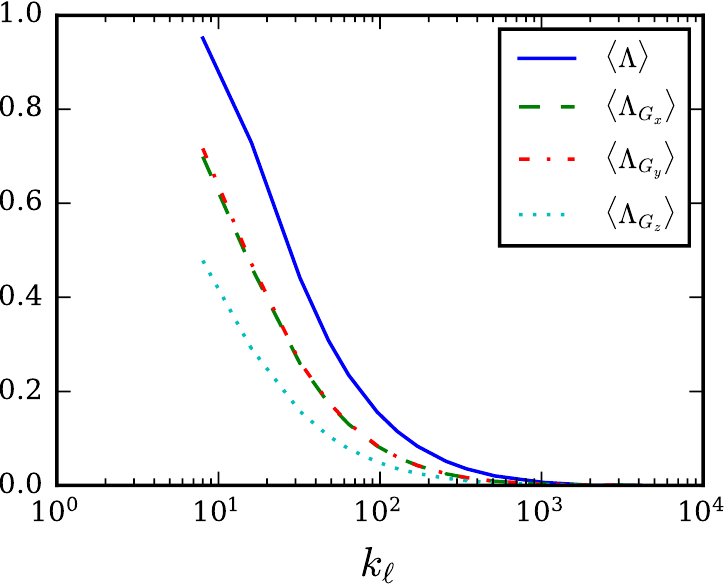} \label{fig:anisotropic_flux_c}} 
\phantom{a}
\subfigure[\footnotesize{Anisotropic $\Pi$ in 3D}]
{\includegraphics[height=2.1in]{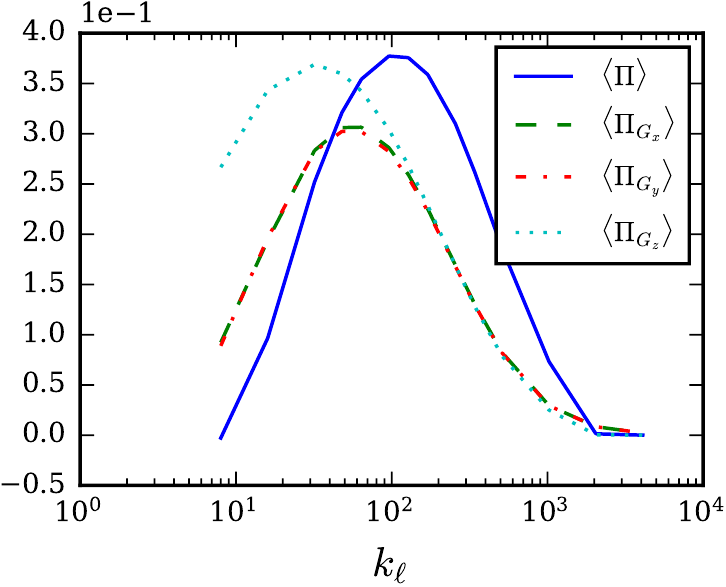} \label{fig:anisotropic_flux_d}} 
\caption{\footnotesize{Probing scale anisotropy of fluxes $\Lambda_G$ and $\Pi_G$ using the 2D4096 and 3D1024 data at $\widehat{t}=4.0$. Filtering wavenumber is $k_\ell=L_z/\ell$. Plots are normalized by $\langle \epsilon^{\textrm{inj}} + P\nabla\cdot \bu\rangle$ at the corresponding time. Energy transfer across horizontal and vertical scales are shown alongside energy transferred isotropically in scale space.} \label{fig:anisotropic_flux}}
\end{minipage}
\end{figure}

Mean anisotropic fluxes of the 2D4096 and 3D1024 data, at time $\widehat{t}=4.0$, are shown in figure \ref{fig:anisotropic_flux}. For 
baropycnal work $\Lambda$, both the 2D and 3D results shown in figures \ref{fig:anisotropic_flux_a} and \ref{fig:anisotropic_flux_c} indicate
that $\langle\Lambda_{G_x}\rangle$ is larger than $\langle\Lambda_{G_z}\rangle$. This implies that most of the potential energy is being transferred to horizontal scales, although nonzero transfer is also creating flow variation in the vertical direction.
The tendency of baropycnal transfer in energizing small-scales along the x-direction is consistent with our understanding of RT instability during the linear stage when potential energy feeds perturbations that vary in the horizontal direction. Figures \ref{fig:anisotropic_flux_a},\ref{fig:anisotropic_flux_c} indicate that baropycnal transfer from potential energy at later times, when the flow becomes nonlinear, also participates in the creation of vertical scales. This is consistent with the association of baropycnal work with vorticity and strain creation (\citeSupp{Aarne19}), which necessarily involve small-scales in the vertical direction. Nevertheless, horizontal scale creation still dominates at late times, with $\langle \Lambda_{G_x}\rangle > \langle \Lambda_{G_z}\rangle$ at all scales. Note that in 3D, we have $\langle \Lambda_{G_x}\rangle \approx \langle \Lambda_{G_y}\rangle$ in figure \ref{fig:anisotropic_flux_c}, indicating
that horizontal scales are being energized isotropically by $\Lambda$, as expected.

Probing anisotropy of deformation work $\Pi$, we can observe significant differences between the 2D case, shown in figure \ref{fig:anisotropic_flux_b}, and the 3D case, shown in 
figure \ref{fig:anisotropic_flux_d}. In 2D, the mean value of $\Pi$ calculated from isotropic kernel is negative, and so is the anisotropic $\Pi_{G_x}$. On the other hand,
mean $\Pi_{G_z}$ attains positive values over a certain range of scales. Since $\Pi=-\OL{\rho}\partial_j\wt{u}_i\wt{\tau}(u_i, u_j)$ is primarily determined by the velocity fields,
we shall focus on the characteristics of velocity fields in RT evolution.
The positive $\langle\Pi_{G_z}\rangle$ measures energy transfer across vertical scales, which goes into sustaining the `mixing fronts,'
which are the two envelopes enclosing the mixing layer in RT turbulence. We verified this from visualizations of $\Pi_{G_z}$ (not shown here). This is similar to the downscale energy transfer that forms and sustains the shock front in a 1D Burgers flow. In our RT flow, the velocity profile along the z-axis comprises of a vertical flow inside the mixing layer pushing against a more quiescent fluid outside. Such fronts necessarily involve small vertical scales that need to be constantly replenished by $\Pi$, otherwise the front would broaden. We find that inside the mixing layer (not shown), away from the fronts, $\Pi_{G_z}<0$, transferring net energy upscale as in a homogeneous 2D flow. However, the contribution from the front regions is stronger such that overall $\langle \Pi_{G_z}\rangle >0$. Despite being positive, $\langle \Pi_{G_z}\rangle$ is much smaller than the net upscale transfer in the horizontal direction measured by $\langle \Pi_{G_x}\rangle$ in figure \ref{fig:anisotropic_flux_b}, which dominates the cross-scale transfer when we treat all directions equally by using isotropic kernels to measure $\langle\Pi\rangle$ in figure \ref{fig:anisotropic_flux_b}.
In the 3D-RT flow, we see in figure \ref{fig:anisotropic_flux_d} that
$\Pi_{G_x}$, $\Pi_{G_y}$, and $\Pi_{G_z}$ are all positive, transferring 
energy downscale. We find that $\Pi_{G_z}$ peaks at slightly larger scales and again plays a role in the formation of the mixing fronts as in 2D, while $\langle\Pi_{G_x}\rangle\approx \langle\Pi_{G_y}\rangle \ge0$, indicating isotropic downscale transfer in the horizontal scales.  
 
 We also note that
$\langle \Pi_{G_z}\rangle > \langle \Pi_{G_x}\rangle$ at all scales, indicating that the `mixing front' propagation plays an important role in the downscale kinetic energy transfer. This aspect of energy transfer was missing in previous studies relying on FFTs in the horizontal while neglecting vertical scales.

\subsection{Filtering Spectra of Kinetic Energy}\label{suppsec:append_spectra}
Here, we compare the `filtering' spectra using three different scale-decompositions to define kinetic energy, following \citeSupp{Zhao18}. In variable density flows, scale decomposition is not as straightforward as in constant density flows. One possible
decomposition is to define large-scale kinetic energy as $\OL\rho_\ell |\OL\bu_\ell|^2/2$,
which has been used in many studies (e.g. \citeSupp{chassaing1985alternative,BodonyLele05,Burton11,KarimiGirimaji17}). Another 
possibility is to define large-scale kinetic energy as $|\OL{(\sqrt{\rho}\bu)}_\ell|^2/2$,
which has also been used extensively in compressible turbulence studies
(e.g. \citeSupp{kida1990energy,CookZhou02,Wang13,Greteetal17}). A third is based on the Favre decomposition
(\citeSupp{Hesselberg26,Favre58a}), which we adopt in this work, and uses
$\OL\rho_\ell|\wt{\bu}_\ell|^2/2$ 
as the definition for large-scale kinetic energy, where
$\wt\bu_\ell(\bx) = \OL{\rho \bu}_\ell/\OL\rho_\ell$.

\citeSupp{Zhao18} demonstrated numerically that these three different quantities, all of which have units of energy, are governed by different dynamics. Specifically, it was shown that for large $\ell$, viscous dissipation is negligible for the Favre decomposition but significant for the other two in a variety of flows; normal 1D shock, 2D-RT and 3D-RT, subject to a dynamic viscosity that is either constant or temperature-dependent. 

Following \citeSupp{Sadek18}, the filtering spectra for the three definitions are:
\begin{eqnarray}
E_{F}(k_\ell)&\equiv& \frac{d}{dk_\ell}\langle \OL\rho_\ell|\wt{\bu}_\ell(\bx)|^2\rangle /2 \label{eq:FavreKEspectrum_App}\\
E_{C}(k_\ell)&\equiv& \frac{d}{dk_\ell}\langle \OL\rho_\ell|\OL{\bu}_\ell(\bx)|^2\rangle /2 \label{eq:ChassaingKEspectrum_App}\\
E_{K}(k_\ell)&\equiv& \frac{d}{dk_\ell}\langle |\OL{(\sqrt{\rho}\bu)}_\ell(\bx)|^2\rangle /2 \label{eq:KidaKEspectrum_App}
\end{eqnarray}
where $k_\ell=L/\ell$, $L$ is the domain size of interest, $\ell$ is the scale we are probing, and $\langle\cdot\rangle$ stands for spatial averaging. Subscript `F' stands for Favre, while `C' and `K'  denote the lead authors of papers in which 
those definitions, to our best knowledge, first appeared (\citeSupp{chassaing1985alternative,kida1990energy}). 

\begin{figure}
\centering 
\begin{minipage}[b]{1.0\textwidth}  
\centering
{\includegraphics[height=0.4\textwidth]{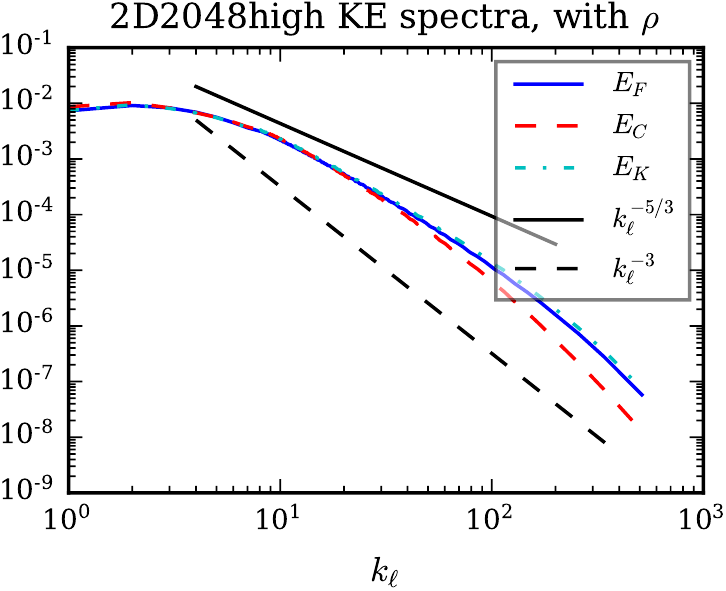} } 
{\includegraphics[height=0.4\textwidth]{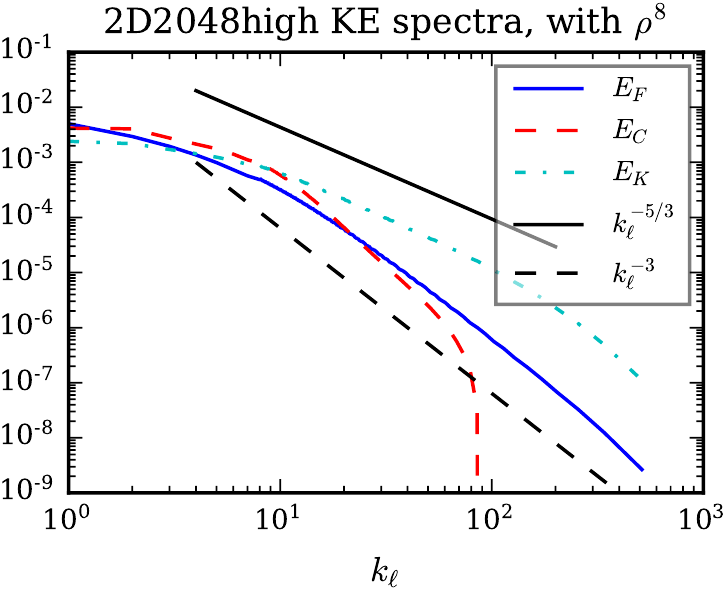} }\\ 
{\includegraphics[height=0.4\textwidth]{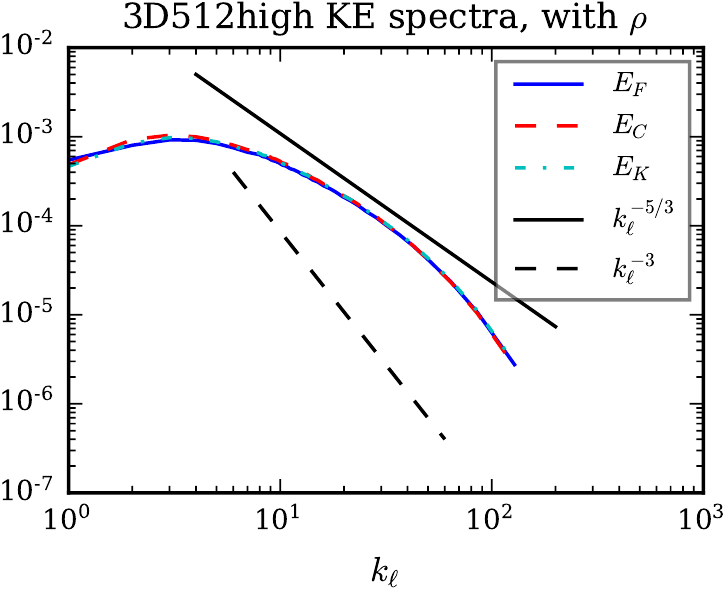} } 
{\includegraphics[height=0.4\textwidth]{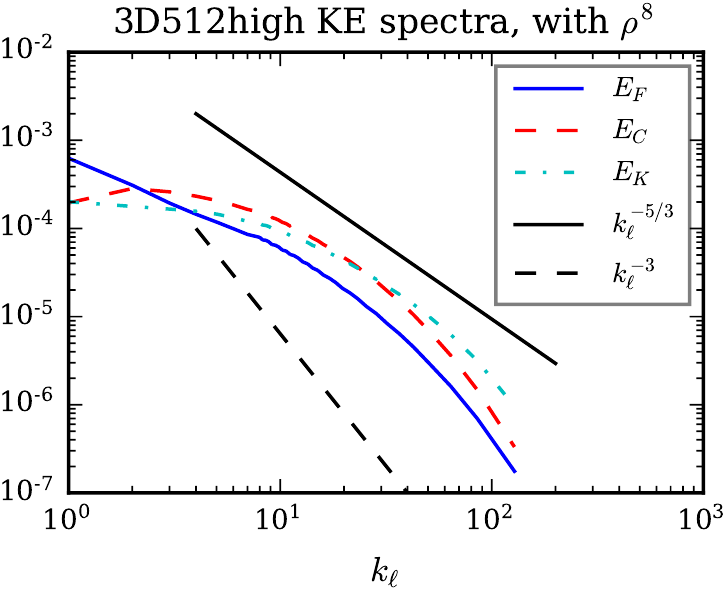} } 
    \caption{\footnotesize{Filtering spectra of kinetic energy using three different scale-decompositions (eqs. \eqref{eq:FavreKEspectrum_App}-\eqref{eq:KidaKEspectrum_App}). Top and bottom rows use 2D-RT and 3D-RT data, respectively. Left column uses the raw density field, $\rho$. Right column usesq $\rho^8$ (normalized to conserve mass) to highlight differences among the definitions under an increased density contrast. Data from 2D2048high was at $\hat{t}=4.4$. Data from 3D512high was at $\hat{t}=4.0$.} \label{fig:spectra_App}}
\end{minipage}
\end{figure}

Figure \ref{fig:spectra_App} compares these three spectra using the high Atwood RT simulations in 2D and 3D. We find discernible differences from the 2D2048high simulation but not from the 3D512high data. We suspect the lack of differences from the 3D512high data may be due to an enhanced microscopic mixing between the heavy and light fluids at late times. We observe (not shown) that differences among the definitions using the 3D512high data were relatively larger at earlier times, albeit still small in absolute terms, especially when compared to those from the 2D2048high data. Microscopic mixing in 2D-RT is much smaller than in 3D-RT as observed in previous studies \citeSupp[e.g.][]{Cabot06POF}, due to weaker small inertial scales as we discussed in the main text above.

With $\mathcal{A}=0.8$, our flows have an initial density ratio of $\rho_h/\rho_l = 9$. Achieving higher ratios in a well-resolved turbulence simulation is computationally challenging, especially in 3D (e.g. \citeSupp{Livescu08}). Yet, many flows of interest have much larger density ratios, reaching of up to 600 laboratory flow experiments (\citeSupp{Read84,DimonteSchneider00}), typically exceeding $10^4-10^5$ in laboratory fusion plasmas (e.g. \citeSupp{Craxtonetal15,Yan16,Zhangetal2020}), and ranging
from $10^6$ to $10^{20}$ in molecular clouds in the interstellar medium (e.g. \citeSupp{Kritsuketal07,Federrathetal10,Panetal16}). The most ubiquitous terrestrial two-fluid mixing is between air and water which have a density ratio of 1000.

Since we are testing possible differences among the definitions, which can be applied to any density and velocity fields regardless of their dynamical origin, 
we synthetically increase the density contrast \citeSupp{Zhao18} in the flows we are analyzing by taking powers of the density, $A\,\rho^8(\bx)$, as a post-processing step, then normalizing the resultant field such that the total mass in the domain is the same as in the original flow, $\langle\rho\rangle = A\,\langle\rho^8\rangle$. Note that this is not based on physical grounds, but serves to highlight differences among the three decompositions under higher density contrasts. 

Figure \ref{fig:spectra_App} (right column) compares these three spectra using $A\,\rho^8(\bx)$ as a density field and original velocity from the high Atwood RT simulations in 2D and 3D. We now find more pronounced differences among the three decompositions, especially from the 2D2048high data; $E_{C}(k_\ell)$ from equation \eqref{eq:ChassaingKEspectrum_App} decays rapidly at high $k_\ell$, while $E_{K}(k_\ell)$ from equation \eqref{eq:KidaKEspectrum_App} becomes shallower than the Favre-based spectrum $E_{F}(k_\ell)$ (equation \eqref{eq:FavreKEspectrum_App}), scaling similar to $k^{-5/3}$ even in 2D.

The purpose of Figure \ref{fig:spectra_App} is to demonstrate that filtering spectra based on different kinetic energy decompositions can differ. It is not our intention here to argue which is more physical, which was the subject of previous studies (\citeSupp{Zhao18,Aluie13}).

\subsection{2D-RT simulation with same initial condition as in 3D1024} \label{suppsec:append_2DRT}
For verification purpose, we performed two additional 2D turbulent RT simulations (see Table \ref{tab:parameter}); run 2D1024 of grid size $1024\times 2048$ and run 2D2048 on a $2048\times 4096$ grid. Both have the same initial perturbations as in 3D1024. Specifically, the vertical velocity field is perturbed at the interface in wavenumber space within the wavenumber annulus $k\in [32, 128]$, with amplitude proportional to $e^{-\frac{1}{c}|k_x^2-80^2|}$, where $c \approx 22.63$ is a normalization constant to further limit the range of effectively perturbed wavenumber.  We shall see that the analysis performed on these results are  consistent with those obtained from run 2D4096 in the main text.

\begin{figure}
\centering
\begin{minipage}[b]{1.0\textwidth}
\centering
    \subfigure{\includegraphics[height=2.05in]{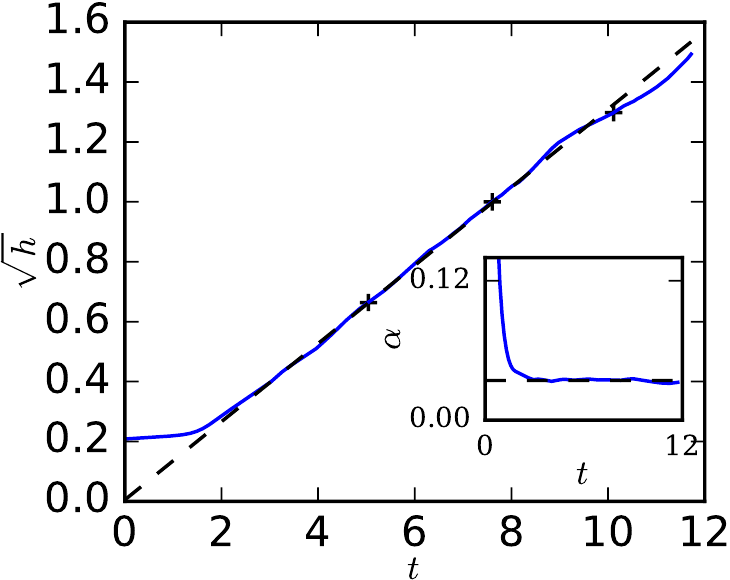}}
    \phantom{}
    \subfigure{\includegraphics[height=2.05in]{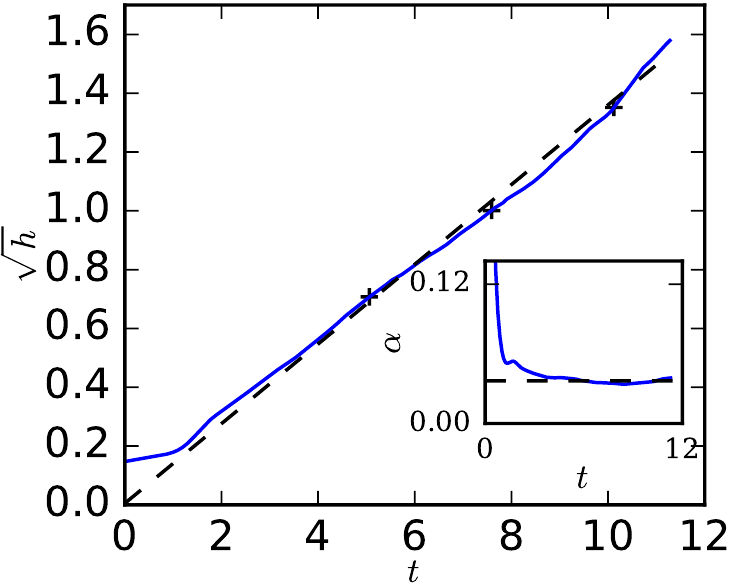}}
    \caption{\footnotesize{Similar to figure \ref{fig:mixingwidth} of the main text, the square root of mixing width $\sqrt{h(t)}$ versus time for the new 2D-RT simulations. Left figure is for 2D1024, with $\alpha=0.034$, and right figure is for 2D2048, with $\alpha=0.037$. The `+' markers correspond to dimensionless time $\widehat{t}=t/\sqrt{\frac{L_x}{\mathcal{A}g}}= 2, 3, 4$. Inset: the compensated plot $\alpha = \frac{h(t)}{\mathcal{A}gt^2}$ versus time, in which the horizontal lines correspond to the $\alpha$ value obtained by linear fit. } \label{fig:appendix_2d_mixing_width}}
\end{minipage}
\end{figure}

\begin{figure}
\centering
\begin{minipage}[b]{1.0\textwidth}
\centering
    \subfigure[\footnotesize{2D1024 instantaneous budget}]
{\includegraphics[height=2.1in]{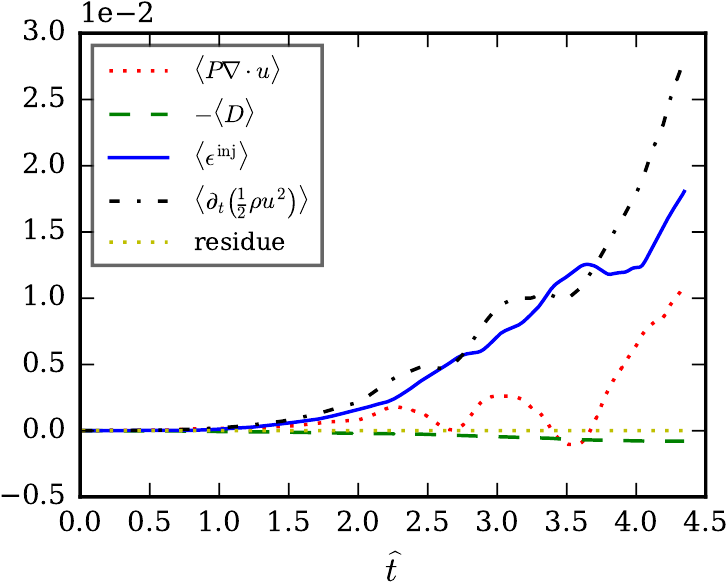}}
\phantom{}
    \subfigure[\footnotesize{2D1024 overall budget}]
{\includegraphics[height=2.1in]{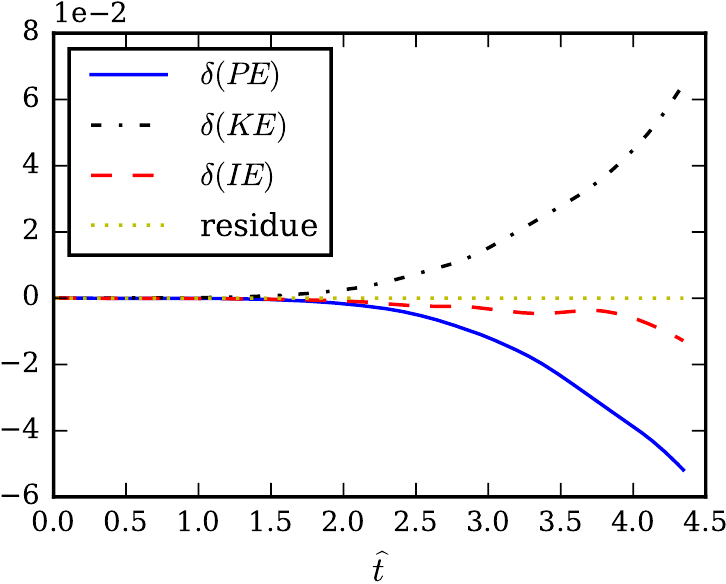}} \\
\subfigure[\footnotesize{2D2048 instantaneous budget}]
{\includegraphics[height=2.1in]{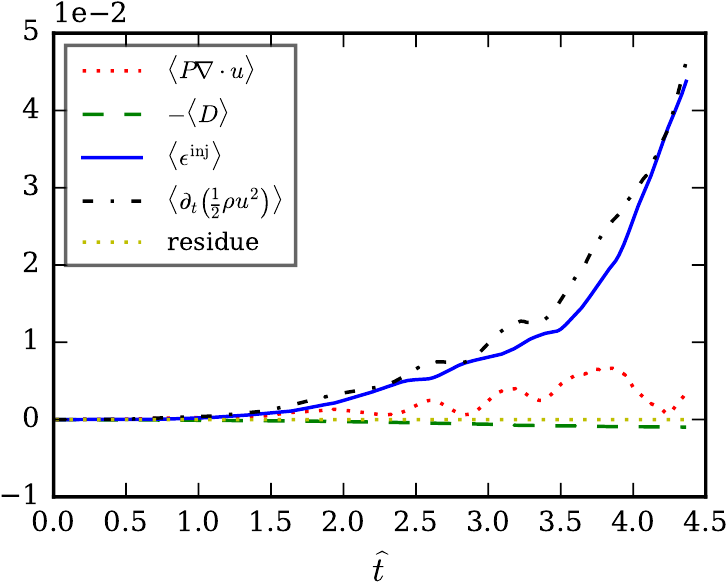}}
\phantom{}
    \subfigure[\footnotesize{2D2048 overall budget}]
{\includegraphics[height=2.1in]{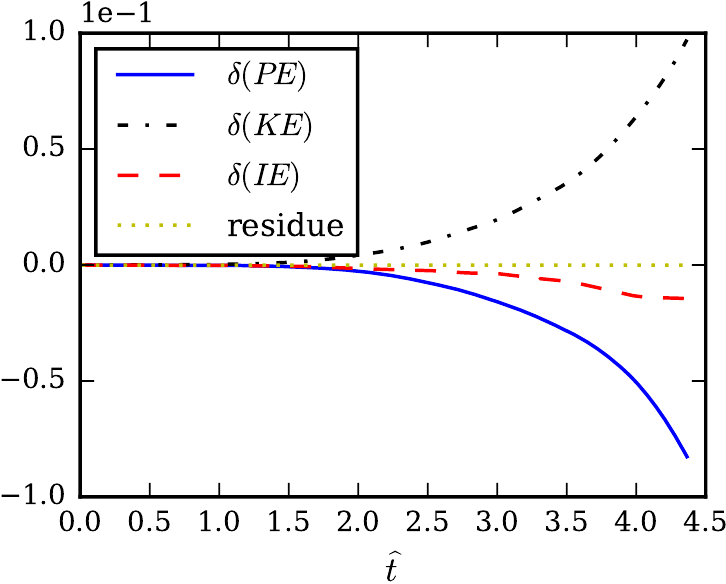}}
    \caption{\footnotesize{Temporal evolution of kinetic energy budget, and overall energy balance for 2D1024 in (a)-(b), and 2D2048 in (c)-(d), similar to figure \ref{fig:ke_detailed_balance_2d} and \ref{fig:ke_budge_2d} of the main text.} \label{fig:appendix_ke_detailed_balance_2d}}
\end{minipage}
\end{figure}

\begin{figure}
\centering
\begin{minipage}[b]{1.0\textwidth}
\centering
    \subfigure{\includegraphics[height=2.1in]{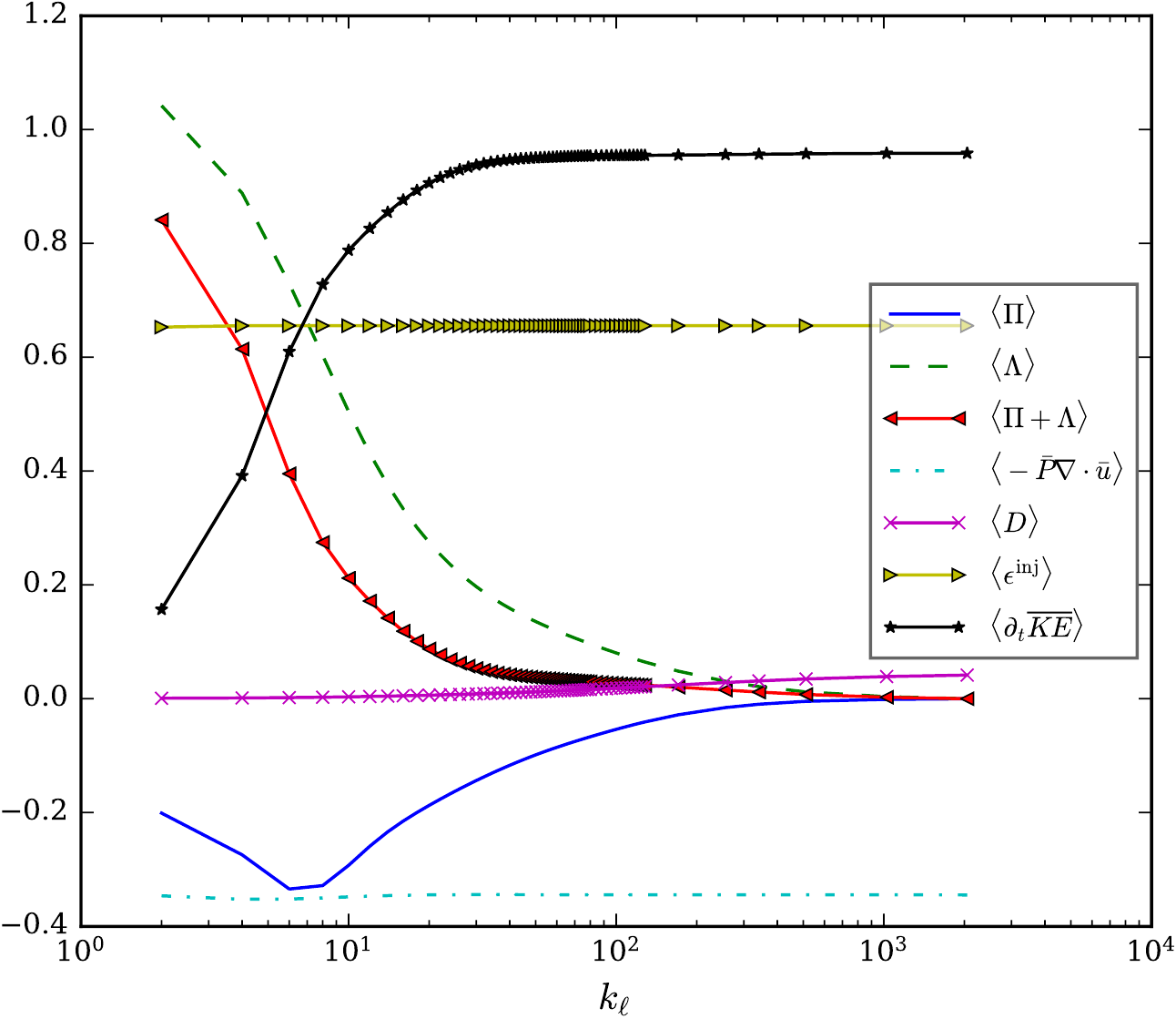}}
    \phantom{}
    \subfigure{\includegraphics[height=2.1in]{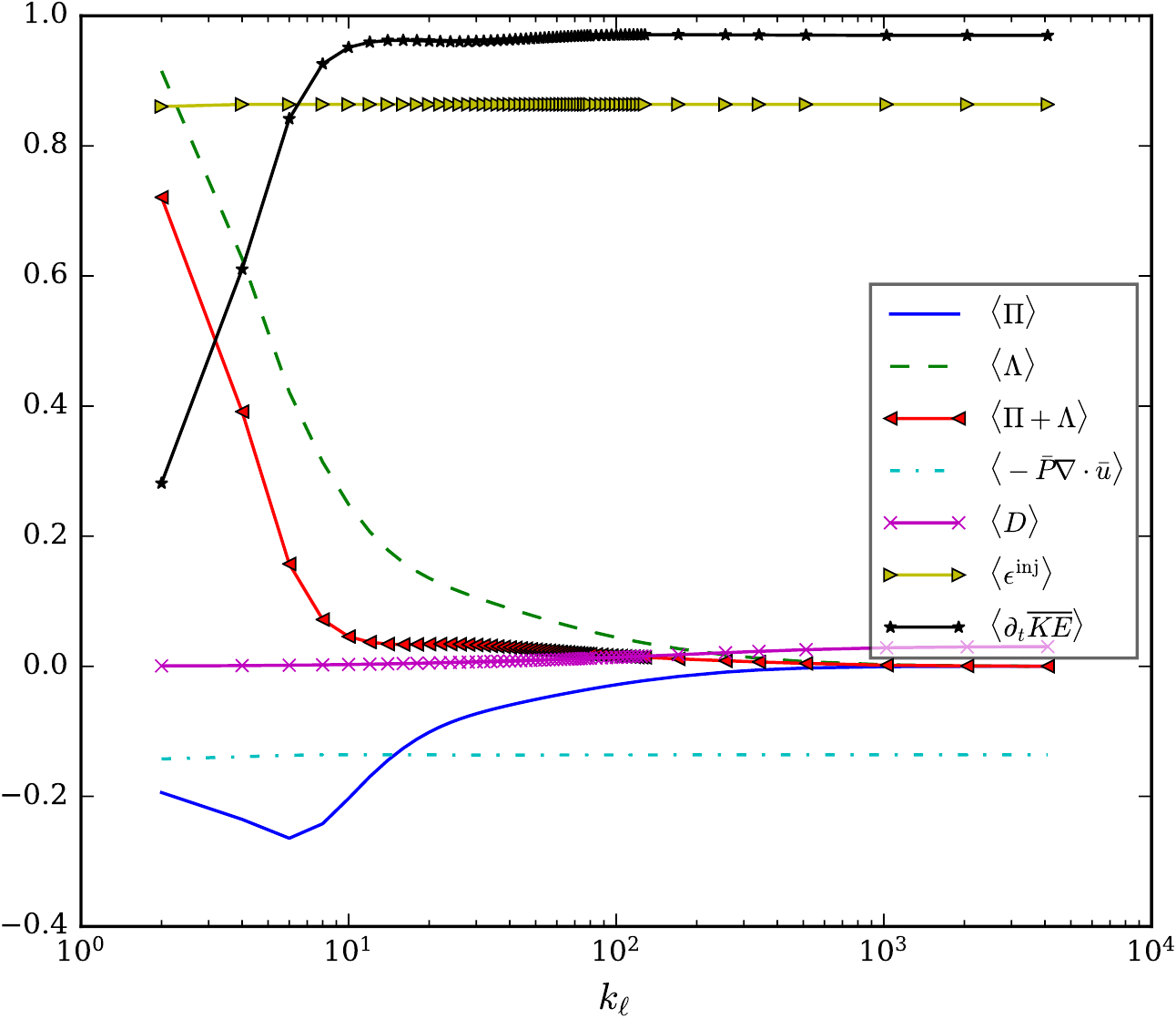}}
    \caption{\footnotesize{Similar to figure \ref{fig:2D4096_cascade} of the main text, mean kinetic energy budget as a function of scale in 2D at dimensionless time $\widehat{t}=4.0$.
    Left figure is for 2D1024 and right figure for 2D2048. The plots are normalized by $\langle \epsilon^{\textrm{inj}} + P\nabla\cdot \bu\rangle$, the available mean  source of kinetic energy. } \label{fig:append_2D1024_cascade}}
\end{minipage}
\end{figure}

\begin{figure}
\centering 
\begin{minipage}[b]{1.0\textwidth}  
\subfigure
{\includegraphics[height=1.3in]{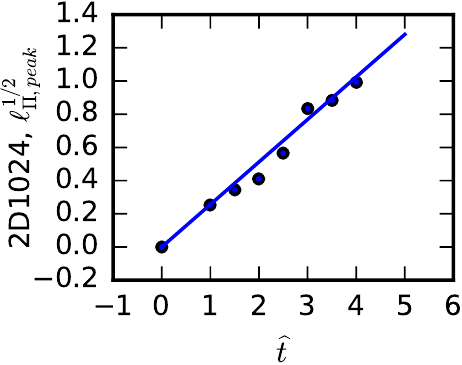} 
\llap{\parbox[b]{1.1in}{{\large (a)}\\\rule{0ex}{0.9in}}}
    \label{fig:appendix_self_similar_flux_1}} 
\subfigure
{\includegraphics[height=1.33in]{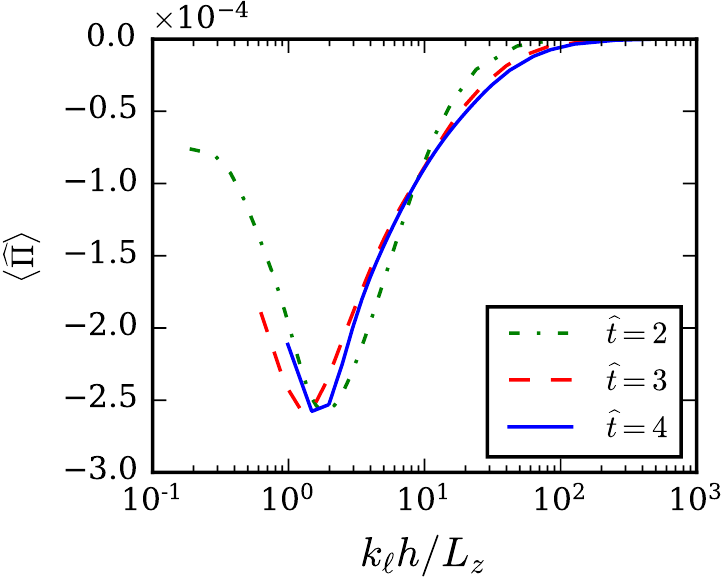} 
\llap{\parbox[b]{1.25in}{{\large (b)}\\\rule{0ex}{0.9in}}}
    \label{fig:appendix_self_similar_flux_2}} 
\subfigure
{\includegraphics[height=1.33in]{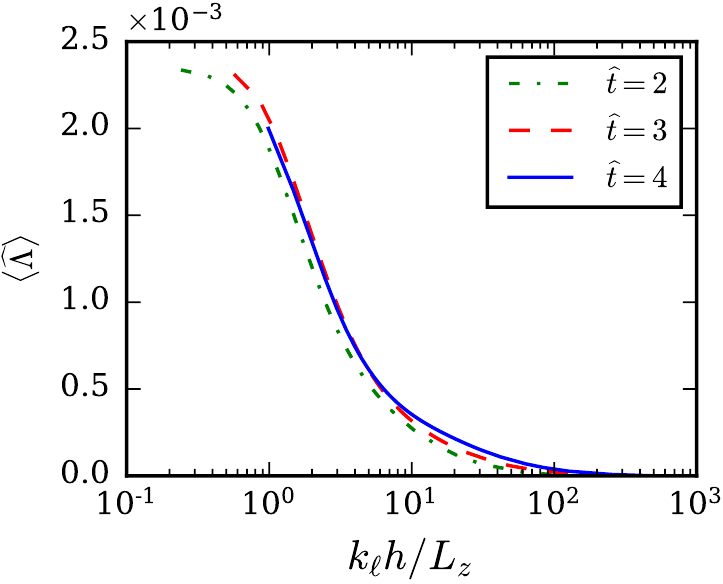} 
\llap{\parbox[b]{1.3in}{{\large (c)}\\\rule{0ex}{0.9in}}}
    \label{fig:appendix_self_similar_flux_3}} \\
\subfigure
{\includegraphics[height=1.3in]{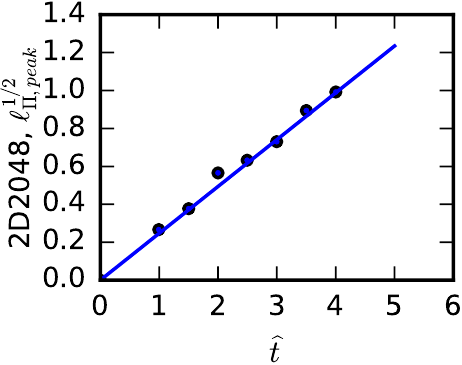} 
\llap{\parbox[b]{1.1in}{{\large (d)}\\\rule{0ex}{0.9in}}}
    \label{fig:appendix_self_similar_flux_1}} 
\subfigure
{\includegraphics[height=1.33in]{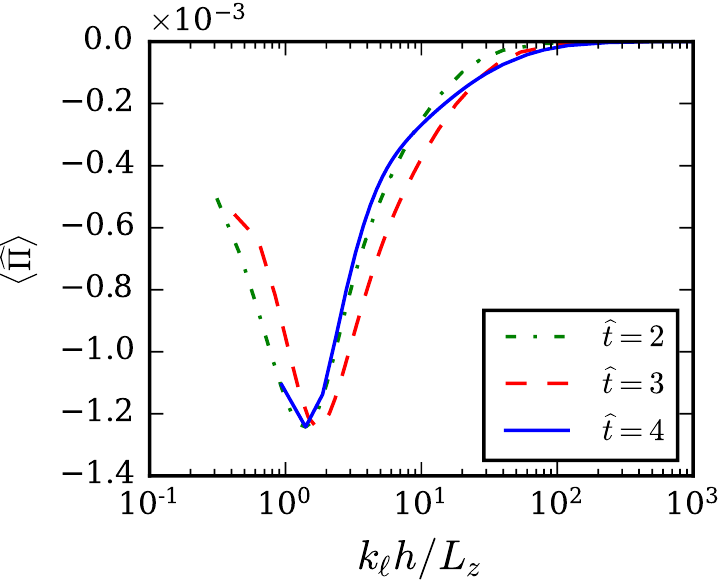} 
\llap{\parbox[b]{1.25in}{{\large (e)}\\\rule{0ex}{0.9in}}}
    \label{fig:appendix_self_similar_flux_2}} 
\subfigure
{\includegraphics[height=1.33in]{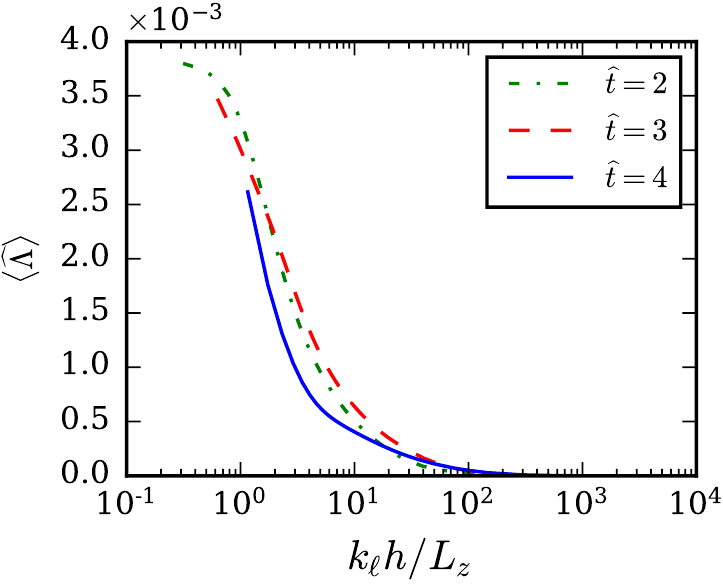} 
\llap{\parbox[b]{1.3in}{{\large (f)}\\\rule{0ex}{0.9in}}}
    \label{fig:appendix_self_similar_flux_3}}
\caption{\footnotesize{Similar to figure \ref{fig:self_similar_flux} of the main text, the temporal self-similarity of turbulent RT fluxes for the new 2D-RT simulations. Panels (a)-(c) are 2D1024 results, and panels (d)-(f) are 2D2048 results.
    Panels (a), (d) plot length scale $\ell_{\Pi, \text{peak}}$ associated with the peak of $\Pi$ versus dimensionless time $\widehat{t}$. Panels (b), (e) show rescaled $\langle\widehat{\Pi}\rangle$ in
    equation (\ref{eq:self_similar_flux}). Panels (c), (f) show rescaled $\langle\widehat{\Lambda}\rangle$.} \label{fig:appendix_self_similar_flux}}
\end{minipage}
\end{figure}

Figure \ref{fig:appendix_2d_mixing_width} shows the mixing width $h(t)$ versus time, with $\alpha=0.034$ for 2D1024, and $\alpha=0.037$ for 2D2048, which are similar to 2D4096 in figure \ref{fig:mixingwidth} with $\alpha=0.036$. Figure \ref{fig:appendix_ke_detailed_balance_2d} shows the corresponding instantaneous and overall kinetic energy budgets, which is similar to the 2D4096 case shown in figure \ref{fig:ke_detailed_balance_2d} and \ref{fig:ke_budge_2d}. The KE budget as a function of scale is shown in figure \ref{fig:append_2D1024_cascade}, which resembles the 2D4096 result in figure \ref{fig:2D4096_cascade} of the main text. The temporal self-similarity of the new 2D1024 and 2D2048 RT fluxes also holds, and is shown in figure \ref{fig:appendix_self_similar_flux}, similar to those in the 2D4096 case in figure \ref{fig:self_similar_flux} of the main text. All the analysis presented here is consistent with the results from run 2D4096 in the main text, implying that the slight difference in initial conditions between 2D4096 and 3D1024 does not affect our conclusions.

\subsection{Another form of kinetic energy budget}\label{suppsec:pathway_comparison}

In the compressible turbulence LES modeling literature, another formulation of the Favre filtered large scale kinetic energy budget, similar to equation (\ref{eq:KE_budget}) of the main text, is often used. In this formulation, pressure dilatation is lumped with the $\Lambda$ term by the relation:
\begin{align}
    -\Lambda_\ell + \OL{P}_\ell\nabla\cdot\OL{\bu}_\ell = \OL{P}\nabla\cdot\wt{\bu}_\ell +
    \nabla\cdot\left[\OL{P}(\OL{\bu}_\ell - \wt{\bu}_\ell)\right]
\lb{eq:AppLambdaPressDilRel}\end{align}
using the identity (\citeSupp{Aluie11c})
$$\wt{\bu}_\ell = \OL{\bu} + \frac{\OL{\tau}(\rho, \bu)}{\OL{\rho}}$$
Thus, the filtered large and small scale kinetic energy equations,
and the filtered internal energy equation are:

\begin{align} \label{eq:KE_budget_another}
\begin{split}
&\partial_t\bar{\rho}_\ell \frac{|\widetilde{\boldsymbol{u}}_\ell |^2}{2}+\nabla\cdot [\cdots] 
    =\bar{P}_\ell \nabla\cdot\widetilde{\boldsymbol{u}}_\ell+ \cdots \\
&\partial_t \frac{\bar{\rho}_\ell \widetilde{\tau}_\ell(u_i, u_i)}{2} + \nabla \cdot [\cdots] 
    = (\OL{P\nabla \cdot \bu} - \bar{P} \nabla \cdot \widetilde{\bu}) + \cdots \\
&\partial_t \overline{\rho e}_\ell + \nabla \cdot [\cdots] 
    =-\bar{P}_\ell \nabla \cdot \widetilde{\bu}_\ell - (\OL{P\nabla\cdot \bu}_\ell - \bar{P}_\ell \nabla \cdot \widetilde{\bu}_\ell)+\cdots
\end{split}
\end{align}
the omitted terms are similar to those in our formulation of the coarse-grained budgets in equation (\ref{eq:KE_budget}) of the main text.
Many previous works on compressible turbulence adopt the framework of equation (\ref{eq:KE_budget_another}), see \citeSupp{Lele94} and the 
references therein.  There are a few exceptions, such as in \citeSupp{Huang95}, in which they preferred to 
have a separate $\Lambda$ based on modelling considerations that density weighted decomposition should only be applied to the convective terms,
and also in (\citeSupp{Aluie11c,Aluie13,Wang13,EyinkDrivas17a}) where it was argued that $\Lambda$ and $\bar{P}\nabla \cdot \bar{\bu}$ are fundamentally different from  a physics standpoint since the former has contributions from scales both larger and smaller than $\ell$ and represents the interaction between different scales, while the latter only involves large scale quantities.

\begin{figure}
\centering
\begin{minipage}[b]{1.0\textwidth}
\centering
    \subfigure[\footnotesize{Energy pathways by equation (\ref{eq:KE_budget})}]
{\includegraphics[height=2.5in]{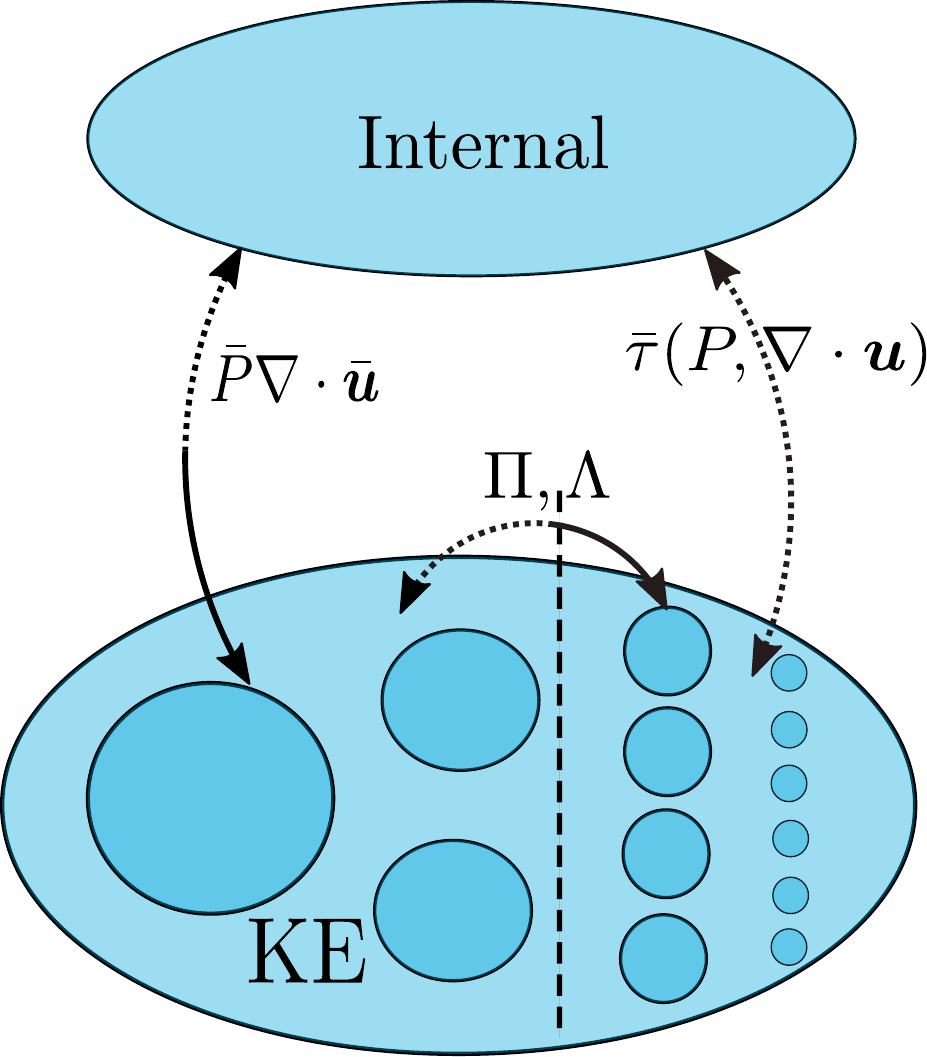}\label{fig:energy_pathways_v1}}
\phantom{}
    \subfigure[\footnotesize{Energy pathways by equation (\ref{eq:KE_budget_another})}]
{\includegraphics[height=2.5in]{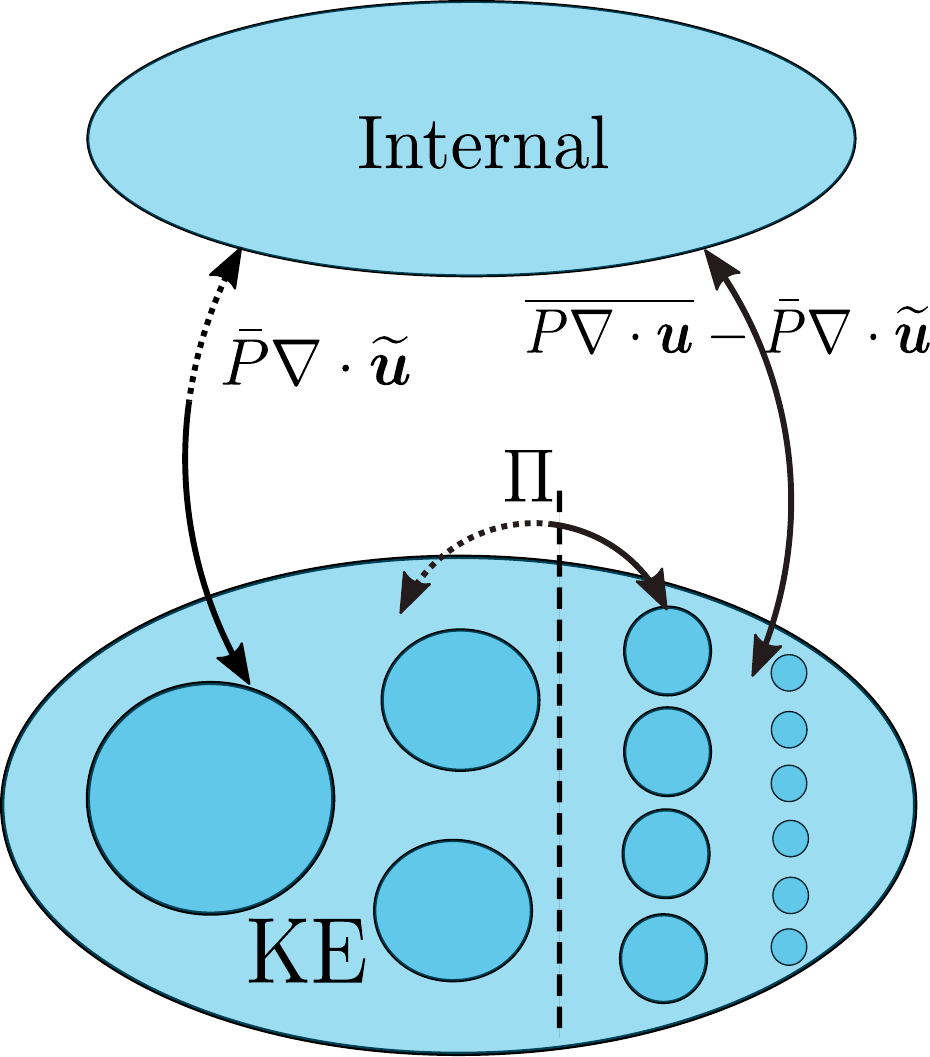}\label{fig:energy_pathways_v2}} 
    \caption{\footnotesize{Two formulations of energy pathways according to equation (\ref{eq:KE_budget}) of the main text and equation (\ref{eq:KE_budget_another}).
    \emph{Kinematically possible} pathways are depicted with dashed arrows, while those that are \emph{dynamically manifested} are depicted with solid arrows.
    (a) Internal energy (IE) is coupled to large scale KE via $\bar{P}\nabla\cdot\bar{\bu}$, and large scale KE transfers to small scale 
    KE via both $\Pi$ and $\Lambda$. At small scales, IE is channeled to KE at the rate $\bar{\tau}(P, \nabla\cdot \bu)$, but this pathway is found to be dynamically inactive (see figures \ref{fig:2D4096_cascade},\ref{fig:3D1024_cascade} and discussion in sections \ref{sec:ke_inertial}-\ref{sec:ke_fluxes} of the main text).
    (b) According to the formulation common in LES, IE is coupled to large scale KE via $\bar{P}\nabla\cdot\widetilde{\bu}$, and large scale KE transfers to small scale KE only via $\Pi$ ($\Lambda$ is absent). At small scales, IE is channeled to KE at the rate $\OL{P\nabla\cdot\bu}-\bar{P}\nabla\cdot \widetilde{\bu}$, which unlike our formulation, is a dynamically significant pathway since its mean equals $ \langle\Lambda\rangle + \langle\bar{\tau}(P, \nabla\cdot \bu)\rangle$ (see equation \eqref{eq:AppLambdaPressDilRel}), incorrectly suggesting that pressure dilatation is important at small scales in RT flows. Other terms such as dissipation and injection are not shown here for clarity. The main difference between the two formulations is that IE is coupled to KE at both large
    and small scales in (b), which can hinder the unraveling of inertial range dynamics.} \label{fig:energy_pathways}}
\end{minipage}
\end{figure}

We remark that these two formulations, equation (\ref{eq:KE_budget})  of the main text and equation (\ref{eq:KE_budget_another}), are mathematically equivalent but represents  different interpretations of the scale processes, which is illustrated schematically in figure \ref{fig:energy_pathways} (in which potential energy 
and dissipation have been omitted for clarity).  In figure \ref{fig:energy_pathways}, there are three channels for energy transfer: i) between internal energy (IE) and large scale kinetic energy (KE),
ii) between IE and small scale KE, and iii) between large and small scale KE. For each channel, the energy transfer is different between
the two interpretations, but the total energy transferred between IE and KE (when summing both large scales and small scales) is the same and equals to $\OL{P \nabla\cdot\bu}$, as can
be readily checked from the sum of the values associated with the arrows in both figures. However, an important distinction is that in the second formulation in figure \ref{fig:energy_pathways_v2},  mean pressure dilatation acts over a much wider range of scales, preventing a simple understanding of the flow in terms of inertial range dynamics.  This is illustrated with solid  arrows in coupling internal energy to KE at scales both large and small, which can hinder the unraveling of inertial range dynamics. We thus advocate for using our formulation in equation (\ref{eq:KE_budget}) of the main text.

\bibliographystyleSupp{jfm}
\bibliographySupp{RT_citation,Compressible_citation,Turbulence_citation,Numerical_citation}

\end{document}